\documentclass[11pt]{article}
\pdfoutput = 1

\usepackage[utf8]{inputenc}
\usepackage{color,graphicx}
\usepackage{epsfig}
\usepackage{amsmath}
\usepackage{verbatim}
\usepackage{amssymb}
\usepackage{mathabx}
\usepackage[mathcal]{eucal}
\usepackage{stmaryrd}
\usepackage{esint}
\usepackage{physics}
\usepackage{amsfonts}
\usepackage{cite}
\usepackage{array}
\usepackage{setspace}
\usepackage{braket}
\usepackage{float}
\usepackage{url}
\usepackage{mathtools}
\usepackage{bbold}
\usepackage{slashed}
\usepackage{tikz}
\usepackage{graphicx}
\usepackage{epstopdf}
\usepackage{subfigure}
\usepackage[labelfont=bf,font={sf}]{caption}
\usepackage{pgfplots}
\usepackage{mathrsfs}
\usepackage{fancybox}
\usepackage{eurosym}
\usepackage{tcolorbox}
\usepackage{tensor}
\allowdisplaybreaks

\usepackage[margin = 2.2cm]{geometry}
\usepackage[ragged]{footmisc}
\usepackage[bookmarks=true,bookmarksnumbered=false,
hyperindex=true,bookmarksopen=true,hyperfigures=true,
colorlinks=true,linkcolor=webblue,citecolor=webgreen,urlcolor=webblue,breaklinks]{hyperref}
\definecolor{webgreen}{rgb}{0, 0.5, 0}
\definecolor{webblue}{rgb}{0, 0, 0.5}
\definecolor{webred}{rgb}{0.5, 0, 0}
\definecolor{darkgreen}{rgb}{0,0.5,0}

\renewcommand{\d}{\mathrm{d}}
\renewcommand{\i}{\mathrm{i}}

\newcommand{\average}[1]{\left\langle #1 \right\rangle}

\def\ben{\begin{equation}}
\def\een{\end{equation}}

\let\a=\alpha \let\b=\beta \let\g=\gamma \let\d=\delta \let\e=\varepsilon
   \let\k=\kappa
\let\l=\lambda     \let\r=v
 \let\t=\tau

\let\w=\omega \let\G=\Gamma

\def\nn{\nonumber}

\def\be{\begin{equation}}
\def\ee{\end{equation}}
\def\ba{\begin{array}}
\def\ea{\end{array}}

\def\mo{\mathcal{O}}

\def\dalemb#1#2{{\vbox{\hrule height .#2pt
       \hbox{\vrule width.#2pt height#1pt \kern#1pt
               \vrule width.#2pt}
       \hrule height.#2pt}}}

\newcommand{\bea}{\begin{eqnarray}}
\newcommand{\eea}{\end{eqnarray}}

\def\md{\mathcal{D}}

\renewcommand{\d}{\mathrm{d}}
\renewcommand{\i}{\mathrm{i}}
\renewcommand{\S}{\textsf{S}_0}

\numberwithin{equation}{section}

\title{Gravity without averaging}

\begin{document}

\thispagestyle{empty}
\begin{center}
    ~\vspace{5mm}

     {\LARGE \bf An integrable road to a perturbative plateau
   }
    
   \vspace{0.4in}
    
    {\bf Andreas Blommaert$^1$, Jorrit Kruthoff$\,^2$ and Shunyu Yao$\,^2$}

    \vspace{0.4in}
    {$^1$SISSA, Via Bonomea 265, 34127 Trieste, Italy\\$^2$Department of Physics, Stanford University, Stanford, CA 94305, USA}
    \vspace{0.1in}
    
    {\tt ablommae@sissa.it, kruthoff@stanford.edu, alfred97@stanford.edu}
\end{center}

\vspace{0.4in}

\begin{abstract}
\noindent As has been known since the 90s, there is an integrable structure underlying two-dimensional gravity theories. Recently, two-dimensional gravity theories have regained an enormous amount of attention, but now in relation with quantum chaos - superficially nothing like integrability. In this paper, we return to the roots and exploit the integrable structure underlying dilaton gravity theories to study a late time, large $e^{S_\text{BH}}$ double scaled limit of the spectral form factor. In this limit, a novel cancellation due to the integrable structure ensures that at each genus $g$ the spectral form factor grows like $T^{2g+1}$, and that the sum over genera converges, realising a perturbative approach to the late-time plateau. Along the way, we clarify various aspects of this integrable structure. In particular, we explain the central role played by ribbon graphs, we discuss intersection theory, and we explain what the relations with dilaton gravity and matrix models are from a more modern holographic perspective.

\end{abstract}

\pagebreak
\setcounter{page}{1}
\tableofcontents

\newpage
\section{Introduction}
Without dynamical gravity, correlation functions measured in black hole backgrounds decay all the way to zero for late times. However, as pointed out initially by Maldacena \cite{Maldacena:2001kr}, in any full theory of quantum gravity this should not be possible. This is because black holes themselves are finite entropy quantum systems.

This predicted behavior for late time correlators was later sharpened by \cite{Cotler:2016fpe}. They considered a toy model of the thermal two point function, known as the spectral form factor ($H$ is the Hamiltonian of the holographically dual quantum system)
\begin{equation}
    Z(\beta+\i T,\beta-\i T)=\Tr(e^{-(\beta+\i T) H})\Tr(e^{-(\beta-\i T) H})=\sum_{i=1}^{\text{dim}(H)}\sum_{j=1}^{\text{dim}(H)}e^{-\beta(E_i+E_j)}e^{-\i T(E_i-E_j)}\,,\label{11}
\end{equation}
and argued that its late-time behavior is universally given by a ramp-and plateau structure (on time scales $\sim e^{S}$ with $S$ the black hole entropy)\footnote{Actually, this simple ramp-and plateau structure is only visible after some time-averaging, or other types of averaging \cite{Cotler:2016fpe}. We will exclusively be concerned with the gravitational interpretation of such smeared quantities in this work.}
\begin{equation}
    Z(\beta+\i T,\beta-\i T)=\int_{0}^{\infty}\d E\,e^{-2\beta E}\,\text{min}(\rho(E),T/2\pi)\,,\quad \rho(E)=e^{S(E)}\,.\label{1.2}
\end{equation}
In particular for $T\to \infty$ this goes to a \emph{non-zero} constant $Z(2\beta)$. That this does not decay all the way to zero indeed follows from the fact that black holes are discrete (or finite entropy) quantum systems: the value $Z(2\beta)$ arises from the terms $i=j$ in the sum \eqref{11}. If we instead had a continuous spectrum those terms would measure zero, so the correlator would instead indeed decay to zero.

It is interesting to ask how the bulk gravitational path integral reproduces this ramp-and plateau structure; one, because they are universal; and two, because the plateau is a signature of microstructure (and hence unitarity) in gravity. In gravity, one computes the spectral form factor by path integrating over all geometries with two asymptotically AdS boundaries (with appropriate boundary conditions implementing the $\beta\pm \i T$). It was found \cite{Saad:2018bqo} that the linear ramp (the $T/2\pi$ piece in \eqref{1.2}) is explained by wormhole geometries connecting both boundaries\footnote{Similarly wormholes were found to be important for understanding late-time correlators \cite{Saad:2019pqd,Blommaert:2019hjr,Iliesiu:2021ari, Kruthoff:2022voq}, the Page curve \cite{Penington:2019kki,Almheiri:2019qdq}, the fate of late-time infalling observers \cite{Stanford:2022fdt} and more.}
\begin{equation}
     Z(\beta+\i T,\beta-\i T)_\text{conn}\supset \quad \begin{tikzpicture}[baseline={([yshift=-.5ex]current bounding box.center)}, scale=0.7]
 \pgftext{\includegraphics[scale=1]{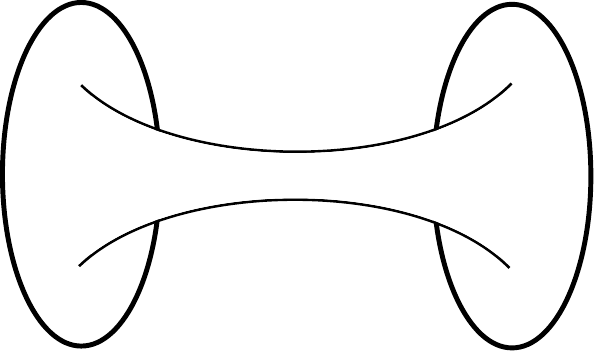}} at (0,0);
     \draw (-0.04, -1.5) node {$1$ wormhole};
     \draw (-2.1, 2.2) node {$\beta+\i T$};
     \draw (2.1, 2.2) node {$\beta-\i T$};
  \end{tikzpicture}\quad.
\end{equation}
The origin of the plateau seems more mysterious, and discussions have been limited largely to scattered comments about D-brane effects in \cite{Saad:2019lba,Blommaert:2019wfy,Altland:2020ccq,Altland:2022xqx}, which lack an obvious truly geometric interpretation.\footnote{In particular, the plateau can be understood as due to a second saddle in a universe field theory description of gravity \cite{Haake:1315494,Altland:2020ccq,Post:2022dfi,Altland:2022xqx,Anous:2020lka}, but this second saddle and the perturbations around it are less geometric, they can not be understood using the gravitational path integral (as far as we know), which describes perturbations around the first saddle.} 

In this work we consider generic models of AdS$_2$ dilaton gravity, of the type studied in \cite{Witten:2020wvy,Maxfield:2020ale,Almheiri:2014cka}
\begin{equation}
    -I=\S\chi +\frac{1}{2}\int \d^2 x\sqrt{g} (\Phi R + W(\Phi))\,.\label{14}
\end{equation}
Extending results of Okuyama-Sakai \cite{Okuyama:2020ncd} (for Airy gravity), and Saad-Shenker-Stanford-Yang-Yao \cite{workinprogressothergroup} (for JT gravity), we show in \textbf{section \ref{sect:eucl}} that in the double scaling limit where $T\to\infty$ and $e^{\S}\to\infty$ with the combination $Te^{-\S}$ held fixed, that the plateau simply follows from the perturbative sum over all genus $g$ wormholes
\begin{equation}
    Z(\beta+\i T,\beta-\i T)_\text{conn}=\sum_{g=0}^\infty e^{-2g\S}\,Z_g(\beta+\i T,\beta-\i T)_\text{conn}\,,
\end{equation}
and \emph{no} non-perturbative (non-geometric) D-brane corrections are required. More precisely, the genus $g$ wormhole amplitude is found to grow \emph{universally} as $T^{2g+1}$
\begin{equation}
    Z_g(\beta+\i T,\beta-\i T)_\text{conn}=\quad \begin{tikzpicture}[baseline={([yshift=-.5ex]current bounding box.center)}, scale=0.7]
 \pgftext{\includegraphics[scale=1]{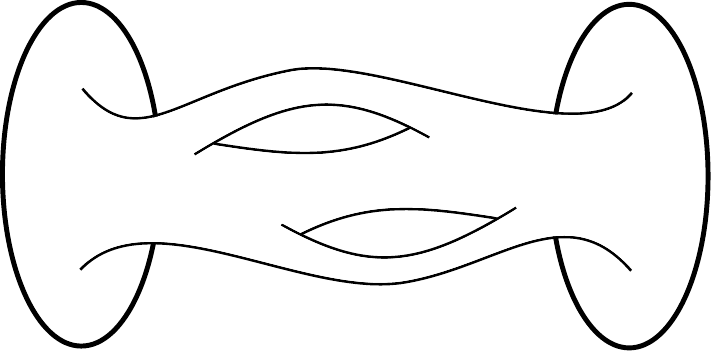}} at (0,0);
     \draw (-0.04, -2) node {genus $g$ wormholes};
     \draw (-2.65, 2.2) node {$\beta+\i T$};
     \draw (2.65, 2.2) node {$\beta-\i T$};
     \draw (0, 0) node {$\dots$};
  \end{tikzpicture}\quad =  P_{g-1}(\beta)\,T^{2g+1}\,,\label{universalseries}
\end{equation}
with $P_{g-1}(\beta)$ a theory-specific degree $g-1$ polynomial. And, the sum over genus reproduces the plateau
\begin{equation}
    \lim_{T\to\infty}\sum_{g=0}^\infty  P_{g-1}(\beta)\,T^{2g+1}e^{-2 g \S}=Z(2\beta)\,.
\end{equation}
For the Airy case of Okuyama-Sakai \cite{Okuyama:2020ncd} this is visible in Fig. \ref{fig:airyconvergenceintro}, other cases are presented in \textbf{appendix \ref{app:a}}.

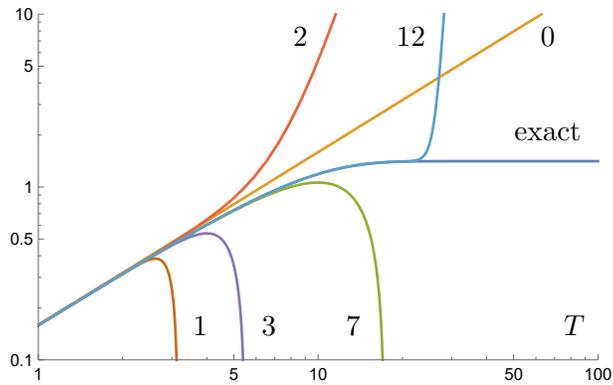
\begin{figure}[t]
    \centering
    \begin{tikzpicture}[baseline={([yshift=-.5ex]current bounding box.center)}, scale=0.7]
 \pgftext{\includegraphics[scale=1]{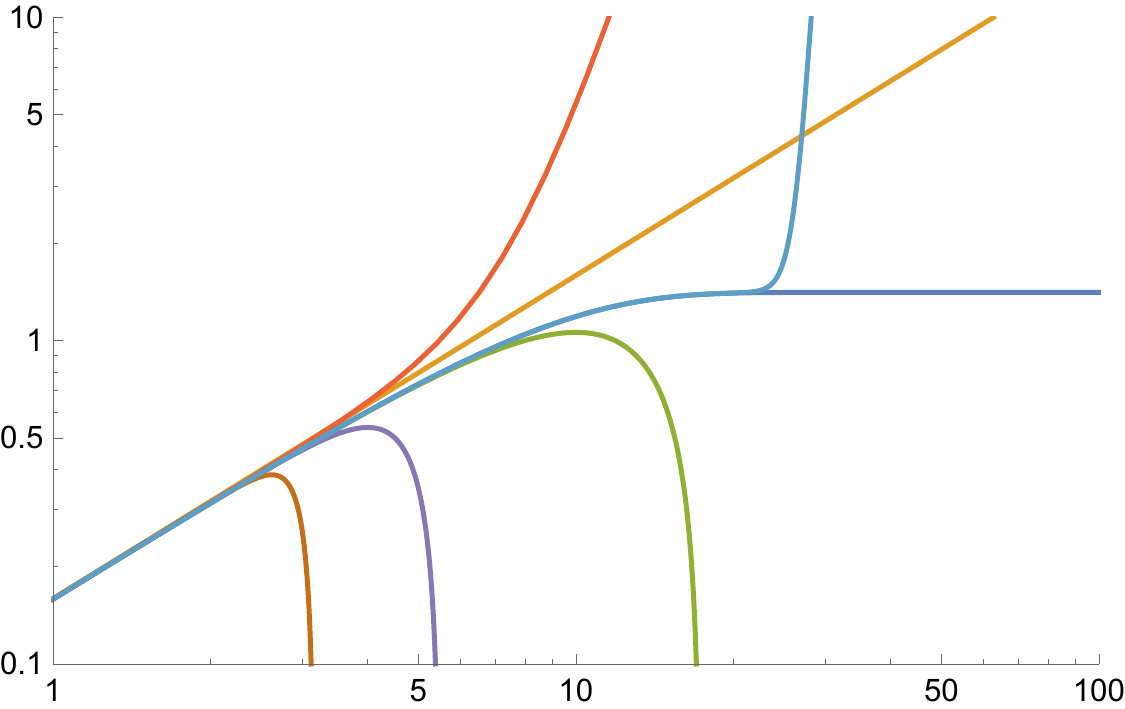}} at (0,0);
     \draw (-0.2, 3) node {$2$};
     \draw (4.5, 3) node {$0$};
     \draw (1.9, 3) node {$12$};
     \draw (-2.1, -2.5) node {$1$};
     \draw (-0.8, -2.5) node {$3$};
     \draw (0.8, -2.5) node {$7$};
     \draw (4.5, 1.2) node {exact};
     \draw (5, -2.5) node {$T$};
  \end{tikzpicture}\quad
    \caption{The double scaling limit of the sum over genus $g$ wormholes in the Airy model, up to $g=g_\text{max}$ (numbers shown) with $e^{\S}=10$ and $\beta=1/2$.}
    \label{fig:airyconvergenceintro}
\end{figure}

The focus of our work is explaining the universal growth $T^{2g+1}$ of the genus $g$ wormhole amplitudes. For fixed theories and fixed $g$ this behavior can be checked manually: genus $g$ amplitudes in the theories \eqref{14} can be computed \cite{Maxfield:2020ale,Witten:2020wvy} using topological recursion \cite{Eynard:2007fi,Eynard:2007kz,Saad:2019lba,Stanford:2019vob}. In particular, topological recursion spits out an even ``volume'' polynomial $V_{g,2}(b_1,b_2)$ of degree $6g-2$. As discussed in \textbf{section \ref{sect:eucl}}, based on dimensional analysis one actually expects a faster growth rate
\begin{equation}
    Z_g(\beta+\i T,\beta-\i T)_\text{conn}\overset{?}{=}\#(g)\,T^{3g}\,.
\end{equation}
The factual case-by-case limitation of the maximal power of $T$ to $T^{2g+1}$ seems miraculous. It depends on very nontrivial cancellations in large sums involving the expansion coefficients of these polynomial volumes $V_{g,2}(b_1,b_2)$. To derive the universality of $T^{2g+1}$ \eqref{universalseries} using the gravitational path integral, we thus need to prove that these cancellations in the volumes happen at any genus $g$, and for any dilaton gravity theory \eqref{14}.

In \textbf{section \ref{sect:intersect}} and \textbf{section \ref{sect:universalcanc}} we derive these cancellations using the relations between our gravity models \eqref{14} and an infinite set of differential equations, known as the KdV hierarchy. In particular, one can use these differential equations to prove \cite{Liu:2007ip,eynard2021natural} some set of cancellations in sums of intersection numbers $\average{\tau_{d_1}\dots \tau_{d_n}}$, which are essentially integrals of products of specific two forms over the moduli space of Riemann surfaces. We then prove that those cancellations in intersection numbers imply the cancellations that we need in the volumes, for all theories \eqref{14}. One key ingredient is a duality between exponentials of operators $\tau_k$ and changes in the dilaton gravity potential, which we find in \textbf{section \ref{sect:interpretation}}
\begin{equation}
    \exp\bigg(\sum_{k=2}^\infty t_k \tau_k\bigg)\quad \Leftrightarrow \quad \exp\bigg(\int\d^2 x\sqrt{g}\sum_{k=2}^\infty \frac{(-1)^k }{(2k-1)!!}\,t_k\,\Phi^{2k}\,e^{-2\pi\Phi} \bigg)\,.
\end{equation}
This can be viewed as an application of the fact that intersection numbers can be viewed as correlators of cusp defects in JT gravity, which we also derive. This also leads to a new understanding of the KdV equations directly in terms of dilaton gravity variables: they express how observables change when one alters certain parameters in the gravitational action.

In \textbf{section \ref{sect:intersect}} we give a gentle introduction to intersection numbers and their relation with gravity, with more intuitive comments gathered in \textbf{appendix \ref{app:c}}, as we did not want to assume that the readers were familiar with these more mathematical constructions.

\subsection*{Relation with other work}
Part of \textbf{section \ref{sect:eucl}} is based upon discussions with Saad, Shenker, Stanford and Yang. In particular, the observation that for Airy and JT gravity equation \eqref{latetime} and \eqref{27} gives the exact spectral form factor in the limit $T\to\infty$ and $e^{\S}\to\infty$ with $Te^{-\S}$ fixed, and that this leads to a series with a non-zero radius of convergence (matched by topological recursion) which gives the plateau \eqref{airyexact} and \eqref{JTexact}, is due to Saad, Stanford, Yang and Yao \cite{workinprogressothergroup}. The fact that equation \eqref{latetime} implies cancellations in Weil-Petersson volumes for JT gravity was independently observed and investigated by \cite{workinprogressregensburg}. 
\section{The plateau from the perturbative sum over wormhole geometries}\label{sect:eucl}

In this section we derive a simple integral representation of the spectral form factor in the limit $T$ and $e^{\S}$ go to infinity keeping their ratio fixed (the $\tau$-scaling limit \cite{Okuyama:2018gfr,Okuyama:2020ncd}), as advertised in the introduction. This integral admits a power series in $T^{2g+1}$, which is reproduced by computing the genus $g$ wormhole amplitudes in gravity. In this $\tau$-scaling limit, the sum over genus converges to the plateau, without the need for non-perturbative (in $e^{\S}$) corrections.

\subsection{Powers of time}\label{sect:folding}
The connected spectral form factor is related with the spectral correlation via a Laplace transform
\begin{equation}
    Z(\beta+\i T,\beta-\i T)_\text{conn}=\int_{-\infty}^{+\infty}\d E_1\int_{-\infty}^{+\infty}\d E_2\,e^{-\beta(E_1+E_2)}e^{\i T(E_1-E_2)}\,\rho(E_1,E_2)_\text{conn}\,.\label{basic}
\end{equation}
We will consider 2d dilaton gravities with an exact dual description as random matrix theories \cite{Saad:2019lba,Witten:2020wvy,Maxfield:2020ale,Mertens:2020hbs}. In the full matrix integral, $\rho(E_1,E_2)_\text{conn}$ is a smooth function of $E_1$ and $E_2$, except for a contact term. In random matrix theory that in the limit $T\to\infty$ and $e^{\S}\to\infty$ with $\tau = Te^{-\S}$ fixed \cite{Haake:1315494,Okuyama:2018gfr,Okuyama:2020ncd,workinprogressothergroup} ($\tau$-scaling limit), this integral simplifies to
\begin{equation}
    Z(\beta+\i T,\beta-\i T)_\text{conn}=\int_{-\infty}^{+\infty}d E_1\int_{-\infty}^{+\infty}d E_2\,e^{-\beta(E_1+E_2)}e^{i T(E_1-E_2)}\,\rho(E_1,E_2)_\text{conn eff}\,,\label{latetime}
\end{equation}
where the effective spectral correlation features the sine kernel \cite{mehta2004random}\footnote{The generic proof of this fact goes through Efetov's non-linear sigma model description of random matrix theory \cite{efetov1983supersymmetry,Haake:1315494}, see also recently \cite{Altland:2020ccq,Belin:2021ibv,Altland:2022xqx,Blommaert:2021fob}. We can alternatively use D-brane calculus as explained in \cite{Saad:2019lba,Blommaert:2019wfy}.}
\begin{equation}
    \rho(E_1,E_2)_\text{conn eff}=\delta(E_1-E_2)\rho_0(E)-\frac{\sin^2(\pi\rho_0(E)(E_1-E_2))}{\pi^2(E_1-E_2)^2}\,.\label{sinekernel}
\end{equation}
Changing variables to $E_1-E_2=\omega$ and $E_1+E_2=2E$ this becomes
\begin{equation}
    Z(\beta+\i T,\beta-\i T)_\text{conn}=\int_{-\infty}^{+\infty}\d E\,e^{-2\beta E}\,\rho_0(E)-\int_{-\infty}^{+\infty}\d E\,e^{-2\beta E}\int_{-\infty}^{+\infty}\d\omega\, e^{\i T \omega}\,\frac{\sin^2(\pi \rho_0(E)\omega)}{\pi^2\omega^2}\,.\label{26}
\end{equation}
The $\omega$-integral is a standard Fourier-transform, which gives the familiar ramp-and plateau \cite{Cotler:2016fpe}
\begin{align}
    Z(\beta+\i T,\beta-\i T)_\text{conn}&=\int_{-\infty}^{+\infty}\d E\,e^{-2\beta E}\,\text{min}(\rho_0(E),T/2\pi)\label{27}\\&=\int_{-\infty}^{+\infty}\d E\,e^{-2\beta E}\,\rho_0(E)-\int_{E(T)}^{+\infty}d E\,e^{-2\beta E}\,(\rho_0(E)-T/2\pi)\,,\quad \rho_0(E(T))=T/2\pi\,,\nonumber
\end{align}
with $E(T)$ determined by solving $\rho_0(E(T)) = T/2\pi$.\footnote{Here we assume $\rho_0(E)$ grows monotonically, in order to have a unique solution. The conclusion \eqref{expaexpa} remains true for non-monotonic spectra though, see appendix \ref{sect:non-monotonic}.} Choosing the energy axis such that $\rho_0(0)=0$,\footnote{Some dilaton gravities have a non-zero threshold energy, but such modifications are straightforward to incorporate.} and using integration by parts, we arrive at
\begin{align}
    Z(\beta+\i T,\beta-\i T)_\text{conn}&=\int_{0}^{E(T)}\d E\,e^{-2\beta E}\,\rho_0(E)+\frac{1}{2\beta}\rho_0(E(T))e^{-2\beta E(T)}\nn\\&=\frac{1}{2\beta}\int_0^{T/2\pi}\d\rho_0\,e^{-2\beta E(\rho_0)}\,.\label{28}
\end{align}
This is the final formula for the spectral form factor in the $\tau$-scaling limit. One might say that we have put non-perturbative information into the calculation at this point, but here we just use the matrix model answers. The objective is to reproduce the final formula from gravity and explain its convergence properties, solely using perturbation theory.

As already alluded to, we consider 2d dilaton gravity \eqref{14} with a matrix integral dual and for these theories the spectrum has an expansion in powers of $E^{k+1/2}$ \cite{brezin1993exactly,douglas1990strings,Gross1989nonperturbative},
\begin{align}\label{spectrum}
    \rho_0(E) = \frac{e^{\S}}{2\pi}\sum_{k=0}^\infty f_{k}\,E^{1/2+k} =  \frac{e^{\S}}{2\pi}\,\bigg(w+\sum_{k=1}^{\infty}f_k\,w^{2k+1}\bigg)\,,
\end{align}
where we've introduced the notation $E = w^2$, and without loss of generality we have set $f_0 = 1$ (by changing $\S$). The function in round brackets will be referred to as $f(w)$ and its inverse (which exits by assumption) can be obtained using the Lagrange inversion theorem. Notably, this has a powers series expansion in odd powers of $f$ too
\begin{equation}
    w(f)=f+\sum_{k=1}^{\infty}w_k\,f^{2k+1}\,.
\end{equation}
and as a result the Taylor series
\begin{equation}
    e^{-2\beta w(f)^2}=1-2\beta f^2+4\pi\beta \sum_{n=2}^\infty (2n+1) P_{n-1}(\beta)\, f^{2 n}\,, 
\end{equation}
is even in $f$. Here $P_n(\beta)$ is a degree $n$ polynomial in $\beta$, which can easily be computed explicitly for any fixed $n$. The tau-scaled spectral form factor \eqref{28} then expands as
\begin{align}
    Z(\beta+\i T,\beta-\i T)_\text{conn}&=\frac{e^{\S}}{4\pi\beta}\int_0^{Te^{-\S}}\d f\,e^{-2\beta E(f)}\nonumber\\&=\sum_{g=0}^\infty  P_{g-1}(\beta)\,T^{2g+1}e^{-2g\S}=\frac{T}{4\pi\beta}-\frac{1}{6\pi}T^3e^{-2\S}+\dots\label{expaexpa}
\end{align}
This is the main result of this section. Computing the tau-scaled limit of the (connected) spectral form factor using the non-perturbatively exact matrix integral formulation of 2d dilaton gravities results in a \emph{universal} series expansion in $T^{2g+1}e^{-2g\S}$. It is tempting to interpret the term at order $e^{-2g\S}$ as the $\tau$-scaling limit of the perturbative genus $g$ wormhole amplitude in gravity. We will confirm below in section \eqref{sect:euclideanwormholes} that this is indeed the case. Let us point out some noteworthy features of this expansion.
\begin{enumerate}
    \item In the limit $T\to\infty$ this series approaches the (leading order) plateau
    \begin{equation}
        \frac{e^{\S}}{4\pi\beta}\int_0^{\infty}\d f\,e^{-2\beta w(f)^2}=\int_0^\infty \d E\,\rho_0(E)\,e^{-2\beta E}=Z_0(2\beta)\,.
    \end{equation}
    Depending on the case, the series expansion can have a finite or an infinite radius of convergence (as function of $T$). However in all cases (with invertible spectrum) the series converges to the exact answer for small enough $T$, and that answer has a unique analytic continuation to all positive $T$. The examples in appendix \ref{app:a} should clarify this. 
    
    So, in the $\tau$-scaling limit, the plateau is \emph{perturbatively} accessible in the genus expansion. Notice also that this is different from Borel resummation, since we are taking a limit where some badly growing terms in the genus $g$ go away.
    \item The first two terms in the expansion \eqref{expaexpa} are theory-independent, meaning they do not depend on the $f_k$. The polynomial $P_n(\beta)$ depends only on $f_2\dots f_n$, so the higher the genus the more we probe the UV part of the spectrum. The sign of each term depends on these $f_k$ (and $\beta$) and are not-universal. 
    \item Theories with a nonzero Hagedorn temperature never reach a plateau when $2\beta<\beta_\text{H}$, because for these cases the would-be plateau $Z_0(2\beta)$ is divergent. We can appreciate this also using
    \be \label{simple}
    \partial_T Z(\beta+\i T,\beta-\i T)_\text{conn} = \frac{1}{4\pi\b}e^{-2\b E(Te^{-\S})}
    \ee
    For Hagedorn spectra $f(E)\sim e^{\beta_\text{H}E}$ one finds $E(f)=\log(f)/\beta_\text{H}+\rm{ constant}$, and hence indeed the $\tau$-scaled spectral form factor grows without bounds for late times. This Hagedorn growth at high energies only occurs for non-local theories, such as string theories.
    \item The sine kernel \eqref{sinekernel} or the associated level repulsion is generally considered to be the hallmark feature of chaotic quantum systems, it is essentially synonymous with random matrix universality. The scaling $T^{2g+1}$ contains the same information as this sine kernel, and we consider it therefore to be the real-time version of random matrix universality. Therefore these should be an argument why in any gravity model (beyond 2d dilaton gravity) there are contributions growing like $T^{2g+1}$ in the gravitational path integral. This explanation we think should be intrinsically Lorentzian. 
    A proposal based on topology changing processes for how these powers of time can be explained, analogous to the double-cone \cite{Saad:2018bqo}, will be presented elsewhere \cite{workinprogress}.
\end{enumerate}

We further discuss this integral and its series expansion \eqref{expaexpa} for several examples, as well as the modifications for non-monotonic spectra, in appendix \ref{app:a}, in order not to disrupt the flow of the paper.
\subsection{Euclidean wormholes and cancellations in volumes}\label{sect:euclideanwormholes}
We now switch gears, and move to the gravitational computation. For this we do an expansion of the connected two-boundary observable in genus
\begin{equation}
    Z(\beta+\i T,\beta-\i T)_\text{conn}=\sum_{g=0}^\infty e^{-2g\S}\,Z_g(\beta+\i T,\beta-\i T)_\text{conn}\,\overset{?}{+}\text{non-perturbative}\,,
\end{equation}
where the genus $g$ gravitational wormhole amplitude is computed as
\begin{equation}
    Z_g(\beta_1,\beta_2)_\text{conn}=\int_0^\infty \d b_1 b_1\, \frac{1}{2\pi^{1/2}{\beta_1}^{1/2}}e^{-\frac{b_1^2}{4\beta_1}}\int_0^\infty \d b_2 b_2\, \frac{1}{2\pi^{1/2}{\beta_2}^{1/2}}e^{-\frac{b_2^2}{4\beta_2}}\,V_{g,2}(b_1,b_2)\,,\label{genusg}
\end{equation}
with $V_{g,2}(b_1,b_2)$ the (deformed) Weil-Petersson volume\footnote{Here we mean a slight generalisation of the Weil-Petersson volumes, which includes cases where we have summed over defects \cite{Maxfield:2020ale, Witten:2020wvy}. One might call these deformed Weil-Petersson volumes.} for a genus $g$ wormhole with (geodesic) boundaries with length $b_1$ and $b_2$. They are symmetric polynomial in $b_1^2$ and $b_2^2$ with a maximum total degree of $3g-1$. The way to compute these polynomials $V_{g,2}(b_1,b_2)$ in practice is using the Eynard-Orantin topological recursion \cite{Eynard:2004mh,Eynard:2007fi,Stanford:2019vob,Saad:2019lba}, with a spectral curve $y(z)$ (originating from \eqref{spectrum}) given by\footnote{Only for the spectral curve $y(z) = \sin(2\pi z)/(4\pi)$ do we obtain the true Weil-Petersson volumes.} 
\begin{equation}
    y(z)=\frac{1}{2}\sum_{k=0}^\infty (-1)^k f_{k}\,z^{1+2 k}\,.\label{speccurve}
\end{equation}
We stress that these polynomials are the result of doing the gravitational path integral over all metrics which are topologically a genus $g$ connected wormhole in the dilaton gravity with genus zero spectrum \eqref{spectrum}. 
In order to compare \eqref{genusg} to \eqref{expaexpa}, we compute \eqref{genusg} explicitly by writing the WP volume as
\begin{equation}
    V_{g,2}(b_1,b_2)=\sum_{d_1,d_2=0}^{d_1+d_2=3g-1}V_{g,2}^{d_1,d_2}\frac{b_1^{2d_1}}{4^{d_1}d_1!}\frac{b_2^{2d_2}}{4^{d_2}d_2!}\,,\label{vexp}
\end{equation}
with some symmetric constants $V_{g,2}^{d_1,d_2}$. We obtain
\begin{equation}
    Z_g(\beta_1,\beta_2)_\text{conn}=\frac{1}{\pi}\sum_{d_1,d_2=0}^{d_1+d_2=3g-1}V_{g,2}^{d_1,d_2}\,\beta_1^{1/2+d_1}\beta_2^{1/2+d_2}\,.\label{222}
\end{equation}
Continuing to Lorentzian signature and putting $\beta=0$ (see below for finite $\beta$) this becomes
\begin{equation}
    Z_g(\i T,-\i T)_\text{conn}=\frac{1}{\pi}\sum_{q=0}^{(3g-1)/2}(-1)^q\,T^{2 q+1}\sum_{d=0}^{2q}(-1)^d\,V_{g,2}^{d,2q-d}\,,\label{223}
\end{equation}
Comparing to \eqref{expaexpa}, we claim that in the $\tau$-scaling limit for generic dilaton gravities we should have
\begin{equation}
    Z_g(\i T,-\i T)_\text{conn}=\,P_{g-1}(0)\,T^{2g+1}\,.
\end{equation}
This is rather surprising because in \eqref{223} we have powers of $T$ that are larger than $2g+1$, and hence should dominate the $\tau$-scaling limit. This indicates a novel cancellation between the various coefficients of the volumes $V_{g,2}(b_1,b_2)$! More precisely we are claiming that for \emph{any} theory
\begin{equation}
     \sum_{d=0}^{2q}(-1)^d\,V_{g,2}^{d,2q-d}=0\,,\quad q>g\,,\quad \sum_{d=0}^{2g}(-1)^d\,V_{g,2}^{d,2g-d}=\pi (-1)^g P_{g-1}(0)\,.\label{simple}
\end{equation}
These cancellations for $q>g$ are quite surprising from the point of view of these polynomials, but they are nevertheless true, as one can check case by case. The simplest example is the genus $g=3$ wormhole with $q=4$. Based on dimensional analysis, one expects a term proportional to $T^9$. However, this term is absent because of the cancellation\footnote{For a list of volumes see for instance \cite{van2008intersection}.}
\begin{equation}
    \frac{1}{2}\sum_{d=0}^{8}(-1)^d\,V_{3,2}^{d,8-d}=\frac{1}{324}-\frac{5}{324}+\frac{77}{1620}-\frac{503}{5670}+\frac{607}{11340}=0\,.\label{cancelbasic}
\end{equation}
This example is theory independent since it only depends on $f_1=1$, but we stress that more in general the $V_{g,2}^{d_1,d_2}$ depend on all $f_k$ such that \eqref{expaexpa} predicts theory-dependent cancellations. One checks easily case by case that cancellations indeed occur, more examples are provided in appendix \ref{app:a}.

To compute the $\tau$-scaled spectral form factor for finite $\beta$, we can introduce symmetric polynomials $e_2=\beta_1\beta_2$ and $e_1=\beta_1+\beta_2$, and reorganize \eqref{222} into
\begin{align}
    Z_g(\beta+\i T,\beta-\i T)_\text{conn}=&\frac{1}{\pi}\sum_{q=0}^{(3g-1)/2}(-1)^q\,e_2^{ q+1/2}\sum_{m=2q}^{3g-1}e_1^{m-2q}\nonumber\\&\qquad \qquad 2\sum_{d=0}^q(-1)^d\,V_{g,2}^{d,m-d}\,\frac{m/2-d}{m-q-d}\frac{(m-q-d)!}{(q-d)!(m-2q)!}\,,\label{z2bdyexp}
\end{align}
where now $e_2=\beta^2+T^2$ and $e_1=2\beta$.\footnote{For the term $m=2q$ the term with $d=q$ receives an extra $1/2$, which we left implicit.} The constraint that this amplitude grows no faster than $T^{2g+1}$ imposes that the coefficient of $e_2^{q+1/2}$ vanishes for $q>g$
\begin{equation}
    \sum_{d=0}^q(-1)^d\,V_{g,2}^{d,m-d}\,\frac{m/2-d}{m-q-d}\frac{(m-q-d)!}{(q-d)!(m-2q)!}=0\,,\quad q>g\,,m\geq 2q\,,\label{genericcancel}
\end{equation}
and furthermore there is a precise theory-\emph{dependent} prediction for $q=g$
\begin{equation}
    2\sum_{m=2g}^{3g-1}(2\beta)^{m-2q}\sum_{d=0}^g(-1)^d\,V_{g,2}^{d,m-d}\,\frac{m/2-d}{m-g-d}\frac{(m-g-d)!}{(g-d)!(m-2g)!}=\pi (-1)^g P_{g-1}(\beta)\,.
\end{equation}
The case $m=2q$ reduces to the $\beta=0$ constraints \eqref{simple} (taking into account the factor $1/2$ mentioned above). The constraints with $m>2q$ represent additional cancellations which we claim are satisfied by the genus $g$ wormhole amplitudes of \emph{all} dilaton gravities (with matrix integral duals).

A happy consequence of the $\tau$-scaling limit is that all terms with $q < g$ are subleading. These terms contain lower powers of $T$, but their coefficients are growing more and more rapidly the lower the power of $T$, ultimately growing like $(2g)!$. This was conjectured in \cite{zograf2008large} and proven in \cite{mirzakhani2015towards}, see also (228) in \cite{Saad:2019lba} as Taylor series in $b^2$. We make some further comments on this in section \ref{sect:5.4}. Including such terms makes the sum over genus very complicated and one needs to resort to a Borel resummation in order to study it. The $\tau$-scaling limit thus gets rid of these intricacies, by selecting only the $q=g$ term at each genus. This is why the series is convergent, in part, and non-perturbative corrections are \emph{not} present.

We can also understand this from Efetov's non-linear sigma model \cite{Haake:1315494,Altland:2020ccq,Altland:2022xqx,efetov1983supersymmetry} (which described the $\tau$-scaling limit) formulation of these quantities. The resolvent is represented by a double integral over variables $s_{11}$ and $s_{22}$. The energy dependence in the action is $e^{\S}E(s_{11}-s_{22})$, and the double integral is dominated by two saddle-points. The expansion around the first saddle grows like $(2g)!$ in the energy domain, whereas the expansion around the second saddle gives a non-perturbative (in $e^{\S}$) correction. We can now inverse Laplace transform this to get a non-linear sigma model computation of $Z(\beta)$. The energy integral gives a delta function $\delta(\beta-e^{\S}(s_{11}-s_{22}))$, which collapses the double integral to a single integral. This single integral, notably, has just one saddle point, expanding around which produces the sum over wormholes. Because there is only one saddle, there are no non-perturbative corrections in the thermal ensemble. This calculation extends to the connected spectral form factor, it becomes a bit messier but the conclusion remains the same.

In summary, there are two reasons why this genus expansion is convergent: the $\tau$-scaling limit gets rid of terms that grow too fast, and the genus expansion in the thermal ensemble is more convergent than that in the microcanonical ensemble.\footnote{One other way to see that perturbation theory in the microcanonical ensemble is more complicated is because the integrals over $b$ with the density of states of the trumpet are not convergent and need to be defined using analytic continuation. This is true in the Airy case, in JT an additional complication is that the volumes themselves contain pieces that grow like $(2g)!$. The $\tau$-scaling limit cures the second, but not the first complication, which is cured by going to the canonical ensemble.}


\section{Cancellations in topological gravity}\label{sect:intersect}

In the previous section we found that there need to be non-trivial cancellations between the coefficients of (deformed) Weil-Petersson volumes. In this section and the next we explain how these relations come about. First we will consider topological gravity, i.e. the Airy model, where many of these cancellations are known \cite{Liu13896,eynard2021natural} in the context of intersection theory. In section \ref{sect:universalcanc} we will use open-closed duality to show that these very cancellations of \cite{Liu13896,eynard2021natural} actually imply the growth $T^{2g+1}e^{-2g\S}$ for all double-scaled matrix integrals, including all models of dilaton gravity \cite{Witten:2020wvy,Maxfield:2020ale,Mertens:2020hbs,Blommaert:2021fob,Blommaert:2021gha,Blommaert:2022ucs}.

We remind the reader that this $T^{2g+1}e^{-2\S}$ is the type of series that we found earlier will converge to the plateau. So, in part, these cancellations explain why the sum over genus $g$ wormholes converges (in the $\t$-scaling limit) to the plateau.

\subsection{Ribbon graphs and intersection numbers}\label{sect:3.1}

In 1991, Witten \cite{Witten:1990hr} showed that topological gravity is related to a matrix model with spectral density 
\begin{equation}
    \rho_0(E)=e^{\S}\frac{E^{1/2}}{2\pi}. \label{topgravspec}
\end{equation}
This spectral density occurs for any dilaton gravity at small enough energy, and is thus an important first case to consider. 

Small energy means that we are at small temperature and so huge asymptotic boundaries. Since the partition functions $Z(\b_1,\dots,\b_n)$ are given by integrals over the (deformed) Weil-Petersson volumes \eqref{genusg}, one might wonder what large $\b_i$ implies for those volumes. It is straightforward to see from the trumpet partition function \cite{Saad:2019lba}
\be 
Z_{\text{trumpet}}(\b,b) = \frac{1}{2\pi^{1/2}\beta^{1/2}}e^{-\frac{b^2}{4\beta}}\,,
\ee
that we now get contributions mostly from the region where $b^2 \sim \b$, hence $b^2$ is large. In other words, we can consider the leading large $b_i$ terms in the (deformed) WP volumes. These are theory-independent terms which have a have a clear geometric interpretation, as we now point out, see also Appendix \ref{app:c4}. 

\begin{figure}[t]
    \centering
    \includegraphics[scale=0.6]{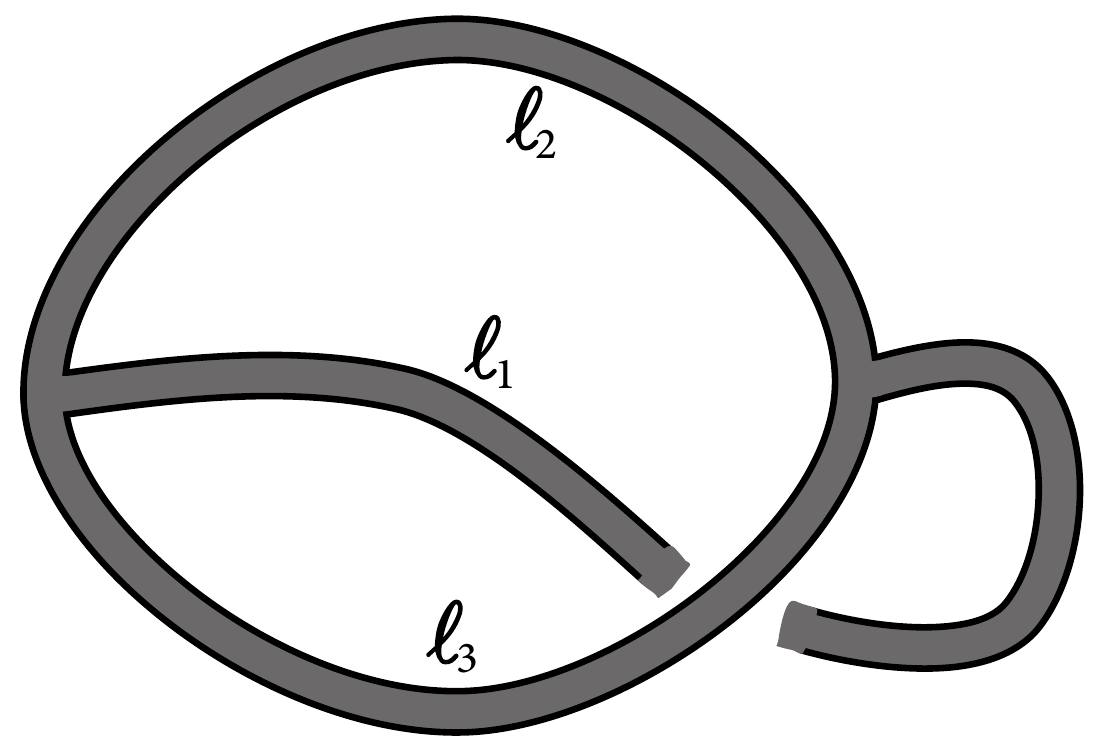}
    \caption{The torus with one puncture $V_{1,1}(b)$ becomes a ribbon graph, when the hole $b$ becomes large. In this limit the moduli space is spanned by the lengths of the ribbons with the constraint $b=2\ell_1+2\ell_2+2\ell_3$, and we integrate with the flat measure to recover $V_{1,1}(b)=b^2/48$.}
    \label{fig:ribbon}
\end{figure}

Since they are theory independent, we can start by considering the undeformed WP volumes, which compute the volumes of moduli spaces of hyperbolic Riemann surfaces with $R=-2$. As the boundaries of these Riemann surfaces are $K=0$ geodesics we can use Gauss-Bonnet to show that the area of such Riemann surfaces of genus $g$ and $n$ geodesic boundary is given by 
\be 
A = 2\pi (2g+n - 2)\,,\label{3.3}
\ee
and hence is a \emph{constant}. This means that when we take the $b_i$ very large, the Riemann surface needs to become a collection of very thin strips glued along trivalent vertices (trivalent because we can consider the pair of pants decomposition of the Riemann surface). In other words, the (deformed or undeformed) WP volumes reduce to the volume of moduli space of trivalent ribbon graphs, see also Fig. \ref{fig:ribbon}.

These volumes are simple to compute, one can parameterize the moduli space of ribbon graphs for a certain graph $\G$ by the lengths $\ell_j$ of the edges of the graph, those lengths are constrained only by the fact that the sum of the lengths $\ell_{j_i}$ of the edges forming a certain boundary $i$ add up to $b_i$. Accounting for the standard symmetry factor in Feynman diagrams one thus obtains
\begin{equation}
    V_{\G_{g,n}} (b_1\dots b_n)=\frac{1}{\abs{\text{Aut}(\G_{g,n})}}\prod_{j=1}^{6g-6+3n}\int_0^\infty \d \ell_j\,\prod_{i=1}^n \delta(b_i-\sum_{j_i}\ell_{j_i})\,.\label{3.4}
\end{equation}
Summing over all ${\G_{g,n}}$ with a certain genus $g$ can then indeed be checked to reproduce the Airy volumes. But how is this related to intersection numbers? The crux of Kontsevich's seminal work \cite{KontsevichModel} was the realization that these integrals can be represented equivalently as (see also \cite{Dijkgraaf:2018vnm,Okuyama:2019xbv})
\begin{equation}
    V_{g,n}(b_1\dots b_n)=\sum_{\G_{g,n}}V_{\G_{g,n}} (b_1\dots b_n)=\int_{\overline{\mathcal{M}}_{g,n}}\exp\bigg(\frac{1}{2}\sum_{i=1}^n b_i^2\, \psi_i\bigg)\,.\label{43}
\end{equation}
The right hand side is in integral over the moduli space of Riemann surfaces. Just like for the ordinary (deformed) WP volumes we include degenerate Riemann surfaces, which results in the so-called Deligne-Mumford compactification of the moduli space of Riemann surfaces $\overline{\mathcal{M}}_{g,n}$. Here the $\psi_i$ denote the first Chern classes $c_1(\mathcal{L}_i)$ of some bundles of one forms $\mathcal{L}_i$ that are constructed as follows, see also appendix \ref{app:c}. Take a punctured Riemann surface and consider the cotangent space at each puncture $x = x_i$. For each $x_i$ these cotangent spaces depend on the moduli of the Riemann surface under consideration. The collection of all those spaces is a bundle $\mathcal{L}_i$ over $\overline{\mathcal{M}}_{g,n}$. 

The key to proving the relation \eqref{43} is to realize that one can choose a specific one form $\alpha$ in $\mathcal{L}_i$ for which the curvature two-form $\d\alpha=c_1(\mathcal{L}_i)$ takes on a simple form \cite{KontsevichModel}. The first step is to use the fact that there is an equivalence between the moduli space of Riemann surfaces and the moduli spaces of ribbon graphs \cite{penner1987decorated,harer1988cohomology,strebel1984quadratic}. Concretely, one can associate to every Riemann surface with $n$ boundaries of lengths $b_i$ a unique ribbon graph with lengths $\ell_j$ with again the sum of $\ell_{j_i}$ constrained to $b_i$. This map is provided by the Jenkins-Strebel quadratic differential \cite{strebel1984quadratic}. So we can exchange the fundamental domain in terms of the Teichmuller coordinates $b_i$ and $\tau_i$ for a sum over ribbon graphs with the simple constraint that the sum of lengths $\ell_{j_i}$ is constrained to $b_i$.\footnote{It is quite remarkable that the moduli space of Riemann surfaces is so simple in these coordinates $\ell_j$. One can think of $\ell_j$ as the propagation times of open strings, and this simplicity of moduli space is roughly why open string field theory is simpler than closed string field theory.} In these coordinates, Kontsevich found a local expression for a one form $\alpha$ whose curvature is constant\footnote{We are suppressing some factors of two and minus signs in the sum, associated for instance with cases where one edges contributing twice to $b_i$, see theorem 3.20 in \cite{do2008intersection}. These are not important for this intuitive argument.}
\begin{equation}
    \psi_i=\d\alpha=c_1(\mathcal{L}_i)=\frac{2}{b_i^2}\sum_{j<k}\d \ell_{j_i}\wedge \d \ell_{k_i}\,.\label{3.6}
\end{equation}
With this equation it is not hard to imagine that writing out the exponential in \eqref{43} generates simply the flat measure in \eqref{3.4}. The actual proof still involves some combinatorics \cite{KontsevichModel}, but the point should be obvious. We want to emphasize that all this aside, the $\psi_i$ are simply two forms on $\overline{\mathcal{M}}_{g,n}$. In terms of Teichmuller coordinates $b_i$ and $\tau_i$ they have rather complicated expressions, but in the $\ell_j$ coordinates they are constants.\footnote{To avoid confusion, the Weil-Petersson measure $\d b_i\wedge \d\tau_i$ becomes only flat in the $\ell_j$ coordinates for large $b_i$ \cite{do2008intersection}. So in general one can view Weil-Petersson volumes as integrating over the moduli space of ribbon graphs, but with a non-flat measure. The integration domain is the same \cite{penner1987decorated,harer1988cohomology,strebel1984quadratic}, but the integrand is different.}

Introducing common notation for the $2k$-forms (the $k$'th power of the Chern classes)
\begin{equation}
    \tau_k=\psi_i^k\,,
\end{equation}
and writing out the exponentials in \eqref{43} we arrive at the relation between (deformed) WP volumes for large $b_i$ and so-called intersection numbers $\average{\tau_{d_1}\dots \tau_{d_n}}$ (for more about this name see appendix \ref{app:c})
\begin{equation}
    V_{g,n}(b_1\dots b_n)=\sum_{d_1=0}^\infty\frac{b_1^{2d_1}}{2^{d_1}d_1!}\dots\sum_{d_n=0}^\infty\frac{b_n^{2d_n}}{2^{d_n}d_n!}\average{\tau_{d_1}\dots \tau_{d_n}}_g\,,\quad \average{\tau_{d_1}\dots \tau_{d_n}}_g=\int_{\overline{\mathcal{M}}_{g,n}}\psi_1^{d_1}\dots\psi_n^{d_n}\,.\label{expvolumes}
\end{equation}
Because first Chern classes $\psi_i$ are two-forms, and the dimension of the space $\overline{\mathcal{M}}_{g,n}$ is $6g-6+2n$, there is an obvious selection rule
\begin{equation}
    \sum_{i=1}^n d_i=3g-3+n\,,\label{selection}
\end{equation}
and from the definition, the correlators $\average{\tau_{d_1}\dots \tau_{d_n}}$ are symmetric under exchanging any two labels. In terms of the $n$-boundary finite temperature partition function we then obtain
\begin{align}
    Z(\beta_1\dots \beta_n)_\text{conn}&=\frac{\beta_1^{1/2}}{\pi^{1/2}}\dots\frac{\beta_n^{1/2}}{\pi^{1/2}}\sum_{g=0}^\infty e^{-2g\S} \sum_{d_1=0}^\infty (2\beta_1)^{d_1}\dots \sum_{d_2=0}^\infty (2\beta_n)^{d_n}\average{\tau_{d_1}\dots \tau_{d_n}}_g\nonumber\\&=\frac{x_1^{1/2}}{(2\pi)^{1/2}}\dots\frac{x_n^{1/2}}{(2\pi)^{1/2}}\,\mathcal{F}(x_1\dots x_n)\,,\quad x_i=2\beta_i e^{-2\S/3}\,,\label{zfromtau}
\end{align}
with the generating functional
\begin{equation}
    \mathcal{F}(x_1\dots x_n)=\sum_{d_1=0}^\infty x_1^{d_1}\dots \sum_{d_n=0}^\infty x_n^{d_n}\average{\tau_{d_1}\dots \tau_{d_n}}\,.\label{58}
\end{equation}
By taking the large $b_i$ limit in Mirzakhani's recursion relations \cite{mirzakhani2007simple,Stanford:2019vob} (either in the answer or in the derivation) one obtains a simple version of topological recursion for ribbon graphs, see appendix \ref{app:c5}. This can be translated to simple recursion relations for intersection numbers $\average{\tau_{d_1}\dots \tau_{d_n}}$, which are the expansion coefficients of ribbon graph volumes \eqref{expvolumes}. One finds eventually equation (25) in \cite{faber2000logarithmic}\footnote{This is quite teadious. We can rewrite the recursion relations of volumes in terms of recursion relations for intersection numbers as in \cite{mulase2006mirzakhani}, but now using the volumes of ribbon graphs and without the $\kappa$ two-forms. What one obtains are the Dijkgraaf-Verlinde$^2$ \cite{dijkgraaf1991topological} version of the recursion relations, equation (7.27) in \cite{Dijkgraaf:1991qh} but with appropriate normalization. These recursion relations are not manifestly the same as the once we wrote above, they correspond with the Virasoro constraint equations in the KdV formalism, whereas our equations above literally follow from the KdV equations themselves. That these two infinite sets of differential equations (and correspondingly, two infinite sets of recursion equations) are equivalent was proven again in \cite{Dijkgraaf:1991qh}, see also \cite{fukuma1993continuum,Witten:1991mn}. We explain these steps more carefully in section \ref{sect:KdV}.}$^,$\footnote{The labels $d_j$ fix the genus $g$ via the selection rule \eqref{selection} and should be chosen such that $g$ is a positive integer (otherwise the correlator trivially vanishes). Applied to this case the selection rule reads
\begin{equation}
    \sum_{j=0}^\infty (j-1)d_j+(k-1)=3g-1\,.
\end{equation}
Similarly the correlators on the right side only get contributions from values $a_j$ and $b_j$ such that $g_1$ ad $g_2$ are non-negative integers. We automatically have $g_1+g_2=g$, as one checks using the selection rules. There are no $e^{\S}$ in these equations, these are just statements about integrals of $2j$ forms over some symplectic manifold.
}
\begin{align}
    &(2k+1)\bigg\langle \tau_0^2\tau_k\prod_{j=0}^\infty \tau_j^{d_j}\bigg\rangle_g=\prod_{j=0}^\infty\sum_{a_j=0}^{d_j} \frac{d_j!}{a_j!(d_j-a_j)!}\bigg\langle \tau_0\tau_{k-1}\prod_{j=0}^\infty \tau_j^{d_j-a_j}\bigg\rangle_{g_1}\bigg\langle \tau_0^3\prod_{j=0}^\infty \tau_j^{a_j}\bigg\rangle_{g_2}\nn\\&\hspace{1cm}+2\prod_{j=0}^\infty \sum_{b_j=0}^{d_j}\frac{d_j!}{b_j!(d_j-b_j)!}\,\bigg\langle \tau_0^2\tau_{k-1}\prod_{j=0}^\infty \tau_j^{d_j-b_j}\bigg\rangle_{g_1} \bigg\langle \tau_0^2\prod_{j=0}^\infty \tau_j^{b_j}\bigg\rangle_{g_2} +\frac{1}{4}\,\bigg\langle \tau_0^4\tau_{k-1}\prod_{j=0}^\infty \tau_j^{d_j}\bigg\rangle_{g-1}\,.\label{KdVrec}
\end{align}
These recursion relations encode the famous Korteweg-de Vries (KdV) hierarchy \cite{Dijkgraaf1991loop}, which is an infinite set of differential equations. The connection to the KdV hierarchy will prove useful in our work, so we summarize it now.
\subsection{KdV equations and intersection numbers}\label{sect:KdV}
The KdV hierarchy consists of an infinite set of differential equations \cite{Witten:1990hr}
\begin{equation}
(2n+1)\partial_{n}\partial_0^2 F=\partial_{n-1}\partial_0 F\partial_0^3 F+2\partial_{n-1}\partial_0^2 F\partial_0^2 F+\frac{1}{4}\partial_{n-1}\partial_0^4 F\,,\quad n\geq 0\,.\label{kdv}
\end{equation}
Here $F$ is a function of an infinite set of parameters $t_0,t_1,t_2\dots$ called KdV times and $\partial_i=\partial/\partial t_i$. Often, these equations are presented in a slightly different way, by introducing functions $R_n$ and $u$ of the KdV times
\begin{equation}
    R_n=\partial_{n-1}\partial_0 F\,,\quad R_1=\partial_0^2 F=u\,.
\end{equation}
By definition on then has
\begin{equation}
    \partial_n u=\partial_0 R_{n+1}\,.
\end{equation}
The functions $R_n$ for $n\geq 0$ are then determined recursively by the KdV hierarchy \eqref{kdv} as functions of $u$
\begin{equation}
    (2n+1)\partial_0 R_{n+1}= (\partial_0 u) R_n + 2 u \partial_0 R_n +\frac{1}{4}\partial_0^3 R_n\,,\label{413}
\end{equation}
the $n=0$ version of this equation
\begin{equation}
    \partial_0 u= R_0 \partial_0 u+2 u \partial_0 R_0+\frac{1}{4}\partial_0^3 R_0
\end{equation}
has the solution $R_0=1$. Using that $R_1=u$, one can then solve the $n=1$ version of \eqref{413} and obtain the solution\footnote{This is the standard Korteweg-de Vries equation $\partial_1u=u\partial_0u+\partial_0^3 u/12$.}
\begin{equation}
    R_2=\frac{1}{2}u^2+\frac{1}{12}\partial_0^2 u\,.
\end{equation}
This can be continued recursively and $n$ and so in principle one can find solutions to the KdV equations in terms of one unknown function $u$. These functions $R_n$ that one obtains this way are (up to potential normalization factors appearing in various papers) known as Gelfand-Dikii polynomials. 

In fact the KdV hierarchy has one more equation, known as the string equation \cite{Witten:1990hr}
\begin{equation}
    \partial_0^2 F=t_0+\sum_{i=0}^\infty
    t_{i+1}\partial_0\partial_i F\,,\label{stringeq}
\end{equation}
which, via the definition of the Gelfand-Dikii polynomials becomes
\begin{equation}
    0=\sum_{n=0}^\infty (t_k-\delta_{k,1})R_k\,.
\end{equation}
Inserting the expressions for the Gelfand-Dikii polynomials in terms of powers of $u$ (and $t_0$ derivatives thereof) this becomes a highly complicated equation for $u$, which is all that remains to be solved of the KdV hierarchy of equations. But what does this have to do with the intersection numbers we discussed previously?

The relation with intersection numbers is that this unknown function $F$ is a generating functional of intersection numbers \cite{KontsevichModel,Witten:1990hr} (see appendix \ref{app:c2} for more details about this)
\begin{equation}
    F=\bigg\langle\exp\bigg(\sum_{k=0}^\infty t_k \tau_k\bigg)\bigg\rangle=\sum_{\{d_i\}} \prod_{i=0}^\infty\frac{t_i^{d_i}}{d_i!}\bigg\langle \prod_{j=0}^\infty \tau_j^{d_j}\bigg\rangle\,.\label{Finter}
\end{equation}
Bearing in mind that in the gravitational description one interprets $\average{\tau_{d_1}\dots \tau_{d_n}}$ as connected correlators \eqref{zfromtau}, it is also natural to consider the generating function of full (not per-se connected) correlators
\begin{equation}
    Z=\exp( F)\,.
\end{equation}
There are approximately a gazillion proofs for \eqref{Finter}, but in the spirit of the discussion we have had thus far we only mention one approach, due to Dijkgraaf-Verlinde$^2$ \cite{Dijkgraaf:1991qh}, and explained nicely by Witten \cite{Witten:1991mn}. The gist is that using standard manipulations one can show that the solution $F$ of the KdV hierarchy also satisfies linear differential equations \cite{Witten:1991mn}
\begin{equation}
    L_n Z=0\,,\quad n\geq -1\,,\label{vir}
\end{equation}
with $L_n$ satisfying the Virasoro algebra\footnote{As a result, you actually only need to consider the equations for $-1\leq n \leq 2$ \cite{Witten:1991mn}.\label{few_n_Needed}}
\be 
[L_m,L_n] = (m-n)L_{m+n} + \delta_{n+m,0}\frac{m(m^2-1)}{12}\,.
\ee
They are explicitly given by
\begin{equation}
    L_n=\frac{1}{2}\sum_{m=-\infty}^{+\infty}\alpha_m \alpha_{n-m+1}+\frac{1}{16}\delta_{n,0}\,, \label{421}
\end{equation}
with creation and annihilation operators
\begin{equation}
\alpha_m=
\begin{cases}
    \frac{1}{2^{1/2}}(2m-1)!!\,\partial_{m-1}\, &m\geq 1
    \\\frac{1}{2^{1/2}}\frac{1}{(-2m-1)!!}\,(t_{-m}-\delta_{m,-1})\, &m\leq 0\,.
\end{cases}    
\end{equation}
Then one can show that these differential equations are equivalent to the topological recursion relations that we found in appendix \ref{app:c5} for the ribbon graph volumes $V_{g,n}(b_1\dots b_n)$. In particular one can write the recursion relations of those volumes in terms of recursion relations for the intersection numbers as in \cite{mulase2006mirzakhani}, the result is the Dijkgraaf-Verlinde$^2$ recursion relation (equation (7.27) in \cite{Dijkgraaf:1991qh} or (4.1) in \cite{Dijkgraaf1991loop})
\begin{align}
    &(2n+3)!!\,\bigg\langle \tau_{n+1}\prod_{i\in X} \tau_{k_i}\bigg\rangle_g =\sum_{j\in X}\frac{(2n+2k_j+1)!!}{(2k_j-1)!!}\,\bigg\langle \tau_{k_j+n}\prod_{i\in X/j} \tau_{k_i}\bigg\rangle_g\nonumber\\&\qquad\qquad\qquad\qquad\qquad\quad\,\,\,+\frac{1}{2}\sum_{m=1}^n(2m-1)!!(2n-2m-1)!!\,\bigg\langle \tau_{m-1}\tau_{n-m}\prod_{i\in X} \tau_{k_i}\bigg\rangle_{g-1}  \nonumber\\&\qquad\qquad+\frac{1}{2}\sum_{g'=0}^g\sum_{X_1\cup X_2=X}\sum_{m=1}^n(2m-1)!!(2n-2m+1)!!\,\bigg\langle \tau_{m-1}\prod_{i_i\in X_1} \tau_{k_{i_1}}\bigg\rangle_{g'}\bigg\langle \tau_{n-m}\prod_{i_2\in X_2} \tau_{k_{i_2}}\bigg\rangle_{g-g'} \label{523}
\end{align}
We remind the reader that these correlators have an implicit genus, because of the selection rule \eqref{selection}. If we assign genus $g$ to the correlators on the first line, one sees that the correlators on the second line have genus $g-1$, and the ones on the final line have genera $g'$ and $g-g'$. It is obvious then where each term comes from in the Mirzakhani recursion relations for $V_{g,n}(b_1\dots b_n)$.

This is identical to the equations that one finds from writing out \eqref{vir}. To see this, one writes $Z$ in terms of $F$, and expands $F$ as in \eqref{Finter} in powers of $t_k$. In such an expansion, a derivative $\partial_k$ is the creator of an extra $\tau_k$ 
\begin{equation}
    \partial_k F=\sum_{\{d_i\}}\prod_{i=0}^\infty \frac{t_i^{d_i}}{d_i!}\bigg\langle \tau_k\prod_{j=0}^\infty \tau_j^{d_j}\bigg\rangle\,,\label{string}
\end{equation}
and multiplication with $t_k$ is like removing or annihilating a $\tau_k$
\begin{equation}
    t_k F=\sum_{d_i}\prod_{i=0}^\infty \frac{t_i^{d_i}}{d_i!}\bigg\langle d_k \tau_k^{d_k-1}\prod_{j\neq k}^\infty \tau_j^{d_j}\bigg\rangle\,.
\end{equation}
Applying this several times one indeed recovers \eqref{523} (by comparing terms with identical powers $t_i^{d_i}$). Before proceeding we mention two special cases of \eqref{523}. The case $n=-1$ is called the string equation, and is equivalent to \eqref{stringeq}\footnote{Here and below $g$ is determined using the selection rule \eqref{selection}, the correlator vanishes whenever $g$ is not a positive integer
\begin{equation}
    \sum_{i=0}^m k_i=3g-2+m\,.
\end{equation}
}
\begin{equation}
    \bigg\langle \tau_{0}\prod_{i=1}^m \tau_{k_i}\bigg\rangle_g =\sum_{j=1}^m\bigg\langle \tau_{k_j-1}\prod_{i\neq j}^m \tau_{k_i}\bigg\rangle_g\,,\label{426}
\end{equation}and the case $n=0$ is called the dilaton equation (use the selection rule \eqref{selection} to simplify the prefactor)
\begin{equation}
    \bigg\langle \tau_{1}\prod_{i=1}^m \tau_{k_i}\bigg\rangle_g =-\chi\,\bigg\langle\prod_{i=1}^m \tau_{k_i}\bigg\rangle_g\,,\quad \chi=2-2g-m\label{427}
\end{equation}
The first term in
\bea
L_{-1}=\frac{1}{4} t_0^2 -\frac{1}{2}\partial_0+\sum_{m=1} \frac{1}{2} t_m \partial_{m-1}\,,
\eea
fixes the fact that $\langle \tau_0^n\rangle=\delta_{n,3}$. Similarly, we can understand the term $1/16$ in \eqref{421} as fixing $\langle \tau_1\rangle=1/24$. Note that using equation \eqref{3.4} one finds $V_{1,1}(b)=b^2/48$ (see also \cite{Stanford:2022fdt} and appendix \ref{app:c}), from which we deduce $\average{\tau_1}=1/24$. Furthermore from $V_{0,3}(b_1,b_2,b_3)=1$ (the moduli space $\mathcal{M}_{0,3}$ is a point) one finds $\average{\tau_0^3}=1$. These are the initial conditions for the recursion relations.

As mentioned in footnote \ref{few_n_Needed}, beyond the $n=-1$ and $n=0$, we only need the $n=1$ and $n=2$ Virasoro conditions. Imposing these conditions results in non-linearities. Whereas before acting with $L_{-1}$ and $L_0$ is first order in the derivatives w.r.t. $t_m$, $L_1$ and $L_2$ contain second order derivatives acting on $Z = e^F$, resulting, for instance, in terms of the form $(\partial_0 F)^2 + \partial_0^2 F$ when converting to the free energy and giving rise to the final term in \eqref{523}.

Instead of expanding the Virasoro constraints \eqref{vir} in intersection numbers, we can also just expand the KdV hierarchy \eqref{kdv} (and string equation). Since both sets of equations carry identical information, so do the resulting recursion relations. One matches in a straightforward manner each consecutive term in \eqref{kdv} with those in the recursion relations \eqref{KdVrec}. We proceed with \eqref{KdVrec} now.
\subsection{Simplest cancellations}\label{sect:simplecancel}
One can use this recursive version of the KdV equations \eqref{KdVrec} to get differential equations for $\mathcal{F}(x_1\dots x_n)$ \eqref{58} \cite{faber2000logarithmic} which can then be solved exactly \cite{okounkov2002generating,faber2000logarithmic}. For $\mathcal{F}(x)$ we start from \eqref{KdVrec} with $d_j=0$
\begin{equation}\label{recursionUnReduced}
    (2n+1)\average{\tau_0^2\tau_n}=\average{\tau_0\tau_{n-1}}\average{\tau_0^3}+2\average{\tau_0^2\tau_{n-1}}\average{\tau_0^2}+\frac{1}{4}\,\average{\tau_0^4\tau_{n-1}}\,,
\end{equation}
which, using the string equation \eqref{426} and $\average{\tau_0^n}=\delta_{n,3}$ simplifies to
\begin{equation}\label{recursionFx}
    2n\average{\tau_{n-2}}-\frac{1}{4}\,\average{\tau_{n-5}}=0\,.
\end{equation}
We can rewrite this recursion relation into a differential equation for $\mathcal{F}(x)$ by basically doing the inverse of what took us from the KdV equations to \eqref{KdVrec}.\footnote{The difference is that here we consider all $n$, but only one insertion from the exponential, whereas in the KdV hierarchy we consider the whole exponential but fixed $n$.} Multiplying with $x^{n-1}$ and summing over $n$ we find
\begin{equation}
    2\partial_x \sum_n x^{n}\average{\tau_{n-2}}-\frac{1}{4}\sum_n x^{n-1} \average{\tau_{n-5}}=0\quad\Rightarrow \quad\bigg(2\partial_x x^2-\frac{1}{4}x^4\bigg)\mathcal{F}(x)=0\,.
\end{equation}
This equation has a unique solution once one impose that the linear term $\mathcal{F}(x)\supset x\average{\tau_1}=x/24$
\begin{equation}\label{Fx}
    \mathcal{F}(x)=\frac{1}{x^2}\exp\bigg( \frac{x^3}{24}\bigg)\,.
\end{equation}
With the relation \eqref{zfromtau} this computes the exact one-boundary partition function for topological gravity with spectrum \eqref{topgravspec}, which reproduces the answer one obtains from double scaling the exactly solvable Gaussian matrix integral \cite{Saad:2019lba}
\begin{equation}
    Z(\beta)=\frac{e^{\S}}{4\pi^{1/2}\beta^{3/2}}\exp\bigg(\frac{1}{3}\,\beta^3 e^{-2\S}\bigg)\,.
\end{equation}

We can obtain the two-boundary partition function in similar manner. For $\mathcal{F}(x_1,x_2)$ one start from \eqref{KdVrec} with $d_j=\delta_{j,q}$
\begin{equation}
    (2n+1)\,\average{\tau_0^2\tau_n\tau_q}=\average{\tau_0\tau_{n-1}}\average{\tau_0^3\tau_q}+\average{\tau_0\tau_{n-1}\tau_q}+2\average{\tau_0^2\tau_{n-1}}\average{\tau_0^2\tau_q}+\frac{1}{4}\average{\tau_0^4\tau_{n-1}\tau_q}\,.\label{533}
\end{equation}
Using the string equation we find the relation
\begin{equation}
    \mathcal{F}(x_1,x_2,0)=\sum_{n=0}^\infty\sum_{q=0}^\infty  x_1^n x_2^q \,\average{\tau_0\tau_n\tau_q}=(x_1+x_2)\sum_{n=0}^\infty\sum_{q=0}^\infty x_1^n x_2^q\, \average{\tau_n\tau_q}=(x_1+x_2)\,\mathcal{F}(x_1,x_2)\,.
\end{equation}
Using tricks like this we find that multiplying \eqref{533} with $x_1^n x_2^q$ and summing over $n$ and $q$ results term by term in
\begin{align}
    &2\partial_1\, x_1(x_1+x_2)\, \mathcal{F}(x_1,x_2,0)-(x_1+x_2)\,\mathcal{F}(x_1,x_2,0)\\&=x_2\,\mathcal{F}(0,x_2,0)\,\mathcal{F}(x_1,0,0)+x_1\,\mathcal{F}(x_1,x_2,0)\nonumber+2x_1\,\mathcal{F}(0,x_2,0)\,\mathcal{F}(x_1,0,0)+\frac{1}{4}x_2(x_1+x_2)^3\,\mathcal{F}(x_1,x_2,0)\,.
\end{align}
Using again the string equation $\mathcal{F}(x_1,0,0)=x_1^2\,\mathcal{F}(x_1)=\exp(x_1^3/24)$ one can rearrange this into
\begin{equation}
    \bigg(1-\frac{1}{4}x_1x_2(x_1+x_2)+2\frac{x_1(x_1+x_2)}{x_2+2x_1}\partial_1 \bigg) \mathcal{F}(x_1,x_2,0)\,\exp(-\frac{x_1^3+x_2^3}{24})=1\,.
\end{equation}
This simplifies tremendously by introducing the coordinate $a=x_1x_2(x_1+x_2)$ and naming the function outside the braces $f(a)$, then Mathematica spits out the unique solution (the boundary condition comes from taking $x_1=0$ and using $\mathcal{F}(0,x_2,0)=\exp(x_2^3/24)$ as above)
\begin{equation}
    \bigg(1-\frac{1}{4}a+2a\partial_a \bigg)f(a)=1\,,\quad f(0)=1\quad \Rightarrow\quad f(a)=\frac{(2\pi)^{1/2}}{a^{1/2}}\exp\bigg(\frac{a}{8}\bigg)\,\text{Erf}\bigg( \frac{a^{1/2}}{2^{3/2}}\bigg)\,.
\end{equation}
From this one then obtains finally the exact answer for $\mathcal{F}(x_1,x_2)$
\begin{equation}\label{Fx1x2}
    \mathcal{F}(x_1,x_2)=\frac{(2\pi)^{1/2}}{(x_1x_2)^{1/2}(x_1+x_2)^{3/2}}\exp\bigg(\frac{1}{24}(x_1+x_2)^3\bigg)\,\text{Erf}\bigg(\frac{1}{2^{3/2}}(x_1x_2(x_1+x_2))^{1/2}\bigg)\,,
\end{equation}
and using \eqref{zfromtau} one finds the exact connected two-boundary partition function \cite{Okuyama:2020ncd}
\begin{equation}
    Z(\beta_1,\beta_2)_\text{conn}=\frac{e^{\S}}{4\pi^{1/2}(\beta_1+\beta_2)^{3/2}}\exp\bigg(\frac{1}{3}(\beta_1+\beta_2)^3e^{-2\S}\bigg)\,\text{Erf}\Big(((\beta_1\beta_2(\beta_1+\beta_2))^{1/2}e^{-\S}\Big)\,.
\end{equation}
Note that this epressions indeed reduces to \eqref{airyexact} in the $\tau$-scaling limit.

\subsection*{A comment on unstable surfaces}

We pause to make one comment about \eqref{Fx} and \eqref{Fx1x2}. Notice that they contain the terms $1/x^2$ and $1/(x_1 + x_2)$, which, from our conversion from the recursion relation to the differential equation should not be there. Nevertheless, these are correct and should be included, they represent the contributions at $(g,n) = (0,1)$ and $(0,2)$, which are degenerate surfaces, i.e they have zero area. In the mathematics literature they go under the name unstable surfaces. The matrix integral calculations naturally include these cases and also in the gravity calculations they should be there. From the recursion relation we can also see that they are there if we continue to negative $n$ and $q$. For instance from \eqref{recursionFx} we see that the boundary condition $\average{\tau_1} = 1/(24)$ determines $\average{\tau_{-2}} = 1$, but $\average{\tau_{n}}$ with $n<-2$ (along side with $n=-1,0$) still vanish. This gives the $1/x^2$ indeed. Furthermore, this also implies that $\average{\tau_0\tau_{-1}} = 1$. 

\subsection*{Cancellations for two boundaries}
As explained in the recent paper by Eynard, Lewanski and Ooms \cite{eynard2021natural}, something interesting happens when we write $\mathcal{F}(x_1,x_2)$ as a series expansion (using the known expansions of Erf and exponentials, see for instance equation (2.13) in \cite{okounkov2002generating}) in elementary symmetric polynomials $e_1=x_1+x_2$ and $e_2=x_1x_2$:
\begin{equation}
    \mathcal{F}(x_1,x_2)=\sum_{g=0}^\infty\sum_{m=0}^{g}e_2^m e_1^{3g-1-2m}\frac{3^m}{24^g m!(g-m)!}\frac{1}{2m+1}(-1)^{m}\,.\label{ffexact}
\end{equation}
The label $g$ refers to genus, because of the weight $e^{-2g\S}$ that each term acquires in terms of $\beta_i$ variables \eqref{zfromtau}. Let us compare this with the generic expansion of $\mathcal{F}(x_1,x_2)$ in elementary symmetric polynomials and intersection numbers, following from \eqref{58} (this is elementary, but slightly teadious to find)
\begin{equation}
    \mathcal{F}(x_1,x_2)=\sum_{g}\sum_{m=0}^{(3g-1)/2}e_2^m e_1^{3g-1-2m}\sum_{p=0}^m
\average{\tau_p\tau_{3g-1-p}}\frac{3g-1-2p}{3g-1-p-m}\frac{(3g-1-p-m)!}{(m-p)!(3g-1-2m)!}(-1)^{p+m}\,.
\end{equation}
Notice that this is identical to the term of top degree $m=3g-1$ in \eqref{z2bdyexp}, which is no accident because topological gravity is the case $t_2, t_3\dots =0$ in section \ref{sect:eucl} in which case only the top degree survives. This should become more clear in section \ref{sect:universalcanc}. We left implicit the factor $1/2$ for the case $p=m=(3g-1)/2$.

The noteworthy observation is that the exact expression \eqref{ffexact} at fixed genus $g$ has a maximal power $e_2^g$, where the selection rule \eqref{selection} of intersection theory allows also for powers $e_2^m$ with $g<m<(3g-1)/2$. This implies highly non-trivial cancellations in intersection numbers 
\begin{equation}
    \sum_{p=0}^m
\average{\tau_p\tau_{3g-1-p}}\frac{3g-1-2p}{3g-1-p-m}\frac{(3g-1-p-m)!}{(m-p)!(3g-1-2m)!}(-1)^{p}=0\,,\quad g<m<(3g-1)/2\,.\label{taucanc1}
\end{equation}
The simplest case $m=(3g-1)/2$ has appeared in the literature before, for instance \cite{Liu13896} (theorem 8)
\be 
\sum_{d_1 + d_2 = 3g-1} \braket{\tau_{d_1}\tau_{d_2}}(-1)^{d_1} = 0\,.
\ee
Now, these cancellations \eqref{taucanc1} are exactly the same as the highest degree cancellations $m=3g-1$ that we predicted for volumes \eqref{genericcancel} associated with generic spectra. This is obvious because the expansion coefficients in \eqref{vexp} for the volumes are exactly identical to the the intersection number in \eqref{expvolumes}
\begin{equation}
    V_{g,d_1,d_2}=2^{d_1+d_2}\average{\tau_{d_1}\tau_{d_2}}_g\,,
\end{equation}
which for topological gravity are nonzero only if $d_1+d_2=3g-1$ \eqref{selection}. So, for topological gravity the only non-trivial cancellations in \eqref{genericcancel} are $m=3g-1$, the other cases are trivially satisfied because of the selection rule \eqref{selection}.

The fact that \eqref{ffexact} has a maximal power $e_2^g$ directly implies that the genus $g$ wormhole in topological gravity grows for late times no faster than
\begin{equation}
    Z_g(\beta+\i T,\beta-\i T)_\text{conn}=\quad \begin{tikzpicture}[baseline={([yshift=-.5ex]current bounding box.center)}, scale=0.7]
 \pgftext{\includegraphics[scale=1]{euclwormmany.pdf}} at (0,0);
     \draw (-0.04, -2) node {genus $g$ wormholes};
     \draw (-2.65, 2.2) node {$\beta+\i T$};
     \draw (2.65, 2.2) node {$\beta-\i T$};
     \draw (0, 0) node {$\dots$};
  \end{tikzpicture}\quad \sim T^{2g+1}e^{-2g\S}\,,\label{545}
\end{equation}
because for late enough times $e_1\sim \beta$ and $e_2\sim T^2$. We have shown that the cancellations in the volumes required to make this happen can be derived directly from the KdV equations. The question then arises whether this is also true for \emph{generic} dilaton gravity theories, or generic spectral curves $\rho_0(E)$. We will demonstrate in section \ref{sect:universalcanc} that this \emph{is} the case, the universal scaling \eqref{545} can be derived using the KdV hierarchy.

To set up that argument, we first introduce a generalization of the cancellations \eqref{ffexact} in $\mathcal{F}(x_1,x_2)$ to the $n$-boundary correlators in topological gravity $\mathcal{F}(x_1\dots x_n)$.
\subsection{Multi-boundary cancellations}
Just like for the cases $\mathcal{F}(x_1)$ and $\mathcal{F}(x_1,x_2)$ that we presented in section \ref{sect:simplecancel}, one can get an exact formula for the $n$-boundary correlators in topological gravity $\mathcal{F}(x_1\dots x_n)$ via the KdV recursion relations. This was shown by Liu and Xu \cite{liu2011n}, who found a recursive formula to compute $\mathcal{F}(x_1\dots x_n)$ from $\mathcal{F}(x_1\dots x_m)$ with $m<n$. The proof, nor the particular equation, is particularly insightful, the point is that precise formulas exist and follow purely from KdV.

These formulas can then be expanded in elementary symmetric polynomials \cite{eynard2021natural}, like we did for the two point function in \eqref{ffexact}. From the intersection number formula \eqref{58} and simple dimensional analysis we get an expansion of the type
\begin{equation}
    \mathcal{F}(x_1\dots x_n)=\sum_{g=0}^\infty \sum_{m_1=0}^\infty\dots \sum_{m_n=0}^\infty e_1^{m_1}e_2^{m_2}\dots e_n^{m_n}\,C_g(m_2\dots m_n)\,,\quad \sum_{j=1}^n j m_j=3g-3+n\,,
\end{equation}
where the constraint on the powers comes from the selection rule \eqref{selection}. For clarity
\begin{equation}
    e_1=\sum_{j=1}^n x_j\,,\quad e_2=\sum_{j_1<j_2=1}^n x_{j_1}x_{j_2}\,,\quad e_3=\sum_{j_1<j_2<j_3=1}^n x_{j_1}x_{j_2}x_{j_3}\,,
\end{equation}
etcetera. Crucially, Eynard, Lewanski and Ooms \cite{eynard2021natural} found that many of the naively allowed expansion coefficients identically vanish
\begin{equation}
    C_g(m_2\dots m_n)=0\,,\quad \sum_{j=2}^n m_j>g\,.\label{548}
\end{equation}
This generalizes the fact that the highest power in \eqref{ffexact} is $e_2^g$. They impressively checked this vanishing for all $n$ and $g\leq 7$, and for all $g$ and $n\leq 3$, using the explicit formulas of Liu and Xu \cite{liu2011n} (see also \cite{okounkov2002generating}) and the KdV equations (in various forms).

Taken at face value, this new constraint \eqref{548} does not seem too constraining on the individual powers of $e_3, e_4\dots$, because the original selection rule does not allow $m_3$ to grow faster than $g$, or $m_4$ faster than $3g/4$ etcetera. So one may wonder if, beyond $m=2$, these cancellations \eqref{548} have any intuitive physical interpretation, we will now demonstrate that they do.
\section{Universal cancellations in dilaton gravity via open-closed duality}\label{sect:universalcanc}
In this section we show that the constraint \eqref{548} on the total power of $e_2,e_3\dots$ implies (and in fact is identical to) the universal maximal growth \eqref{545} for all double scaled matrix models (with square root edges \eqref{spectrum}), not just topological gravity. This late-time wormhole universality, in turn, is key for the emergence of the plateau, which, in turn, is gravity's way of saying it is a discrete quantum system \cite{Cotler:2016fpe}.

Readers familiar with how those models are described using the KdV hierarchy may skip to section \ref{subsect:universalcanc}, for didactic purposes we will go slower.
\subsection{KdV equations around different backgrounds}\label{sect:4.1}
We have learned that because of the relation of the volumes with $\psi_i$ classes for topological gravity \eqref{43}
\begin{equation}
    V_{g,n}(b_1\dots b_n)=\int_{\overline{\mathcal{M}}_{g,n}}\exp\bigg(\frac{1}{2}\sum_{i=1}^n b_i^2\, \psi_i\bigg)=\sum_{d_1=0}^\infty\frac{b_1^{2d_1}}{2^{d_1}d_1!}\dots\sum_{d_n=0}^\infty\frac{b_n^{2d_n}}{2^{d_n}d_n!}\average{\tau_{d_1}\dots \tau_{d_n}}_g\,,
\end{equation}
that $n$-boundary partition functions can be expressed in terms of intersection numbers, as in \eqref{zfromtau}. 

More in general, there is such a relation for all double scaled matrix integrals with a leading order spectrum of the type \eqref{spectrum} 
\begin{equation}
    \rho_0(E)=\frac{e^{\S}}{2\pi}\sum_{k=0}^\infty f_k\,E^{k+1/2}\,,\quad f_0=1\,.\label{52}
\end{equation}
In this more general case the volumes are computed in terms of intersection numbers as
\begin{equation}
    V_{g,n}(b_1\dots b_n)=\sum_{d_1=0}^\infty\frac{b_1^{2d_1}}{2^{d_1}d_1!}\dots\sum_{d_n=0}^\infty\frac{b_n^{2d_n}}{2^{d_n}d_n!}\bigg\langle\tau_{d_1}\dots \tau_{d_n}\exp\bigg(\sum_{k=2}^\infty \g_k \tau_k\bigg)\bigg\rangle_g\,,\label{53}
\end{equation}
with a relation $f_k(\g_i)$ that we have yet to determine. Notice that we can get the correlators on the right from the same generating functional $F$ that satisfies the KdV hierarchy \eqref{Finter}, by expanding around a different set of KdV times $t_{k\,\text{total}}=\g_k+t_k$
\begin{align}
     F=\bigg\langle\exp\bigg(\sum_{k=0}^\infty t_{k\,\text{total}} \tau_k\bigg)\bigg\rangle&=\prod_{i=0}^\infty \sum_{d_i}\frac{t_i^{d_i}}{d_i!}\bigg\langle \prod_{j=0}^\infty \tau_j^{d_j}\exp\bigg(\sum_{k=2}^\infty \g_k \tau_k\bigg)\bigg\rangle=\prod_{i=0}^\infty \sum_{d_i}\frac{t_i^{d_i}}{d_i!}\bigg\langle \prod_{j=0}^\infty \tau_j^{d_j}\bigg\rangle_\text{$\g_k$}\,,\label{fexpansions}
\end{align}
where in the last equality we have introduced shorthand notation $\average{\dots}_{\g_k}$ for correlators in the presence of an extra exponential of $\tau_k$ operators.

There are several ways to appreciate that \eqref{53} calculates the correlators of a double scaled matrix integral \cite{Saad:2019lba} with a spectral curve of the type \eqref{52} (or some model of two dimensional (dilaton) gravity). One way is to write out the Virasoro constraints \eqref{vir}; but with the more generally correct form for the creation and annihilation operators (we can always set $\g_0=\g_1=0$)
\begin{equation}
\alpha_m=
\begin{cases}
    \frac{1}{2^{1/2}}(2m-1)!!\,\partial_{m-1}\, &m\geq 1
    \\\frac{1}{2^{1/2}}\frac{1}{(-2m-1)!!}\,(t_{-m}-\delta_{m,-1}+\g_{-m})\, &m\leq 0\,.
\end{cases}    \label{55}
\end{equation}
We can use the definition of $F$ \eqref{fexpansions} to see that we still have the property that a derivative with respect to $t_k$ is the creator of an $\tau_k$ 
\begin{equation}
    \partial_k F=\sum_{\{d_i\}}\prod_{i=0}^\infty \frac{t_i^{d_i}}{d_i!}\bigg\langle \tau_k\prod_{j=0}^\infty \tau_j^{d_j}\bigg\rangle_\text{$\g_k$}\,,
\end{equation}
and furthermore multiplication with $t_k$ is still like removing or annihilating a $\tau_k$
\begin{equation}
    t_k F=\sum_{\{d_i\}}\prod_{i=0}^\infty \frac{t_i^{d_i}}{d_i!}\bigg\langle d_k \tau_k^{d_k-1}\prod_{j\neq k}^\infty \tau_j^{d_j}\bigg\rangle_\text{$\g_k$}\,.
\end{equation}
Then we can write out the generators $L_n$ \eqref{421} and use $Z=\exp(F)$ to obtain recursion relations that are analogous to the Dijkgraaf-Verlinde$^2$ ones \eqref{523} but which depend explicitly on the $\g_k$, because $\g_k$ appears in \eqref{55}. These recursive equations to compute $\average{\tau_{d_1}\dots \tau_{d_n}}_{\g_k}$ can then be recovered alternatively from the topological recursion \cite{Eynard_2009} relations between volumes with spectral curve \eqref{52}, demonstrating the equivalence. 

We emphasize again that we are always dealing with the same function $F$, the same KdV hierarchy \eqref{kdv} and the same string equation (which features the invariant $t_{k\,\text{total}}=t_k+\g_k$)
\begin{equation}
    0=\sum_{k=0}^\infty (t_k+\g_k-\delta_{k,1})R_k\,,
\end{equation}
which in a correlation function reduces to the generalization of \eqref{426}
\begin{equation}
    \bigg\langle\tau_0\prod_{i=1}^n\tau_{k_i}\bigg\rangle_\text{$\g_k$}-\sum_{j=1}^\infty \g_{j+1}\bigg\langle\tau_j\prod_{i=1}^n\tau_{k_i}\bigg\rangle_\text{$\g_k$}=\sum_{m=1}^n\bigg\langle \tau_{k_m-1}\prod_{i\neq m}^n \tau_{k_i}\bigg\rangle_\text{$\g_k$}\,.\label{stringeqtaugen}
\end{equation}
For multi-critical points with $E^{k+1/2}$ edge only $\g_{k+1}$ is nonzero and we recover the $n=0$ case of formula (4.1) in \cite{Dijkgraaf1991loop} (the formulas are more symmetric if we view the $-\delta_{k,1}$ as $\g_1=-1$ instead, which we will adopt below).

Let us now compute the spectrum $\rho_0(E)$ that follows from \eqref{fexpansions}. For this we need to be a bit more clever with the genus counting parameter and define \cite{Okuyama:2019xbv} 
\begin{equation}
    F(e^{\S},\g_k)=\sum_{g=0}^\infty e^{-2g\S}\average{\exp\bigg(\sum_{k=0}^\infty \g_k\tau_k\bigg)}_g\,,\quad R_{k+1}(e^{\S},\g_k)=\partial_0\,\partial_k\, F(e^{\S},\g_k)\,,
\end{equation}
with again $u(e^{\S},\g_k)=R_1(e^{\S},\g_k)$. Using the selection rule \eqref{selection} we observe that this is related to $F$ via $F(e^{\S},\g_k)=e^{-2\S} F(\g_k e^{\S 2(1-k)/3})$. Using these relations one rewrites the KdV recursion relations for $R_n$ \eqref{413} as (all equations below implicitly have the functions of $u(e^{\S},\g_k)$ that we just defined)
\begin{equation}
    (2n+1)\partial_0 R_{n+1}=\partial_0 u R_n+2 u \partial_0 R_n +e^{-2\S}\frac{1}{4}\partial_0^3 R_n\,,\quad 0=\sum_{k=0}^\infty \g_k R_k\,.
\end{equation}
This redefinition may seem a bit like wasted energy, but the benifit is that we can now very easily solve the recursion relation to leading order in $e^{\S}$, because we can neglect the final term. With seed $R_0=1$, one finds $R_k=u^k/k!$ and the leading order string equation becomes quite simple
\begin{equation}
    -\g_0=\sum_{k=1}^\infty \g_k\frac{u^k}{k!}=\mathcal{G}(u)\,.
\end{equation}
One can then use this simple structure to find $Z(\beta)$ to leading order as follows \cite{Okuyama:2019xbv,Johnson:2019eik,Johnson:2020exp,Johnson:2020heh}. In all the expressions we have written down so far $t_0$ was set to zero, but this makes using the KdV hierarchy impossible since there are $\partial_0$s everywhere. Thus we need to consider a partition function $Z(\b,t_0)$ with non-zero $t_0$. For that we need to use the contributions from unstable surfaces discussed below \eqref{Fx1x2}. In particular, consider
\begin{align}
    \partial_0 Z(\beta,t_0) = e^{\S}\,\frac{\beta^{1/2}}{\pi^{1/2}}\sum_d (2\beta)^d \average{\tau_0\tau_d}_{\g_k} =e^{\S}\,\frac{1}{2\pi^{1/2}\beta^{1/2}}\sum_{d=0}^\infty (2\beta)^d R_d
    =e^{\S}\,\frac{e^{2\beta u}}{2\pi^{1/2}\beta^{1/2}}\label{partialZ}\,,
\end{align}
where we used that a derivative wrt to $t_0$ inserts a factor of $\tau_0$\footnote{To be very precise here: we work to first order in perturbation theory in $t_0$, which will be enough as we will later set $t_0$ to zero again.} the definition $R_{k+1}=\partial_0\,\partial_k\,F$ and that $R_d = u^d/d!$. Integrating \eqref{partialZ}, using that $\g_0=-\mathcal{G}(u)$ such that $\d \g_0=-\mathcal{G}'(u)\d u$, and putting $t_0=0$ again one obtains the leading order solution \cite{Okuyama:2019xbv}
\begin{equation}
    Z(\beta)= \frac{e^{\S}}{2\pi^{1/2} \b^{1/2}} \int_{-2 u_0}^{\infty} \d u\,\mathcal{G}'(-u/2)\,e^{- \b u } \,,\label{515}
\end{equation}
with $u_0$ the smallest positive solution to $\mathcal{G}(u_0) = 0$. The spectral density is then found to be
\begin{equation}
    \rho_0(E) = \frac{e^{\S}}{\pi}\int_{E_0}^{E}\d u \frac{\mathcal{G}'(-u/2)}{\sqrt{E-u}} \,.\label{4.16}
\end{equation}
with $E_0=2u_0$. In the case of JT we have 
\be 
\mathcal{G}(-u/2) = \frac{u^{1/2}}{2\pi}I_1(2\pi u^{1/2}),
\ee
which gives $E_0 = 0$ and the density to be the sinh. More in general, one finds an expansion of $\rho_0(E)$ in terms of $E^{k+1/2}$ as in \eqref{spectrum} but with expansion coefficients that depend on $\g_k$ non-linearly because $E_0$ depends non-linearly on these couplings. To sum things up: \eqref{53} computes volumes of two dimensional (dilaton) gravity theories, and the leading order spectrum can be found more or less directly from the KdV equations.\footnote{Some extra overall $e^{E_0 \beta}$ in the integral \eqref{expaexpa} does not change any of the statements about the power series.}

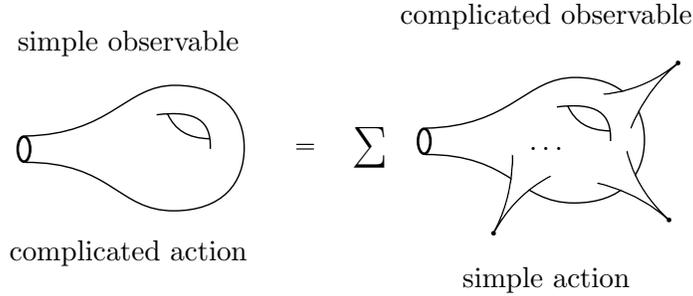
\begin{figure}[t]
    \centering
    \begin{equation}
    \quad \begin{tikzpicture}[baseline={([yshift=-.5ex]current bounding box.center)}, scale=0.7]
 \pgftext{\includegraphics[scale=1]{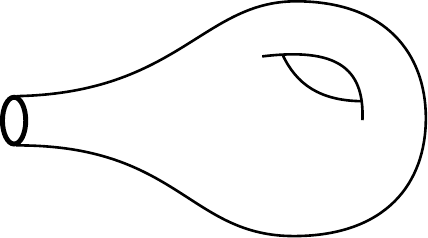}} at (0,0);
     \draw (-0.04, -2) node {complicated action};
     \draw (-0.04, 2) node {simple observable};
  \end{tikzpicture}\quad=\quad\sum \begin{tikzpicture}[baseline={([yshift=-.5ex]current bounding box.center)}, scale=0.7]
 \pgftext{\includegraphics[scale=1]{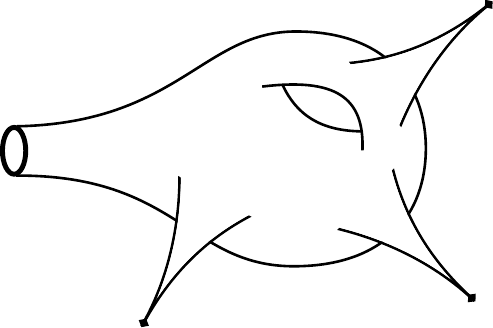}} at (0,0);
     \draw (-0.04, -2.5) node {simple action};
     \draw (-0.04, 2.5) node {complicated observable};
     \draw (0, 0) node {$\dots$};
  \end{tikzpicture}\quad\nonumber
\end{equation}
    \caption{Open-closed duality in a nutshell. Here we have in mind computing $V_{1,1}(b)$ in the theory with an exponential of $\tau_k$ inserted in the action, see \eqref{69} below. Either we use the complicated action \eqref{69} (left); or we expand out the deformations (right), in which case we have a sum over cusp-defects but we just use the simple JT gravity action at the cost of computing a more complicated observable with many boundaries (or operators). Alternatively, in the context of this section, we could think of the left picture as JT gravity and the right picture as Airy with an exponential of \eqref{520} inserted. The idea is always the same.}
    \label{fig:openclosed}
\end{figure}

Before we proceed let us make two comments that will be important later.
\begin{enumerate}
    \item JT gravity corresponds to the case
    \begin{equation}
        \g_{n+1}=(-1)^{n+1}\frac{(2\pi^2)^n}{n!}\,,
    \end{equation}
    which indeed reproduces the series expansion of $\rho_0(E)=e^{\S}\sinh(2\pi E^{1/2})/4\pi^2$. Of course we also have the known expression for the Weil-Petersson symplectic form \cite{Dijkgraaf:2018vnm}
    \begin{equation}
        V_{g,n}(b_1\dots b_n)=\int_{\overline{\mathcal{M}}_{g,n}}\exp\bigg(2\pi^2\kappa+\frac{1}{2}\sum_{i=1}^n b_i^2\, \psi_i\bigg)\,,
    \end{equation}
    where the $\kappa$ class represents the Weil-Petersson two form on punctured Riemann surfaces. If we compare this with our current construction we recover the known duality \cite{Dijkgraaf:2018vnm,eynard2011recursion}
    \begin{equation}
        e^{2\pi^2 \kappa} \quad \Leftrightarrow\quad \exp\left(\sum_{k=2}^\infty (-1)^k\frac{(2\pi^2)^{k-1}}{(k-1)!}\tau_k\right)\,,\label{520}
    \end{equation}
    which is to be understood as holding within expectation values.
    \item This relation between an exponential of $\tau_k$ operators and matrix integrals with different spectral densities should be considered an example of open-closed (string) duality. For instance in \eqref{53} the left-hand side is the ``closed'' picture, we compute $n$-boundary amplitudes in a specific gravity theory. But, the right-hand side is a sum over $m\geq n$-boundary amplitudes in topological gravity
    \begin{equation}\label{openInter}
        \average{\tau_{d_1}\dots \tau_{d_n}}_{\g_k}=\average{\tau_{d_1}\dots \tau_{d_n}}+\sum_{k_1=2}^\infty \g_{k_1} \average{\tau_{d_1}\dots \tau_{d_n}\tau_{k_1}}+\frac{1}{2}\sum_{k_1=2}^\infty\sum_{k_1=2}^\infty\g_{k_1}\g_{k_2} \average{\tau_{d_1}\dots \tau_{d_n}\tau_{k_1}\tau_{k_2}}+\dots 
    \end{equation}
    This should be viewed as the ``open'' description \cite{Gaiotto:2003yb}. The identity \eqref{520} should be interpreted in the same way. The $\kappa$ two forms are integrated over $\overline{\mathcal{M}}_{g,n}$, whereas the $\tau_k$ correspond with Chern classes $c_1(L_a)^{\wedge k}$ associated with $m-n$ extra marked points $x_a$ integrated over $\overline{\mathcal{M}}_{g,m}$ with $m\geq n$. In string language, the exponential of $\tau_k$'s is like inserting D-branes (exponentials of boundaries), see also Fig. \ref{fig:openclosed}.
    \item Another side of the coin is the relation with ribbon graphs. We demonstrated in section \ref{sect:3.1} that topological gravity can be thought of in the same way as JT gravity, but where the higher genus Riemann surfaces one glues to are ribbon graphs with large lengths $b_i$. In turn, the full JT theory also has a ribbon graph interpretation; actually two, related by open-closed duality. 
    
    The open side of the interpretation is essentially rewriting \eqref{openInter} in terms of ribbon graphs using Kontsevich's insights, which we explained in section \ref{sect:3.1}. The crucial thing there is then that the JT ribbon graphs in the open description are sums of the usual (Airy) ribbon graphs, but with additional faces that are weighted by the $\g_i$ instead of the $b_i^2/2$. 
    
    The closed perspective is that one takes the usual (Airy) ribbon graphs, but when computing the volume of moduli space one uses not the flat measure, rather the $\kappa$ class implies a non-trivial measure
    \be \label{muJT}
    \mu(\{\hat{\ell}_i\},\{b_i\})_{\rm JT} = \exp\bigg(2\pi^2\kappa+\frac{1}{2}\sum_{i=1}^n b_i^2\, \psi_i\bigg) = \exp \bigg( - \sum_{i<j}^{6g-6+2n} W^{-1}_{ij} \d \hat{\ell}_i \wedge \d \hat{\ell}_j  \bigg)\,,
    \ee
    with $W_{ij}$ the matrix 
    \be 
    W_{ij} = \sum_{p \in \g_i \cap \g_j} \cos \theta_p\,.  
    \ee
    Here $\g_i$ are $6g-6+2n$ distinct simple closed geodesics with length $\hat{\ell}_i$, and $\theta_p$ is the intersection angle between two of them. These $\hat{\ell}_i$ are crucially not the same as the lengths $\ell_j$ we encountered around equation \eqref{3.4}. For $\hat{\ell}_i$ we picked a set of simple closed geodesics $\g_i$ on the surface, whereas in \eqref{3.4} we looked at the edges of a ribbon graph associated to the surface. How to construct the $\g_i$ from the ribbon graph is non-trivial. From the ribbon graph, as explained in \cite{do2008intersection} there is an algorithmic way of constructing the simple closed geodesics (these can be disjoint unions of curves also) but since the ribbon graph is trivalent\footnote{The moduli space of ribbon graphs of other valency have dimension strickly less than $6g-6+2n$.} one will get $6g-6+3n$ geodesics, which are $n$ too many. One can eliminate them by noticing that the analogue of the matrix $W_{ij}$ will have rank $6g-6+2n$ and so there is a set of rows and colums one can select to get the right geodesics. This is thus rather cumbersome, but luckily at large $b_i$ the $\ell_i$ and $\hat{\ell}_i$ have a simple relation \cite{do2008intersection}, resulting in \eqref{3.6}.
    
    At any rate, the closed picture is thus more complicated only in the sense that we would have to write \eqref{muJT} in terms of the edge lengths $\ell_i$ of the usual ribbon graphs. 
    
    Another point to note is that the relation with the perhaps more familiar measure in terms of the Fenchel-Nielsen length and twist coordinates $(b_j,\tau_j)$ is also not easy. One would have to write the twist and length variables in terms of the edge variables $\ell_j$ of ribbon graphs. The relation with the $\hat{\ell}_i$ is a bit easier, but then one still needs to convert $\hat{\ell}_i$ to the $\ell_i$ variables, which as explained above, is hard.
\end{enumerate}
The point is that you can compute complicated things in a simple theory (many-boundary observables in topological gravity) and learn about simple things in a complicated theory (few-boundary observables in generic two-dimensional gravity models). Eynard, Lewanski and Ooms have such a complicated claim \eqref{548} in a simple theory, we will recast this into a simple claim \eqref{expaexpa} in complicated theories.
\subsection{Universal cancellations in all theories}\label{subsect:universalcanc}
Consider now the two-boundary correlator in a generic background $\g_k$, or for any double scaled matrix integral with square root spectral edge. According to \eqref{zfromtau} and \eqref{53} we have
\begin{align}
    Z(\beta_1,\beta_2)_\text{conn}&=\frac{\beta_1^{1/2}}{\pi^{1/2}}\,\frac{\beta_2^{1/2}}{\pi^{1/2}}\,\mathcal{F}(2\beta_1,2\beta_2)_{\g_k}=\frac{\beta_1^{1/2}}{\pi^{1/2}}\,\frac{\beta_2^{1/2}}{\pi^{1/2}}\sum_{g=0}^\infty e^{-2g\S}\,\mathcal{F}_g(2\b_1,2\b_2)_{\g_k}\,,\label{zfromtaugen}
\end{align}
with the generating functional
\begin{equation}
    \mathcal{F}_g(x_1,x_2)_{\g_k}=\sum_{d_1=0}^\infty\sum_{d_2=0}^\infty x_1^{d_1} x_2^{d_2}\bigg\langle\tau_{d_1} \tau_{d_2}\exp\bigg(\sum_{k=2}^\infty \g_k \tau_k\bigg)\bigg\rangle_g\,.\label{4.22}
\end{equation}
Now let us introduce a new infinite set of variables $y_i(t_k)$ such that 
\begin{equation}
    \g_k=\sum_{i=0}^\infty  y_i^k\,.
\end{equation}
Then expanding out the generating functional gives
\begin{equation}
    \mathcal{F}_g(x_1,x_2)_{\g_k}=\sum_{m=0}^\infty \frac{1}{m!}\sum_{i_1=0}^\infty\dots \sum_{i_m=0}^\infty \mathcal{F}_g(x_1,x_2,y_{i_1}\dots y_{i_m})\,,
\end{equation}
which features the $n\geq 2$-boundary correlators of topological gravity
\begin{equation}
    \mathcal{F}_g(x_1\dots x_n)=\sum_{d_1=0}^\infty\dots \sum_{d_n=0}^\infty x_1^{d_1}\dots  x_n^{d_n}\average{\tau_{d_1}\dots \tau_{d_n}}_g=\sum_{m_1=0}^\infty\dots \sum_{m_n=0}^\infty e_1^{m_1}e_2^{m_2}\dots e_n^{m_n}\,C_g(m_2\dots m_n)\,,
\end{equation}
which as we discussed around \eqref{548} has a constrained expansion in elementary symmetric polynomials
\begin{equation}
    C_g(m_2\dots m_n)=0\,,\quad \sum_{j=2}^n m_j>g\,.\label{548bis}
\end{equation}
For late times in the spectral form factor computation $x_1\sim \i T$ and $x_2\sim -\i T$, but none of the $y_i$ scale in any way with $T$, these are just parameters inherent to the theory. This means that $e_2\sim e_3\sim \dots \sim T^2$ and therefore
\begin{equation}
    e_1^{m_1}e_2^{m_2}\dots e_n^{m_n}\sim T^{2\sum_{j=2}^n m_j}\,.\label{4.27}
\end{equation}
The power of $T$ that appears here is precisely the expression that was constrained by the KdV equations to be upper bound by $2 g$. Hence we have derived that $\mathcal{F}_g(x_1,x_2)_{\g_k}$ grows no faster than $T^{2g}$ for generic $\g_k$
\begin{equation}
    C_g(m_2\dots m_n)=0\,,\quad \sum_{j=2}^n m_j>g\quad \Rightarrow \quad  Z_g(\beta+\i T,\beta-\i T)_\text{conn}\sim T^{2g+1}e^{-2g\S}\,.
\end{equation}
In summary, we have used the KdV equations to prove a universal late-time scaling behavior for genus $g$ wormhole amplitudes in generic double scaled matrix integrals (or two dimensional gravity models). We explained in section \ref{sect:eucl} why this behavior was key for the emergence of the plateau.

In fact we believe that the arrow works both ways
\begin{equation}
    C_g(m_2\dots m_n)=0\,,\quad \sum_{j=2}^n m_j>g\quad \Leftrightarrow \quad  Z_g(\beta+\i T,\beta-\i T)_\text{conn}\sim T^{2g+1}e^{-2g\S}\,,\label{iff}
\end{equation}
in other words this new type of wormhole universality at late times also implies all of the cancellations that Eynard, Lewanski and Ooms \cite{eynard2021natural} found. We demonstrate this in appendix \ref{app:reverse}.
\section{KdV equations in dilaton gravity and the matrix integral}\label{sect:interpretation}
In the previous sections we primarily focused on the intersection theory and its integrable structure by itself, but there is actually a natural interpretation in dilaton gravity and the matrix integral of all this stuff as well. The general message that we want to convey here is that the KdV equations seem to have a lot of applications in gravity, it seems like we have just scratched the surface.

In \textbf{section \ref{sect:KdVdilatongravity}} we prove that the $\tau_k$ can be interpreted as local operators in dilaton gravity which create cusps (sharp defects \cite{Witten:2020wvy,Maxfield:2020ale,Mertens:2019tcm}). KdV backgrounds with different $\g_k$ correspond with inserting a gas of such cusps in JT gravity, this changes the dilaton potential, and in that sense the different KdV backgrounds are different dilaton gravity models. 

The KdV equations mostly describe how observables in dilaton gravity change under changes in the action, the answer is basically that we compute things with a new spectral curve. Likewise, the string-and dilaton equations have an interpretation in dilaton gravity as transformation rules for observables under changes in the action. They reduce to analogues of the Dijkgraaf-Verlinde$^2$ loop equations \cite{Dijkgraaf:1991qh}.

In \textbf{section \ref{sect:matrix}} we present a clean matrix integral dual for the $\tau_k$ operators, improving on equations that appeared in \cite{Maldacena:2004sn, Gross1989nonperturbative,Blommaert:2021gha,DOUGLAS1990176,douglas1990strings}.
\subsection{Dilaton gravity}\label{sect:KdVdilatongravity}
We start with remembering the relation \eqref{53} between Weil-Petersson volumes and $\tau_k$ correlators in the JT gravity background \eqref{520}
\begin{equation}
    V_{g,n}(b_1\dots b_n)=\sum_{d_1=0}^\infty\frac{b_1^{2d_1}}{2^{d_1}d_1!}\dots\sum_{d_n=0}^\infty\frac{b_n^{2d_n}}{2^{d_n}d_n!}\average{\tau_{d_1}\dots\tau_{d_n}}_\kappa\,.\label{61}
\end{equation}
Furthermore \cite{tan2006generalizations,do2009weil,do2011moduli,Cotler:2019nbi,Maxfield:2020ale} we also know that correlators of defects with opening angle $\alpha_i<\pi$ in JT gravity are analytic continuations of Weil-Petersson volumes to $b_i=\i\alpha_i$
\begin{equation}
    \average{\mo_\text{D}(\alpha_1)\dots \mo_\text{D}(\alpha_n)}_{g\,\text{conn}}= V_{g,n}(\i \alpha_1\dots \i \alpha_n)=\sum_{d_1=0}^\infty \frac{(-1)^{d_1}}{2^{d_1}d_1!}\,\alpha_1^{2d_1}\dots\sum_{d_1=0}^\infty \frac{(-1)^{d_n}}{2^{d_n}d_n!}\,\alpha_1^{2d_n}\average{\tau_{d_1}\dots\tau_{d_n}}_\kappa \,,\label{62}
\end{equation}
with $\mo_\text{D}(\alpha)$ the dilaton gravity operator that creates a defect with opening angle $\alpha$\footnote{One can use several quantization schemes for defects in JT gravity \cite{Witten:2020wvy,Turiaci:2020fjj}, leading to slightly different expressions for the defect operator. We are using the most intuitive conventions of \cite{Witten:2020wvy}. The first equality in \eqref{64} is scheme-independent, the second equality can be directly modified to a different scheme. The scheme of \cite{Turiaci:2020fjj} may be more appropriate for doing semiclassics with the action in \eqref{69}.} \cite{Mertens:2019tcm} (using Gauss-Bonnet one can check that this indeed creates the expected angular defect \cite{Mertens:2019tcm,Louko:1995jw}) 
\begin{equation}
    \mo_\text{D}(\alpha)\quad \Leftrightarrow\quad\int \d^2 x \sqrt{g}\, e^{-(2\pi-\alpha) \Phi}\quad \Leftrightarrow \quad \begin{tikzpicture}[baseline={([yshift=-.5ex]current bounding box.center)}, scale=0.7]
 \pgftext{\includegraphics[scale=1]{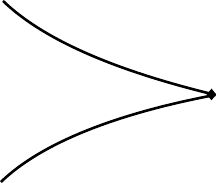}} at (0,0);
     \draw (-0.04, -1.5) node {conical defect};
     \draw (-0.5, 0) node {$\a$};
     \draw (-2,0) node {$\dots$};
  \end{tikzpicture}\quad\label{63}
\end{equation}
Equation \eqref{62} basically means that this defect operator is a generating functional of $\tau_k$ operators
\begin{equation}
    \tau_k\quad \Leftrightarrow\quad (-1)^k \frac{2^k k!}{(2k)!}\,\partial_\alpha^{2k}\,\mo_\text{D}(\alpha)\rvert_{\alpha=0}\quad \Leftrightarrow\quad \frac{(-1)^k}{(2k-1)!!}\,\int\d x^2\sqrt{g}\,\Phi^{2k}\,e^{-2\pi\Phi}\,,\label{64}
\end{equation}
where in the second equality we inserted the dilaton gravity meaning of $\mo_\text{D}(\alpha)$. Another way of stating this is that we have the obvious identity
\begin{equation}
    \int \d^2 x \sqrt{g}\, e^{-(2\pi-\alpha) \Phi}\quad \Leftrightarrow \quad \exp\bigg(\frac{1}{2}(\i\alpha )^2\psi_\text{extra}\bigg)\,,\quad \alpha<\pi\,.\label{65}
\end{equation}
Here $\psi_{\text{extra}}$ is the first Chern class attached to the extra marked point, constructed in a similar way as \eqref{43}. The same equation \eqref{64} is found by working directly with geodesic boundaries and expanding around $b=0$ whilst using the dilaton gravity description of a geodesic boundary \cite{Blommaert:2021fob}\footnote{The reader should not dwell on the $e^{-\S}$ prefactor here, which just signifies that we choose to count this as a hole rather than a local operator topologically. For defects and $\tau_k$ operator insertions there is no such prefactor.}
\begin{equation}
    \mo_\text{G}(b)\quad \Leftrightarrow\quad e^{-\S}\int \d^2 x \sqrt{g}\, e^{-2\pi \Phi} \cos\left(b\Phi\right)\quad \Leftrightarrow\quad \begin{tikzpicture}[baseline={([yshift=-.5ex]current bounding box.center)}, scale=0.7]
 \pgftext{\includegraphics[scale=1]{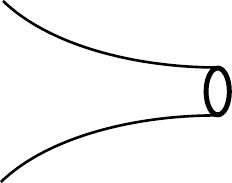}} at (0,0);
     \draw (-0.04, -1.5) node {geodesic boundary};
     \draw (1.75, 0) node {$b$};
     \draw (-2,0) node {$\dots$};
  \end{tikzpicture}\quad\label{expansion}
\end{equation}

We learn from \eqref{64} that the $\tau_k$ operators should be interpreted in JT gravity as cusps (infinitely sharp defects $\alpha=0$) with an additional insertion of $\Phi^{2k}$ at the cusps. This fits nicely with the stringy intuition that local vertex operators (here $\tau_k$) can be viewed as closed string states coming in from $\infty$, the infinity is the fact that the cusp is infinitely sharp and the state is specified by the $\Phi^{2k}$
\begin{equation}
    \tau_k\quad \Leftrightarrow\quad \frac{(-1)^k}{(2k-1)!!}\,\int\d x^2\sqrt{g}\,\Phi^{2k}\,e^{-2\pi\Phi}\quad \Leftrightarrow \quad \begin{tikzpicture}[baseline={([yshift=-.5ex]current bounding box.center)}, scale=0.7]
 \pgftext{\includegraphics[scale=1]{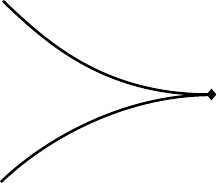}} at (0,0);
     \draw (-0.04, -1.5) node {(nearly) cusp defect};
     \draw (2.5, 0) node {$\a=0$};
     \draw (-1.5,0) node {$\dots$};
  \end{tikzpicture}\quad\label{5.7}
\end{equation}

Suppose now that we take the KdV hierarchy and we consider the following total background $\g_k$
\begin{equation}
    \exp\bigg(2\pi^2\kappa+\sum_{k=2}^\infty t_k \tau_k\bigg)=\exp\bigg(\sum_{k=2}^\infty \g_k \tau_k\bigg)\,,
\end{equation}
where the relation between $\g_k$ and $t_k$ follows from the expansion of the $\kappa$ class in $\tau_k$ operators \eqref{520}. Because the $2\pi^2\kappa$ is specifying JT gravity, we can view the remainder of the exponential as an operator insertion in JT gravity. Using the dictionary \eqref{64} we find that, because the $t_k \tau_k$ are in the exponential, we obtain a deformation of the JT gravity dilaton potential
\begin{equation}
    \exp\bigg(2\pi^2\kappa+\sum_{k=2}^\infty t_k \tau_k\bigg)\quad \Leftrightarrow \quad \exp\bigg(\S\chi+\frac{1}{2}\int\d^2 x\sqrt{g} \,\Phi (R+2)+ \int\d^2 x\sqrt{g}\sum_{k=2}^\infty \frac{(-1)^k }{(2k-1)!!}\,t_k\,\Phi^{2k}\,e^{-2\pi\Phi} \bigg)\,.\label{69}
\end{equation}
These deformations span the class of dilaton gravities that were discussed in \cite{Maxfield:2020ale,Witten:2020ert,Witten:2020wvy}. This formula is exactly true as long as the deformation decays for large $\Phi$ no slower than $e^{-\pi \Phi}$. 

Consider now the string equation \eqref{stringeqtaugen} which for JT gravity $\g_{j+1}=-(-2\pi^2)^j/j!$ becomes
\begin{equation}
    \sum_{a=0}^\infty \frac{(-2\pi^2)^a}{a!}\bigg\langle\tau_a\prod_{i=1}^n\tau_{k_i}\bigg\rangle_\k=\sum_{m=1}^n\bigg\langle \tau_{k_m-1}\prod_{i\neq m}^n \tau_{k_i}\bigg\rangle_\k\,.
\end{equation}
Note that the operator that is being ``inserted'' on the left (replacing the role of the so-called puncture operator $\tau_0$ in topological gravity) is related through \eqref{62} with a Weil-Petersson volume evaluated at $\alpha=2\pi$. We can then rewrite the string equation by summing over the other $\tau_k$ in the correlator with appropriate prefactors, such that they correspond with inserting geodesic boundaries too as in \eqref{61}. One finds that the string equation becomes
\begin{equation}
    V_{g,n+1}(2\pi \i,\,b_1\dots b_n)=\int_0^{b_1} \d a_1 a_1\,V_{g,n}(a_1,b_2\dots b_n)+\dots \int_0^{b_n} \d a_n a_n\,V_{g,n}(b_1\dots b_{m-1},a_n)\,.\label{rel1}
\end{equation}
One can indeed check this explicitly, for instance
\begin{equation}
    V_{1.2}(2\pi \i,b)=\frac{b^4}{192}+\pi^2\frac{b^2}{24}=\int_0^b \d a a\,V_{1,1}(a)\,,\quad V_{1,1}(a)=\frac{a^2}{48}+\frac{\pi^2}{12}\,.
\end{equation}
This relation \eqref{rel1} should be viewed as the first term in the Taylor expansion of Mirzakhani's recursion around $b=2\pi \i$. Indeed, Mirzakhani's recursion relations are \cite{mirzakhani2007simple,Stanford:2019vob}
\begin{align}
b\, V_{g, n+1}(b, B)=\,&\frac{1}{2} \int_{0}^{\infty} \d b' b'\,\d b'' b''\,D(b,b',b'')\bigg( V_{g-1,n+2}(b',b'',B)+\sum_{\text{stable}} V_{h_1,n_1}(b',B_1) V_{h_2,n_2}(b'',B_2)\bigg) \nonumber\\
&+\sum_{i=1}^{n} \int_{0}^{\infty} \d b' b'\,(b-T(b,b', b_i))\, V_{g,n-1}(b', B/b_i)\,.\label{recmirza}
\end{align}
with 
\be 
D(b,b',b'') = 2 \log\left(\frac{e^{b/2} + e^{(b'+b'')/2}}{e^{-b/2} + e^{(b'+b'')/2}}\right),\quad T(b,b',b'') = \log \left( \frac{\cosh\frac{b''}{2} + \cosh\frac{b-b'}{2}}{\cosh\frac{b''}{2} + \cosh\frac{b+b'}{2}} \right).
\ee
Inserting $b=2\pi\i$ (and being careful with branchcuts upon doing the analytic continuation from real $b$) one obtains 
\begin{equation}
    D(2\pi \i,b^{\prime},b^{\prime \prime})=0\,,\quad T(2\pi\i,b',b_k)=2\pi\i\,\theta(b'-b_k)\,,\label{616}
\end{equation}
which reduces the recursion relation \eqref{recmirza} directly to \eqref{rel1}. The other Virasoro constraints \eqref{vir} in the KdV hierarchy compute subleading terms in the Taylor expansion of the volumes around $b=2\pi\i$. 

For instance, working out the $L_0$ constraint (or dilaton equation) for JT gravity gives
\begin{equation}
    \sum_{a=0}^\infty(2a+3) \frac{(-2\pi^2)^a}{a!}\bigg\langle\tau_{a+1}\prod_{i=1}^n\tau_{k_i}\bigg\rangle_\k=\sum_{m=1}^n(2k_i+1)\bigg\langle\prod_{i=1}^n \tau_{k_i}\bigg\rangle_\k\,,
\end{equation}
which in terms of volumes becomes
\begin{equation}
    \bigg(\frac{2}{b}\partial_b+\partial_b^2\bigg)\, V_{g,n+1}(b=2\pi\i ,b_1\dots b_n)=\bigg(n+\sum_{i=1}^n b_i\partial_{b_i}\bigg)\,V_{g,n}(b_1\dots b_n)\,,\label{dildil}
\end{equation}
this relation one can again check explicitly, for instance for $V_{1,2}(b,b_1)$ and $V_{1,1}(b_1)$. In terms of partition functions \eqref{zfromtau} the string equation \eqref{rel1} becomes
\begin{equation}
    Z(b=2\pi i,\,\beta_1\dots \beta_n)_\text{conn}=2\sum_{i=1}^n \beta_i\,Z(\beta_1\dots\beta_n)_\text{conn}\,,
\end{equation}
and the dilaton equation \eqref{dildil} becomes
\begin{equation}
    \bigg(\frac{2}{b}\partial_b+\partial_b^2\bigg)\,Z(b=2\pi i,\,\beta_1\dots \beta_n)_\text{conn}=2\sum_{i=1}^n \beta_i\partial_{\beta_i}\,Z(\beta_1\dots\beta_n)_\text{conn}\,.
\end{equation}
These identities are the equivalent of what Dijkgraaf-Verlinde$^2$ \cite{Dijkgraaf:1991qh} called loop equations for topological gravity. In the latter case the right hand side of the dilaton equation is also equal to $-3\chi Z(\beta_1\dots\beta_n)_\text{conn}$, as in \eqref{427}. This is not true for JT gravity, because the selection rule \eqref{selection} is violated by the $\kappa$ classes in the exponential. However, it turns out \cite{do2008intersection} we also have an equation
\begin{equation}
    \frac{1}{b}\partial_b\,Z(b=2\pi i,\,\beta_1\dots \beta_n)_\text{conn}=-\chi\, Z(\beta_1\dots\beta_n)_\text{conn}\,,\label{619}
\end{equation}
such that the dilaton equation simplifies to
\begin{equation}
    \partial_b^2\,Z(b=2\pi i,\,\beta_1\dots \beta_n)_\text{conn}=\bigg(2\sum_{i=1}^n \beta_i\partial_{\beta_i}-2\chi\bigg)\, Z(\beta_1\dots\beta_n)_\text{conn}\,.\label{620}
\end{equation}

We are interested in understanding these equations (and if possible the other $L_n$ constraints) directly from the dilaton gravity path integral, in the spirit of having a better gravity understanding of the KdV hierarchy. Based on equation \eqref{65} and
\begin{equation}
    \sum_{a=0}^\infty\frac{(-2\pi^2)^a}{a!}\tau_a = \exp\bigg(\frac{1}{2}(2\pi\i)^2\psi_\text{extra}\bigg)\,,
\end{equation}
one might think that this corresponds with a blunt defect $\alpha=2\pi$, which is essentially an area operator (or a marked point which is integrated over spacetime, without creating any source of curvature in that spacetime as backreaction)
\begin{equation}
    \sum_{a=0}^\infty\frac{(-2\pi^2)^a}{a!}\tau_a\quad \overset{?}{\Leftrightarrow} \quad \int\d^2 x\sqrt{g}=A\quad \Leftrightarrow \quad \text{marked point without curvature source}\,.\label{grainofsalt}
\end{equation}
This turns out to be \emph{almost} correct, but not quite. Correlators of blunt defects $\alpha=2\pi$ were discussed around equation (4.23) and (4.18) in \cite{Turiaci:2020fjj}. The JT gravity connected $n$-boundary path integral with $m$ such defects inserted results in\footnote{If we are more careful with prefactors in \cite{Turiaci:2020fjj} we see that actually the relevant operator is $\e A$, then the Weil-Petersson volumes contribute $\epsilon\, 2\pi \chi=0$ and the trumpets contribute $\epsilon \int \d u \sqrt{h}=\beta_i$ indeed. We use the quantization scheme of \cite{Turiaci:2020fjj} here because it is more suitable for making contact with semiclassics, in that scheme for $\alpha\sim 2\pi$ there is roughly an extra prefactor $2\pi-\alpha$ in \eqref{63}, this is the $2\pi/\g(1-\alpha/2\pi)$ in \eqref{5.29} when $\alpha$ is close to $2\pi$, and $\epsilon=2\pi-\alpha$.}
\begin{equation}
    \int_{\b_1\dots\b_n\,\rm conn} \mathcal{D}g \mathcal{D}\Phi \,\bigg(\int\d^2 x\sqrt{g}\bigg)^m e^{-I_{\rm JT}[g,\Phi]} =\bigg(2\sum_{i=1}^n \beta_i\bigg)^mZ(\beta_1\dots \beta_n)_\text{conn}\,,
\end{equation}
which exponentiates indeed to equation (4.23) in \cite{Turiaci:2020fjj}
\begin{equation}
    \int_{\b_1\dots\b_n\,\rm conn} \mathcal{D}g \mathcal{D}\Phi \,\exp\bigg(\lambda\int\d^2 x\sqrt{g}\bigg) e^{-I_{\rm JT}[g,\Phi]}=\exp\bigg(2\lambda\sum_{i=1}^n \beta_i\bigg)Z(\beta_1\dots \beta_n)_\text{conn}\,,
\end{equation}
meaning we have shifted the overall energy scale with $E_0=2\lambda$. Our operator reproduces this behavior, up to contact terms. Indeed we have for instance (using again the string equation)
\begin{equation}
    \sum_{a_1=0}^\infty\frac{(-2\pi^2)^{a_1}}{a_1!}\bigg\langle\tau_{a_1} \sum_{a_2=0}^\infty\frac{(-2\pi^2)^{a_2}}{a_2!}\tau_{a_2}\bigg\rangle_{\k}=\bigg\langle\ \sum_{a_2=0}^\infty\frac{(-2\pi^2)^{a_2+1}}{(a_2+1)!}\tau_{a_2}\bigg\rangle_{\k}\,,
\end{equation}
which translated to Weil-Petersson volumes means
\begin{equation}
    V_{g,2}(2\pi\i,2\pi\i)=\int_0^{2\pi i}\d a a\,V_{g,1}(a)\neq 0\,.
\end{equation}
These contributes from two operators coming into contact give corrections such as the second term in
\begin{equation}
    Z(b=2\pi \i,b=2\pi \i,\beta_1\dots \beta_n)_\text{conn}=\bigg(2\sum_{i=1}^n \beta_i\bigg)^2 Z(\beta_1\dots \beta_n)_\text{conn}+\int_0^{2\pi i}\d a a\,Z(b=a,\beta_1\dots\beta_n)_\text{conn}\,,
\end{equation}
which are absent for the blunt defects $\alpha=2\pi$. In contrast, for topological gravity \cite{dijkgraaf1991topological} the operator on the left-hand side of the string equation is $\tau_0$, and there are then no contact terms because there is no $\tau_{-1}$. This is why in topological gravity $\tau_0$ is exactly the area operator, also called puncture operator $P$.

The blunt defects $\alpha=2\pi$ in JT gravity likewise have no contact terms \cite{Turiaci:2020fjj} (in hyperbolic geometry this is because there is no geodesic that only surrounds a number of these blunt defects, but no handles). Our operators do, because they are built up out of an infinite number of sharp cusps. Their amplitudes are by construction analytic continuations of Weil-Petersson volumes, the blunt defects differ from this by the subtraction of contact terms. Therefore for $\alpha>\pi$ we rather have a relation like
\begin{equation}
    \exp\bigg(\frac{1}{2}(\i\alpha )^2\psi_\text{extra}\bigg)\quad \text{minus contact terms}\quad \Leftrightarrow \quad \int\d^2 x\sqrt{g}\,\frac{2\pi}{\g(1-\alpha/2\pi)}e^{-(2\pi-\alpha)\Phi}\,,\quad \alpha>\pi\,.\label{5.29}
\end{equation}
So a description of blunt defects in terms of cohomology is a but more subtle, we understand that such a description is currently under construction \cite{joaquinlorenztoap}.

Note also that on the path integral level, the fact that we get a simple shift in the energy can also be understood from the fact that adding the area integral $\e A$ to the JT action can be removed by shifting the dilaton, which in turn causes a boundary term to appear that induces the shift in energy. 

In summary, we can trust the dilaton gravity interpretation of the $\tau_k$ in \eqref{64} completely (also when inserting these $\tau_k$ in the action), as long as we do not insert infinite sums that conspire to defects with $\alpha<\pi$, at which point contact terms (dis)appear. 

We end this section with some further comments and open questions.

\begin{enumerate}
    \item It would be nice to have an exact description of our infinite sum \eqref{grainofsalt} in dilaton gravity, then one could write up the dilaton gravity action for all minimal models (by turning off the KdV times $\g_k$ \eqref{520} that took us from topological gravity to JT in the first place).\footnote{See \cite{Goel:2020yxl,Mertens:2020hbs,Mertens:2020pfe} for some (not obviously related) progress in that direction.} There is a description of topological gravity as pure Einstein-Hilbert gravity \cite{Dijkgraaf:1991qh} (section 7.2), where the other minimal models have corrections in the action of the type (notice that equation \eqref{427} is then obvious via Gauss-Bonnet)
    \begin{equation}
        \tau_n\sim \int \d^2 x \sqrt{g} R^n\,.
    \end{equation}
    It would be interesting to see how that worldsheet metric $g$ and the JT gravity metric-and dilaton are related, particularly because there are no AdS$_2$ asymptotics in the first description.
    \item It would be interesting to have some dilaton gravity understanding of the operator on the left-hand side of the dilaton equation(s) \eqref{619} and \eqref{620}, analogous to how the string equation involves an area operator \eqref{grainofsalt}, in the spirit of having a better gravity understanding of the KdV hierarchy. In the quantization scheme of \cite{Maxfield:2020ale} we can immediately understand \eqref{619}\footnote{The $-A/2\pi$ comes from the $\a^{-1} \partial_\alpha$ working on the $2\pi-\alpha$ prefactor in the defect operator $\mo_\text{D}(\alpha)$ in the quantization scheme of \cite{Maxfield:2020ale}.}
    \begin{equation}
        \sum_{a=0}^\infty \frac{(-2\pi^2)^a}{a!}\,\tau_{a+1}\quad \text{minus contact terms}\quad \Leftrightarrow \quad \int\d^2 x\sqrt{g}\,\frac{\epsilon\Phi-1}{2\pi}=\frac{L-A}{2\pi}=-\chi\,,
    \end{equation}
    where we used the boundary conditions on $\Phi$ to show that the boundary contribution (the length $L$) to the area $A$ cancels in this observable, and the bulk part of the area evaluates indeed to $2\pi\chi$. To be clear, in the $\Leftrightarrow$ we used our dictionary \eqref{5.7}, and then we used the JT gravity description to find that the operator on the left computes the Euler character. This precisely reproduces the equation \eqref{619}, which can be viewed as a subset of the KdV equations.
    
    The operator in the dilaton equation \eqref{620} generates a (infinitesimal) metric rescaling (hence the name dilaton equation), which boils down to rescaling the boundary lengths
    \begin{equation}
        \sum_{m=0}^\infty\frac{\lambda^m}{m!} \bigg(2\sum_{i=1}^n \beta_i\partial_{\beta_i}-2\chi\bigg)^m\, Z(\beta_1\dots\beta_n)_\text{conn}=e^{-2\l\chi}\,Z(\beta_1 e^{2\l}\dots\beta_n e^{2\l})_\text{conn}\,.
    \end{equation}
    We can appreciate this by translating the operator in the left hand side of \eqref{620} to dilaton gravity variables, which indeed gives the generator of metric rescalings in the JT gravity action (for the $\chi$ part we remind ourselves of the relation between $\S$ and $\Phi_0$, see for instance \cite{Yang:2018gdb})
    \begin{equation}
        \sum_{a=0}^\infty(2a+1) \frac{(-2\pi^2)^a}{a!}\,\tau_{a+1}\quad \text{minus contact terms}\quad \Leftrightarrow \quad \int\d^2 x\sqrt{g}\,2\,\Phi + \int_\partial \d u\sqrt{h}\,\Phi\,.
    \end{equation}
    So the simplest $L_{-1}$ and $L_0$ of the KdV equations have JT gravity interpretations as describing the covariant transformation rules of observables under certain changes of parameters in the action. This makes sense, as the KdV equations essentially state how $F$ changes with the KdV times $t_k$. The other KdV equations $L_n Z=0$ with $n>0$ are essentially equivalent to \eqref{69} in combination with the statement that these deformed theories are matrix integrals with different spectral curves $\rho_0(E)$ \cite{Maxfield:2020ale}. These constraints state how observables change under changes in the dilaton gravity action with $t_k$ and $k>1$, namely by changing the spectral curve.
    
    The string-and dilaton equation are special cases describing the flow under changes of $t_0$ and $t_1$. This means that given the dictionary \eqref{5.7} one can derive the KdV hierarchy from the JT gravity path integral formulation, essentially.
    \item Let us emphasize that without summing over the $\tau_k$, that the KdV equations should be interpreted as recursion relations between cusps in JT gravity on closed manifolds. Without knowledge of explicit Weil-Petersson volumes these relations seem miraculous. It would be interesting to find an independent dilaton gravity calculation of even the simplest examples, for instance the correlator of one cusp on the torus
    \begin{equation}
        \int_\text{torus} \md g\,\md \Phi\,e^{-I_\text{JT}}\, \int \d^2 x \sqrt{g}\, e^{-2\pi \Phi}\,\Phi^{2k}=\quad \begin{tikzpicture}[baseline={([yshift=-.5ex]current bounding box.center)}, scale=0.7]
 \pgftext{\includegraphics[scale=1]{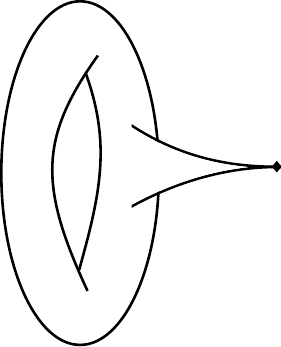}} at (0,0);
     \draw (1.9,0.05) node {$k$};
  \end{tikzpicture}\quad=0\,,\quad k\neq 1\,.
    \end{equation}
    Intersection theory predicts in particular that this vanishes for all $k\neq 1$ and it would be interesting to have a direct dilaton gravity argument for this.
    
    Remember furthermore the relation \eqref{65}
    \begin{equation}
        \int \d^2 x \sqrt{g}\, e^{-(2\pi-\alpha) \Phi}\quad \Leftrightarrow \quad \exp\bigg(\frac{1}{2}(\i\alpha )^2\psi_\text{extra}\bigg)\,,\quad \alpha<\pi\,,
    \end{equation}
    and notice that the right-hand side is even in $\alpha$. This implies that correlators of cusps multiplied with odd powers $\Phi^{2k+1}$ vanish identically in JT gravity (on any surface, and regardless of other operators)
    \begin{equation}
        \int \d^2 x \sqrt{g}\, e^{-2\pi \Phi}\,\Phi^{2k+1}=0\,.\label{636}
    \end{equation}
    This too is mysterious without explicit knowledge of the volumes, it boils down to understanding why the Weil-Petersson volumes are even polynomials of $b_i^2$. One can view this as a redundancy, or a null deformation of the type discussed in \cite{Blommaert:2022ucs}. Indeed, one can add terms like \eqref{636} to the JT action as in \eqref{69} without affecting any observables. So theories whose actions differ by terms of the type \eqref{636} are completely equivalent, and having them or not having them is a gauge choice. As a simple check on this, notice that a potential $U(\Phi)$ as in \eqref{636} is in the kernel of equation (1.4) in \cite{Witten:2020wvy}, the spectral curve (and thus all observables) remain unaffected. This redundancy is also why \eqref{expansion} is an analytic continuation of \eqref{65}, one can replace the cosine by either one of the exponentials without changing any observables.
    
    It is natural to wonder if there is some contour-integral argument for \eqref{636}, directly in the dilaton gravity formulation (without going through the volumes).
\end{enumerate}
\subsection{Matrix integral}\label{sect:matrix}
As the final element in the web of dualities between intersection numbers, matrix integrals and dilaton gravities we present the matrix integral interpretation of the $\psi$-classes. We claim that every insertion of $\tau_k$ corresponds with the following counterclockwise contour integral around the real axis (a contour above and below the real axis, the latter is denoted by $R$)
\begin{equation}
    \tau_k \quad \Leftrightarrow\quad  \frac{(-1)^{k+1}}{\i}\frac{1}{(2k+1)!!}\oint_R \d E\, E^{k+1/2}\Tr \delta(H-E)=\frac{(-1)^{k+1}}{\i}\frac{1}{(2k+1)!!}\oint_R \d E\, E^{k+1/2}\,\rho(E)\,.\label{oper}
\end{equation}
We will derive this equation momentarily, but let us first give a few basic checks that it is correct, and show how this equation is useful in practice. (To our knowledge, this equation and its practical use are new, whereas versions of \eqref{656} were known.)

As a first check notice that $\rho_0(E)$ has no poles on the real axis, for instance for JT gravity $\rho_0(E)=e^{\S}\sinh(2\pi E^{1/2})/4\pi^2$ for positive energies and zero otherwise. So we recover the statement that $\average{\t_k}_0=0$, which is true for generic backgrounds $\g_p$. In intersection theory this follows from the selection rule \eqref{selection}, which for $g=0$ we can rewrite as
\begin{equation}
    \sum_{i=1}^n (k_i-1)+k=-2\,,\quad k_i\geq 2\,,\label{638}
\end{equation}
where the $k_i$ come from expanding out the background and $k_i\geq 2$ because $\g_0=\g_1=0$ (in particular, $\g_0=0$ is important). The left-hand side is non-negative so this is never satisfied, so $\average{\t_k}_0=0$ indeed.\footnote{Notice that the cases $k\leq-2$ can and do give nonzero answers at genus zero, as we found around equation \eqref{Fx}.}

At non-zero genus we obtain finite answer, because $E^{k+1/2}\rho(E)_g$ generically has poles at $E=0$. To appreciate how this arises we can consider the inverse Laplace transform of \eqref{genusg}
\begin{equation}
    Z_g(\beta)=\int_0^\infty \d E\, e^{-\beta E}\,\rho_g(E)\quad \Rightarrow\quad \rho_g(E)=\int_0^\infty \d b b\, \frac{1}{2\pi E^{1/2}}\cos(b E^{1/2})V_{g,1}(b)\,,
\end{equation}
where the volume is an even polynomial \eqref{vexp}
\begin{equation}
    V_{g,1}(b)=\sum_{d=0}^\infty V_{g,d}\,\frac{b^{2d}}{4^d d!}\quad \Rightarrow\quad \rho_g(E)=\frac{1}{2\pi}\sum_{d=0}^\infty V_{g,d}\,(-1)^{d+1} \frac{(2d+1)!!}{2^d}E^{-d-3/2}\,.
\end{equation}
Using this expressions and picking up the pole at the origin we obtain
\begin{align}
    \average{\tau_k}_g &=\frac{(-1)^{k+1}}{\i}\frac{1}{(2k+1)!!}\oint_R \d E\, E^{k+1/2} \rho(E)_g\\&=\sum_{d=0}^\infty V_{g,d} (-1)^{k+d} \frac{1}{2^d}\frac{1}{2\pi\i}\oint_0 \d E\,E^{-1+k-d}=\frac{1}{2^k}V_{g,k}\,,
\end{align}
which means that the operator \eqref{oper} computes the expansion coefficients of the volumes. In other words our matrix integral definition \eqref{oper} is recovering the expansion of volumes in intersection numbers \eqref{53} (in a generic background $\g_k$)
\begin{equation}
    V_{g,n}(b_1\dots \b_n)=\sum_{d_1=0}^\infty\frac{b_1^{2d_1}}{2^{d_1}d_1!}\dots\sum_{d_n=0}^\infty\frac{b_n^{2d_n}}{2^{d_n}d_n!}\average{\tau_{d_1}\dots \tau_{d_n}}_g\,.
\end{equation}
The equation above was for $n=1$ but it is obvious that this extends to generic $n$.

As a further check on this we can compute the disk with a puncture. This example involves picking up the pole at $E_1=E_2$ of the genus zero wormhole, equation (139) in \cite{Saad:2019lba},
\begin{align}
    \average{\rho(M)\tau_k}_{0\,\text{conn}}&=\frac{(-1)^{k+1}}{\i}\frac{1}{(2k+1)!!}\oint_R \d E\, E^{k+1/2} \average{\rho(M)\rho(E)}_{0\,\text{conn}}\\&=\frac{(-1)^{k}}{\i}\frac{1}{(2k+1)!!}\frac{1}{4\pi^2M^{1/2}}\oint_M \d E\, E^{k}(E+M)\frac{1}{(E-M)^2}\nonumber\\
    &=\frac{1}{2\pi}\frac{(-1)^k}{(2k-1)!!}M^{k-1/2}\,,
\end{align}
such that
\begin{equation}
    \average{Z(\beta)\tau_k}_{0\,\text{conn}}=\frac{1}{2\pi}\frac{(-1)^k}{(2k-1)!!}\int_0^\infty \d M\,e^{-\beta M}\, M^{k-1/2}=\frac{(-1)^k}{2^{k+1}\pi^{1/2}}\,\beta^{-1/2-k}\,.
\end{equation}
We can compare this with the prediction that one gets by viewing the trumpet as a generating function for these one-point functions, along the lines of \eqref{expansion}
\begin{align}
    Z_\text{trumpet}(\beta,b)= \frac{1}{2\pi^{1/2}\beta^{1/2}}e^{-\frac{b^2}{4\beta}}&=\sum_{k=0}^\infty \frac{b^{2k}}{2^k k!}\frac{(-1)^k}{2^{k+1}\pi^{1/2}}\,\beta^{-1/2-k}=\sum_{k=0}^\infty \frac{b^{2k}}{2^k k!}\average{Z(\beta)\tau_k}_{0\,\text{conn}}\,,
\end{align}
which indeed gives the same answer on the nose. Finally we also recover the statement that the cylinder amplitude vanishes in intersection theory (or the genus zero two point function)\footnote{Actually, if we include the contribution from unstable surfaces, we should continue to all integers $n$ and $k$, in which case we get $\average{\tau_{-1-k} \tau_k}_0 = (-1)^{k}$. This is consistent with the $1/(x_1 + x_2)$ in \eqref{Fx1x2}.}
\begin{align}
    \average{\tau_n\tau_k}_{0}&=\frac{(-1)^{n+1}}{i}\frac{1}{(2n+1)!!}\oint_R \d E\, E^{n+1/2} \average{\rho(E)\tau_k}_{0\,\text{conn}}\nonumber\\&=\frac{(-1)^{n+1+k}}{2\pi i}\frac{1}{(2n+1)!!(2k-1)!!}\oint_R \d E E^{n+k}=0\,.\label{5.47}
\end{align}
For generic backgrounds $\g_k$ this follows from the same logic as around \eqref{638} (but with $-1$ on the right). 

Now we explain how to derive this equation \eqref{oper} for $\tau_k$. The first step is proving an identity that appears in \cite{Maldacena:2004sn, Gross1989nonperturbative}, which holds miraculously for each arbitrary (but fixed) number $E_0$
\begin{equation}
    \Tr\bigg(\frac{1}{y-H}\bigg)-\frac{\Tr(1)}{y}=\sum_{k=1}^\infty (-1)^{k+1}(-y-E_0)^{-k-1/2}(-y+E_0)^{-1/2}\Tr((H+E_0)^{k-1/2}(H-E_0)^{1/2})_+\,.\label{649}
\end{equation}
The branchcuts of the roots are chosen in the usual way, we have $(-y-E_0)^{-k-1/2}>0$ when $y<-E_0$ and $(-y+E_0)^{-1/2}>0$ when $y<E_0$. Let us prove this equation for $y>E_0$ and real, after collecting all the signs it simplifies slightly
\begin{align}
    \Tr\bigg(\frac{1}{y-H}\bigg)-\frac{\Tr(1)}{y}&=\sum_{k=1}^\infty (y+E_0)^{-k-1/2}(y-E_0)^{-1/2}\Tr((H+E_0)^{k-1/2}(H-E_0)^{1/2})_+\nonumber\\&=\sum_{q=1}^\infty\frac{1}{y^{q+1}}\,\Tr(H^q)\,.
\end{align}
The $+$ means we should expand in powers of $1/H$ and keep only the terms with positive powers of $H$ in the resulting series. To check this, one explicitly expands the binomials in $1/H$, and then rearranges the resulting triple sum to collect the terms that multiply $\Tr(H^q)$, for fixed $q$. The double sum at fixed $q$ surprisingly spits out $1/y^{q+1}$, and we recover the large $y$ expansion of the left-hand side \cite{Blommaert:2021gha}.

Now we can consider the following double scaling limit of \eqref{649} where we take $y=-E_0-z^2$, shift also $H$ by $E_0$ and send $E_0\to\infty$
\begin{equation}
    \Tr\bigg(\frac{1}{z^2+H}\bigg)-\frac{\Tr(1)}{E_0+z^2}=\sum_{k=1}^\infty (-1)^{k}z^{-2k-1}\,W_{k-1}(H)\,,\label{651}
\end{equation}
where we have introduced the so called scaling polynomials of \cite{Gross1989nonperturbative}
\begin{equation}
    W_{k-1}(H)=\lim_{E_0\to\infty} (2E_0)^{-1/2}\Tr(H^{k-1/2}(H-2E_0)^{1/2})_+\,.\label{652}
\end{equation}
The $+$ is confusing in this context, because it is unclear what the expansion in powers of $H$ means upon double scaling. Below we will give a more correct formula that does not involve the subscript $+$ anymore, but is purely based on a contour integral, and therefore does make sense in the double scaling continuum limit.

Using the definition of an FZZT brane \cite{Maldacena:2004sn,Saad:2019lba,Fateev:2000ik,Ponsot:2001ng,Goel:2020yxl,Mertens:2020hbs,Mertens:2020pfe,Hosomichi:2008th,Kostov:2002uq,Okuyama:2021eju,Teschner:2000md,Okuyama:2021eju,Blommaert:2019wfy,Blommaert:2021etf,Blommaert:2021fob}
\begin{equation}
    \mo_\text{FZZT}(z)=\Tr \log(z^2+H)-\Tr(1)\log(z^2)=\int_\infty^{z}\d w\,2w \bigg(\Tr\bigg(\frac{1}{w^2+H}\bigg)-\frac{\Tr(1)}{w^2}\bigg)\label{653}
\end{equation}
and its relation with a geodesic boundary \cite{Blommaert:2021fob} (we are not giving any topological weight to FZZT branes here, this is an irrelevant choice)
\begin{equation}
    \mo_\text{G}(b)=-\frac{e^{-\S}}{2\pi \i}\int_{-\i\infty}^{+\i\infty}\d z \,e^{b z}\, \mo_\text{FZZT}(z)\,,
\end{equation}
we obtain an expansion of the geodesic boundary in scaling polynomials\footnote{It is important that the FZZT branes actually have their poles at $z=-\epsilon\pm \i \sqrt{H}$ \cite{Blommaert:2021gha}, such that in reality the integrand contains $(z+\epsilon)^{-2k}$. Since $b>0$ we can close the contour to $z=-\infty$, and there is no pole at infinity because of the second term on the first line.}
\begin{align}
    \mo_\text{G}(b)&=\frac{e^{-\S}}{b}\frac{1}{2\pi \i}\int_{-\i\infty}^{+\i\infty}\d z\, e^{b z}\, 2z\bigg(\Tr\bigg(\frac{1}{z^2+H}\bigg)-\frac{\Tr(1)}{E_0+z^2}\bigg)\nonumber\\&=2\,e^{-\S}\sum_{k=1}^\infty (-1)^kW_{k-1}(H)\frac{1}{b}\frac{1}{2\pi \i}\int_{-\i\infty}^{+\i\infty}\d z e^{b z}z^{-2k}\nonumber\\
    &=e^{-\S}\sum_{k=0}^\infty (-1)^{k+1}\frac{b^{2k}}{(2k)!}\frac{2}{2k+1}W_k(H)\,.
\end{align}

Comparing with the expansion of a geodesic boundary in cusp defects \eqref{expansion}, we find the identification of the $\tau_k$ insertions with the scaling polynomials \cite{Gross1989nonperturbative}
\begin{equation}
    \mo_\text{G}(b)=e^{-\S}\sum_{k=0}^\infty \frac{b^{2k}}{2^k k!}\,\tau_k\quad \Rightarrow \quad \tau_k\quad \Leftrightarrow\quad (-1)^{k+1}\frac{2}{(2k+1)!!}W_k(H)\,.\label{656}
\end{equation}
Now the question is what the correct implementation of $W_k(H)$ in the double scaling limit is. In \eqref{652} the last square root $(H-2E_0)^{1/2}$ is on the branchcut, so depending on whether $H$ takes values slightly above or below the real axis we get respectively a factor $\pm \i$. We thus have two possible contours for the eigenvalues of $H$ in this observables. We claim that the correct combination is the average of both contours
\begin{equation}
    W_k(H)= \frac{\i}{2} \int_{R+\i\epsilon}\d E\, E^{k+1/2}\Tr \delta(H-E)-\frac{\i}{2}\int_{R-\i\epsilon}\d E\, E^{k+1/2}\Tr \delta(H-E) 
\end{equation}
We can flip the orientation of the first contour, so that this becomes a counterclockwise contour integral around the real axis, which picks up any poles that might occur on the real axis in the integrand
\begin{equation}
    W_k(H)=\frac{1}{2\i}\oint_R \d E\, E^{k+1/2}\Tr \delta(H-E)\,.
\end{equation}
This reproduces our original claim \eqref{oper}, which we have independently proven to be correct above, by reproducing all intersection numbers from it.

To close off this section we remark how this relates to the Kontsevich matrix integral \cite{KontsevichModel} and the equation \eqref{69}. We have
\begin{equation}
    \prod_{i=1}^\infty\det(1+H/z_i^2)=\exp\bigg(\sum_{i=1}^\infty \mo_\text{FZZT}(z_i) \bigg)\quad \Leftrightarrow\quad \exp\bigg(\sum_{k=2}^\infty t_k(z_j)\,\tau_k\bigg)\,,\quad t_k(z_j)=-(2k-1)!!\sum_{i=1}^\infty z_i^{-2k-1}\,,
\end{equation}
where in the first equality we used $\det(A)=\exp(\Tr(\log(A)))$, and in the second equality we computed the $w$ integral in \eqref{653} using the expansion in scaling polynomials \eqref{651}, and identified the $\tau_k$'s using \eqref{656}. Kontsevich proved that this identity is correct, namely inserting the left hand side in a matrix integral with a spectral curve corresponding to KdV times $\g_k$ generates a matrix integral whose spectral curve corresponds with KdV times $\g_k+t_k(z_j)$. He did this by proving that the right-hand side equals $F$ for those KdV times (with $F$ the function that satisfies the KdV hierarchy \eqref{kdv}), see also appendix \ref{app:c2}.

In essence, this relation is the raison d'etre of the scaling polynomials, we can start with a Gaussian matrix integral (which double scales to topological gravity \cite{Saad:2019lba}) and turn on some scaling polynomials in the action to get a double scaled theory whose spectral curve has $t_k(z_j)$ turned on.
\section{Concluding remarks}
We end this paper with three comments.

In \textbf{section \ref{sect:5.4}} we use the KdV equations \eqref{KdVrec} to demonstrate that the $(2g)!$ growth in volumes $V_{g,2}(b_1,b_2)$ for generic models comes from terms in the polynomials with order one powers of $b_1^2$ and $b_2^2$. This is part of the reason why the genus expansion converges in the $\tau$-scaling limit.

In \textbf{section \ref{sect:6.2}} we contemplate the multi-boundary generalization of our discussion, in particular we consider the multi-boundary generalization of the $\tau$-scaled spectral form factor and discuss cancellations.

In \textbf{section \ref{sect:lorentzian}} we explain the logical need for a Lorentzian interpretation of the cancellations and the universal scaling behavior \eqref{545} of late-time genus $g$ wormholes, and preview how this comes about by thinking about Lorentzian topology change in JT gravity \cite{workinprogress}.
\subsection{Open-closed duality and factorial growth}\label{sect:5.4}
As a final application of the KdV equations in gravity we want to give an intuitive argument explaining that the famous $(2g)!$ growth in volumes $V_{g,2}(b_1,b_2)$ for generic models of dilaton gravity comes from terms in the polynomials with order one powers of $b_1^2$ and $b_2^2$ (so not scaling with $g$). Therefore they will not survive in our $T\to\infty$ limit, which acts at each genus $g$ individually because we also send $e^{\S}\to\infty$. This is part of the explanation why the genus expansion is convergent in our setup, see section \ref{sect:euclideanwormholes}.

For concreteness we consider the constant term in the polynomials \eqref{53}
\begin{equation}
    V_{g,2}(0,0)=\bigg\langle \tau_0^2 \exp\bigg(\sum_{k=2}^\infty \g_k\tau_k\bigg)\bigg\rangle_g\,.
\end{equation}
We would like to argue that this grows as $(2g)!$ for large genus $g$, and that the terms with large powers of $b_1^2$ and $b_2^2$ do not. The basic intuition is that for small powers of $b_i^2$, following the selection rule \eqref{selection}, we can get many, many low-dimensional forms such as $\tau_2$ coming out of the exponential, and it makes sense that such correlators with many, many operators would grow factorially. On the other hand, for large powers of $b_i^2$ the selection rule \eqref{selection} does not allow for that many operators to come down from the exponential (for the maximal power no operators come from the exponential and we are effectively computing Airy volumes again). With relatively fewer operators, it is hard to imagine factorial growth.

We believe this intuition should universally hold, but to make our point concrete we will here focus on the $(2,3)$ minimal string with spectrum \cite{Maldacena:2004sn,Saad:2019lba}
\begin{equation}
    2\pi e^{-\S}\rho_0(E)=E^{1/2}+\frac{t_2}{3}E^{3/2}\,.\label{minmal}
\end{equation}
This has an eigenvalue instanton saddle near $E=-3/t_2$ in the matrix integral and therefore, following the techniques of section 5.6 in \cite{Saad:2019lba} one expects factorial growth of the type (in this section we ignore for simplicity of presentation most sub-exponential $g$-dependence)
\begin{equation}
    V_{g,2}(0,0)=\langle \tau_0^2 \exp(t_2\tau_2)\rangle_g\sim (2g)!\, t_2^{3g}\,,
\end{equation}
where we used \eqref{4.16} and the fact that $u_0=0$ for this deformation. This arises from intersection theory as follows. Using the selection rule \eqref{selection} we immediately obtain the correct scaling with $t_2$
\begin{equation}
    V_{g,2}(0,0)=\frac{t_2^{3g-1}}{(3g-1)!}\langle\tau_0^2\tau_2^{3g-1}\rangle_g\,.
\end{equation}
Now we want to prove that the correlator $\langle\tau_0^2\tau_2^{3g-1}\rangle_g$ for large genus grows as $(5g)!$, such that combined with the $1/(3g)!$ we get the predicted double factorial growth. For this we can use the happy fact that we can rewrite the KdV recursion \eqref{KdVrec} entirely as a recursion relation for
\begin{equation}
    f(g)=\langle\tau_0\tau_1\tau_2^{3g-2}\rangle_g=\frac{1}{3g-1}\langle\tau_0^2\tau_2^{3g-1}\rangle_g\,,
\end{equation}
by considering $k=2$ and $d_2=3g-2$ with all other $d_j=0$, and repeatedly using the string-and dilaton equations \eqref{426} and \eqref{427} to eliminate excess $\tau_0$ and $\tau_1$'s in all terms. For large genus, the dominant contributions to \eqref{KdVrec} in this setup come from the first, and the last term. One can check this by first assuming that this is true, then one finds the recursion relation
\begin{equation}
    f(g)\sim g^5 f(g-1)\quad \Rightarrow\quad f(g)\sim (5g)!\,.
\end{equation}
One then checks that with this $g!^5$ growth the other terms in the recursion relation become subleading indeed. Being more careful with all the prefactors and sub-factorial $g$-dependence one can recover the full prediction of the eigenvalue instanton from this recursion relation \eqref{KdVrec}.
\subsection{Multi-boundary generalization}\label{sect:6.2}
Let us briefly comment on the multi-boundary generalization of the story that we have presented here.\footnote{We thank Douglas Stanford for discussions on this.} For instance, we could consider the three-boundary generalization of the spectral form factor
\begin{equation}
    Z(\beta+\i T_1,\beta+\i T_2,\beta+\i T_3)_\text{conn}\,,\quad T_1+T_2+T_3=0\,,
\end{equation}
in the generalized $\tau$-scaling limit where $\tau_1=T_1e^{-\S}$ and $\tau_2=T_2e^{-\S}$ remain finite (and their difference remains finite as well). The triple energy integral in \eqref{basic} is then dominated by $E_1$, $E_2$ and $E_3$ all close together, and we can use the three-boundary generalization of the sine-kernel \eqref{sinekernel}, which can be found for instance in \cite{Blommaert:2019wfy}. Doing the the Fourier transforms over the small energy differences explicitly, one finds a generalization of the ramp-and plateau \eqref{27}
\begin{equation}
    Z(\beta+\i T_1,\beta+\i T_2,\beta-\i T_1-\i T_2)_\text{conn}=\int_0^\infty \d E\,e^{-3\beta E}\,\begin{cases}
    0\, &\frac{T_1}{2\pi}+\frac{T_2}{2\pi}<\rho_0(E)
    \\\frac{T_1}{2\pi}+\frac{T_2}{2\pi}-\rho_0(E)\, &\frac{T_1}{2\pi}<\rho_0(E)<\frac{T_1}{2\pi}+\frac{T_2}{2\pi}
    \\\frac{T_2}{2\pi}\, &\frac{T_2}{2\pi}<\rho_0(E)<\frac{T_1}{2\pi}\\\rho_0(E)\, &\rho_0(E)<\frac{T_2}{2\pi}\,,
\end{cases}
\end{equation}
where we have specialized to $T_2<T_1$. Following the steps that led to \eqref{expaexpa}, this can be massaged into
\begin{equation}
    Z(\beta+\i T_1,\beta+\i T_2,\beta-\i T_1-\i T_2)_\text{conn}=\frac{e^{\S}}{6\pi\beta}\int_0^{\tau_2}\d f\,e^{-3\beta E(f)}-\frac{e^{\S}}{6\pi\beta}\int_{\tau_1}^{\tau_1+\tau_2}\d f\,e^{-3\beta E(f)}\,,\label{5.68}
\end{equation}
for instance for topological gravity $E(f)=f^2$ this is simply the sum of three Erf functions. For infinite times the first term produces the generalized plateau value $Z(3\beta)$, which comes from terms in the triple sum where all energies coincide. However, notice that if one Taylor expands this in $e^{-\S}$ that we again get a series in powers of $e^{-2\S}$ starting at $e^{-2\S}$, whereas the three-boundary connected amplitudes scale as $e^{-(2g+1)\S}$ with $g$ the genus. So, unlike for the two-boundary spectral form factor we do not see any obvious way to reproduce this tau-scaled answer from the sum over wormhole geometries with three boundaries, at least for now.

In some sense this emphasizes how nice it is that this worked for two boundaries, but on the other hand it also shows that we might still need D-brane effects to understand more complicated phenomena.

One thing that does happen for any number of boundaries is that the cancellations in intersection theory \eqref{548} constrain the volumes in a non-trivial manner for all theories. Namely with the constraint $T_1+\dots+T_n=0$ we find the generalization of \eqref{4.27} for any of the times $T_i\to\infty$
\begin{equation}
    e_1^{m_1}e_2^{m_2}\dots e_n^{m_n}\sim T_i^{2\sum_{j=2}^n m_j}\,.\label{4.27}
\end{equation}
The power of $T$ that appears here is precisely the expression that was constrained by the KdV equations to be upper bound by $2 g$, therefore the maximal power of any of the times $T_i$ is constrained
\begin{equation}
    Z_g(\beta+\i T_1\dots\beta+\i T_n)_\text{conn}\sim T_i^{2g+1}e^{-(2g+n-2)\S}\,,\quad T_1+\dots+T_n=0\,.
\end{equation}
On the other hand dimensional analysis of the volumes for generic dilaton gravities would again suggest that naively we could get powers up to $T_i^{3g-2+n}$. So there are also major cancellations in $V_{g,n}(b_1\dots b_n)$ for generic dilaton gravity models. It remains to be seen whether or not those are related with explaining a generalization of the ramp-and plateau such as \eqref{5.68}. That would be interesting.
\subsection{Universal powers of time via Lorentzian topology change}\label{sect:lorentzian}
We want to stress that recovering the universal growth $T^{2g+1}$ is highly non-trivial from the Euclidean gravitational path integral at genus $g$. In 2d dilaton gravity, we understand why this happens, namely because of known cancellations in intersection numbers. But the scaling of wormhole amplitudes with $T^{2g+1}$ for late times should hold for essentially any gravity model in a regime dominated by black holes, because of random matrix universality \cite{Haake:1315494}. 

It is likely that one could argue that late-time physics is effectively dominated by the near-horizon region of black holes, where one often recovers JT gravity. Nevertheless we do not think an explanation in terms of intersection numbers for real-life black holes is fully satisfactory.

Instead, we view the fact that the scaling $T^{2g+1}$ is highly non-trivial in Euclidean signature (requiring quite miraculous, exact cancellations) as a sign that we should really be looking for a Lorentzian picture. After all, without Lorentzian large times there is nothing in the genus $g$ amplitudes alerting us of any cancellations. Because of the universality, we think the Lorentzian picture should be sufficiently simple (unlike the Euclidean story), in the sense that one could imagine a generalization to higher dimensional black holes.

We believe that we have found such an interpretation by thinking about Lorentzian topology change. The idea is that we can build Lorentzian wormhole geometries using the crotch singularities of Louko-Sorkin \cite{Louko:1995jw,workinprogressmisha}.\footnote{For other recent appearances of singular spacetimes in Lorentzian signature see for instance \cite{Marolf:2022ybi}.} For late enough times, the locations of the crotches (these are the places where baby universes or wormholes are born, and die) should behave approximately as zero modes. These $2g$ zero modes give a volume factor $T^{2g}$, which along with the usual factor $T$ from the rotational zero mode of the double cone \cite{Saad:2018bqo} reproduces the universal $T^{2g+1}$.

We will present this picture elsewhere in more detail, with concrete calculations in JT gravity \cite{workinprogress}.
\section*{Acknowledgments}
We thank Alex Altland, Jan Boruch, Fabian Haneder, Thomas Mertens, Klaus Richter, Phil Saad, Steve Shenker, Julian Sonner, Douglas Stanford, Juan-Diego Urbina, Misha Usatyuk, Torsten Weber and Zhenbin Yang for useful discussions. We also thank Fabian Haneder, Klaus Richter, Juan-Diego Urbina and Torsten Weber for sharing a version of their draft. AB was supported in part by a BEAF fellowship, by the SITP at Stanford, and by the ERC-COG Grant NP-QFT No. 864583. JK is supported by the Simons Foundation. SY is supported in part by NSF grant PHY-1720397 and by a Clark fellowship at SITP.

\appendix


\section{Examples}\label{app:a}

We invite interested readers to take their favorite spectral curve $y(z)$ of the form \eqref{speccurve}, compute the genus $g$ wormhole using topological recursion, and compare the result with the expansion \eqref{expaexpa}. It is satisfactory to see that they match, especially given how non-trivial the cancellations \eqref{genericcancel} are.

It is quite elementary to proof this experimentally for fixed genus and generic spectral curve \eqref{speccurve} by computing both sides for generic $f_k$, but we leave explicit expressions out to spare the reader's eyes. Instead let us go through a few basic examples where we also have an analytic formula for the inverse of the spectrum $E(f)$.

\subsection*{Topological gravity}
\begin{figure}[t]
    \centering
    \begin{tikzpicture}[baseline={([yshift=-.5ex]current bounding box.center)}, scale=0.7]
 \pgftext{\includegraphics[scale=1]{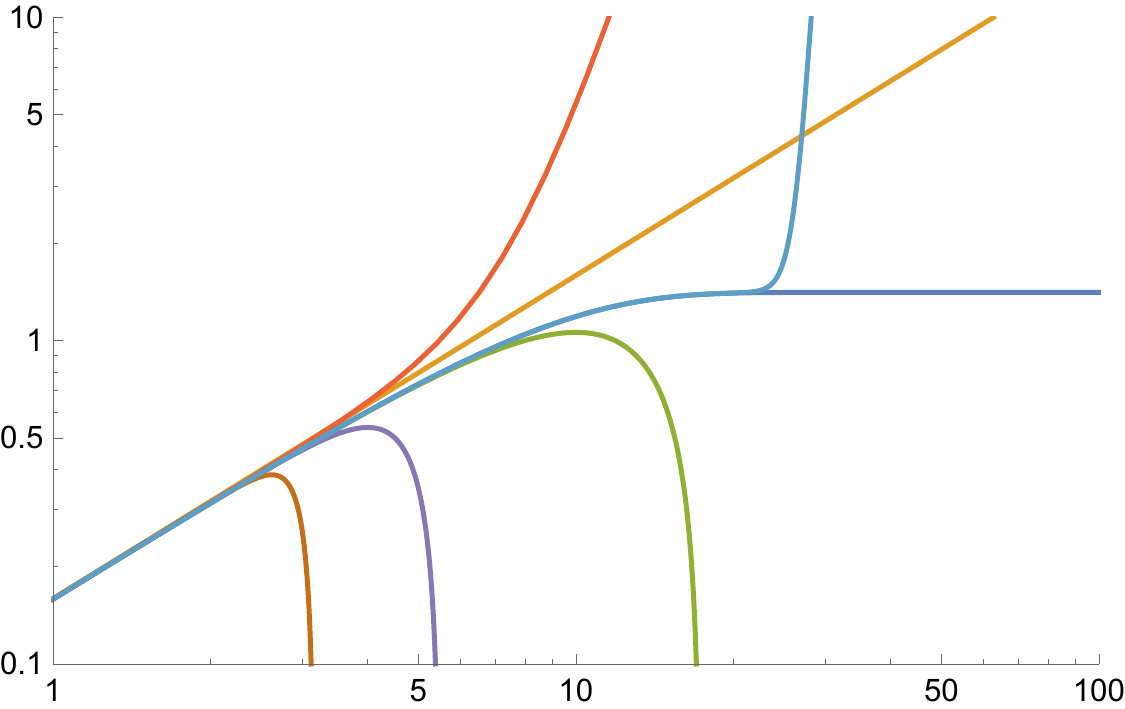}} at (0,0);
     \draw (-0.2, 3) node {$2$};
     \draw (4.5, 3) node {$0$};
     \draw (1.9, 3) node {$12$};
     \draw (-2.1, -2.5) node {$1$};
     \draw (-0.8, -2.5) node {$3$};
     \draw (0.8, -2.5) node {$7$};
     \draw (4.5, 1.2) node {exact};
     \draw (5, -2.5) node {$T$};
  \end{tikzpicture}\quad
    \caption{The $\tau$-scaling limit of the sum over genus $g$ wormholes in the Airy model \eqref{airyexpa} up to $g=g_\text{max}$ (numbers shown) with $e^{\S}=10$ and $\beta=1/2$. This series converges to the exact expression \eqref{airyexact} for all $T$ (infinite radius of convergence) and in particular it converges to the plateau.}
    \label{fig:airyconvergence}
\end{figure}

The Airy model has spectral curve $\rho_0(E)=e^{\S}E^{1/2}/2\pi$, so the inverse spectrum is $E(f)=f^2$ and \eqref{28} evaluates to (we introduce $\tau=Te^{-\S}$)\cite{Okuyama:2020ncd,workinprogressothergroup}
\begin{align}
    Z(\beta+\i T,\beta-\i T)_\text{conn}&=\frac{e^{\S}}{4\pi\beta}\int_0^{\tau}\d f\,e^{-2\beta E(f)}=\frac{e^{\S}}{2^{7/2}\pi^{1/2}\beta^{3/2}}\,\text{Erf}\left(2^{1/2}\beta^{1/2}T e^{-\S}\right)\,\label{airyexact}
\end{align}
Using Erf$(\infty)=1$ we see the plateau value $Z_0(2\beta)$ or the Airy model. The (convergent) Taylor expansion is
\begin{equation}
    Z(\beta+\i T,\beta-\i T)_\text{conn}=\frac{1}{4\pi\beta}\,T-\frac{1}{6\pi}\,T^3e^{-2\S}+\frac{\beta}{10\pi}\,T^5\, e^{-4\S}-\frac{\beta^2}{21\pi}\,T^7\, e^{-6\S}+\frac{\beta^3}{54\pi}\,T^9\,e^{-8\S}\dots\label{airyexpa}
\end{equation}
This expansion is recovered on the nose using topological recursion with the spectral curve $y(z)=z/2$, see Fig. \ref{fig:airyconvergence}.
The volumes to be used in \eqref{genusg} in this case are
\begin{align}
    V_{1,2}(b_1,b_2) &= \frac{b_1^4 + 2b_2^2b_1^2 + b_2^4}{192}\nonumber\\
    V_{2,2}(b_1,b_2) &= \frac{b_1^{10}+15 b_2^2 b_1^8+58 b_2^4 b_1^6+58 b_2^6 b_1^4+15 b_2^8 b_1^2+b_2^{10}}{4423680}\nonumber\\
    V_{3,2}(b_1,b_2) &= \frac{5 b_1^{16}+200 b_2^2 b_1^{14}+2156 b_2^4 b_1^{12}+8048 b_2^6 b_1^{10}+12140 b_2^8 b_1^8+\text{symmetric}}{4280706662400}\label{airyvol}
\end{align}
In the final term in \eqref{airyexpa}, a possible $T^9 e^{-6\S}$ does not appear because of the cancellation \eqref{cancelbasic}. Notice that the cancellations are exact, in line with the statement that \eqref{sinekernel} is effectively exact in the $\t$-scaling limit. At genus $4$ there is another cancellation responsible for the lack of an $T^{11} e^{-8\S}$ term, and so on.
\subsection*{JT gravity}
For JT gravity the spectrum is $\rho_0(E)=e^{\S}\sinh(2\pi E^{1/2})/4\pi^2$ \cite{Stanford:2017thb,Maldacena:2016hyu,Kitaev:2017awl,Mertens:2017mtv} so the inverse spectrum is $E(f)=\text{arcsinh}^2\left(2\pi f\right)/4\pi^2$. Using
\begin{align}
    \int_0^{E(\tau)}\d E\, \partial_E f(E)\,e^{-2\beta E}&=\int_0^{E(\tau)^{1/2}}\d w\, \partial_w f(w)\,e^{-2\beta w^2}=\int_0^{E(\tau)^{1/2}}\d w\,\cosh(2\pi w)\,e^{-2\beta w^2}\,,
\end{align}
one computes the exact tau-scaled connected spectral form factor \eqref{expaexpa} as \cite{workinprogressothergroup}
\begin{align}
    \nonumber Z(\beta+\i T,\beta-\i T)_\text{conn}=\,&\frac{e^{\S}}{2^{9/2}\pi^{1/2}\beta^{3/2}}\,e^{\frac{\pi^2}{2\beta}}\,\text{Erf}\left(2^{1/2}\beta^{1/2}\Big(\text{arcsinh}\left(2\pi Te^{-\S}\right)/2\pi-\pi/2\beta\Big)\right)\\&-\frac{e^{\S}}{2^{9/2}\pi^{1/2}\beta^{3/2}}\,e^{\frac{\pi^2}{2\beta}}\,\text{Erf}\left(2^{1/2}\beta^{1/2}\Big(\text{arcsinh}\left(-2\pi Te^{-\S}\right)/2\pi-\pi/2\beta\Big)\right)\,.\label{JTexact}
\end{align}
We recognize the JT disk for $T=\infty$ using Erf$(\pm\infty)=\pm 1$. For large $\beta$ this reduces to the Airy answer since at low energies the spectrum reduces to the Airy spectrum. The Taylor series in $T^{2g+1}$ starts with
\begin{equation}
    Z(\beta+\i T,\beta-\i T)_\text{conn}=\frac{1}{4\pi\beta}\,T-\frac{1}{6\pi}\,T^3\,e^{-2\S}+\frac{3\beta+4\pi^2}{30\pi}\,T^5\, e^{-4\S}-\frac{15\beta^2+60\pi^2\beta+64\pi^2}{315\pi}\,T^7\, e^{-6\S}+\dots\label{218}
\end{equation}
These are the same coefficients found by taking the Weil-Peterson volumes, doing the Laplace transforms with the trumpet partition functions \eqref{genusg}, and then taking the $\t$-scaling limit. The first such volume is
\begin{align} 
V_{1,2}(b_1,b_2) &= \frac{1}{192}(4\pi^2 + b_1^2 + b_2^2)(12\pi^2 + b_1^2 + b_2^2)
\end{align}
The higher genus examples you should keep hidden away in your Mathematica code. One feature worth noting is that for $b_1\gg 1$ and $b_2\gg 1$ these reduce to the Airy volumes \eqref{airyvol}, see section \ref{sect:3.1}. Notice again there is no $T^9 e^{-6\S}$ because of the cancellation \eqref{cancelbasic}.

\begin{figure}[t]
    \centering
    \begin{tikzpicture}[baseline={([yshift=-.5ex]current bounding box.center)}, scale=0.7]
 \pgftext{\includegraphics[scale=1]{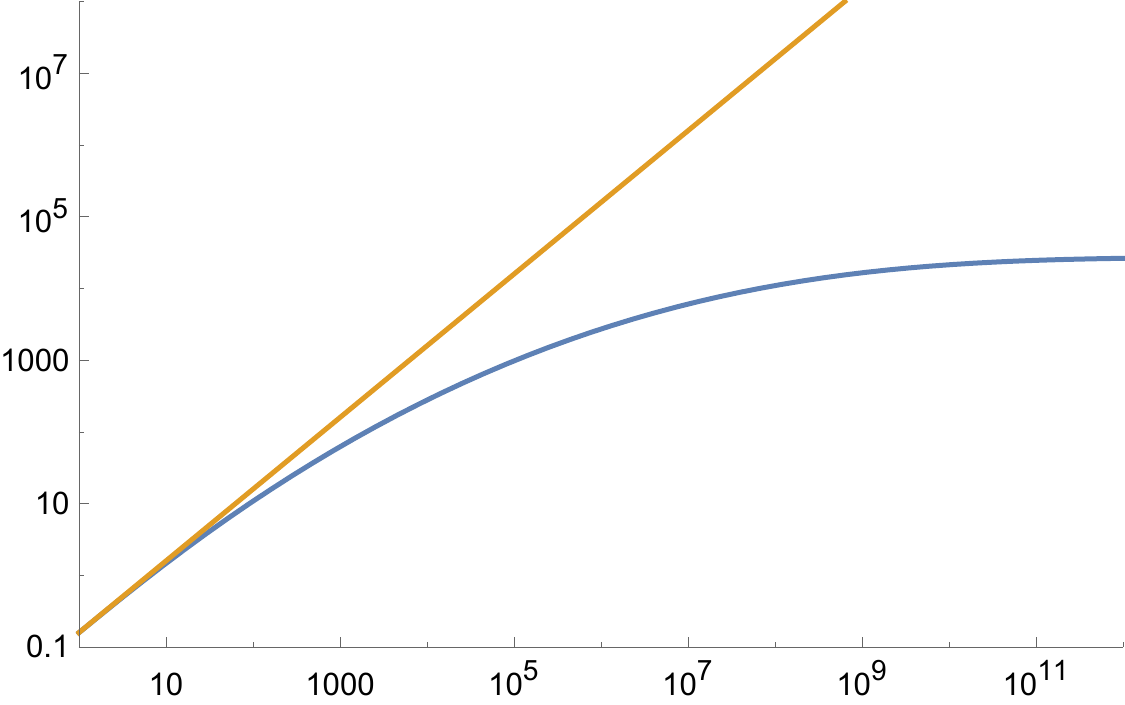}} at (0,0);
     \draw (1.5, 3) node {$0$};
     \draw (4.5, 1.5) node {all genus};
     \draw (5, -2.5) node {$T$};
  \end{tikzpicture}\quad
    \caption{The $\tau$-scaling limit of the sum over genus $g$ wormholes in JT gravity \eqref{JTexact} with $e^{\S}=10$ and $\beta=1/2$. We have not shown individual terms in the series, because it converges only when $Te^{-\S}<1/2\pi$, which is a very short time on this plot. Importantly though, the sum over all genus uniquely gives the exact answer \eqref{JTexact}. The JT plateau is much higher and hence reached much later than the Airy plateau, because of the exponentially larger spectral density.}
    \label{fig:JTplateau}
\end{figure}

In this case plotting the first few terms of \eqref{218} does not immediately converge to the plateau. The reason is that the Taylor expansion of \eqref{JTexact} has a finite radius of convergence $Te^{-\S}<1/2\pi$. This in turn is happens because $E(f)$ has a branchcut for $2\pi\i f<-1$, so the Taylor series is only guaranteed to converge in the disk with radius $1/2\pi$. This same argument applies to the exact answer \eqref{JTexact} because of the arcsinh$(2\pi Te^{-\S})$. But for $Te^{-\S}<1/2\pi$ the series converges uniquely to \eqref{JTexact}, and it's analytic continuation to all $T>0$ is obviously real and smooth, and includes the plateau, see Fig. \ref{fig:JTplateau}.

\subsection*{Other minimal strings}
As final example for the the fact that the cancellations \eqref{genericcancel} occur for all spectra, and that the resulting perturbative genus expansion \eqref{expaexpa} converges to the plateau, we consider
\begin{equation}
    f(E)=2\pi e^{-\S}\rho_0(E)=E^{1/2}+\frac{t_2}{3}E^{3/2}\,.\label{minmal}
\end{equation}
This is an example of the $(2,p)$ minimal string with odd $p$ for which the spectrum is \cite{Saad:2019lba,Maldacena:2004sn}
\begin{equation}
    2\pi e^{-\S}\rho_0(E)=\frac{(2\kappa)^{1/2}}{p}\sinh\bigg(\frac{p}{2}\text{arccosh}\bigg(1+\frac{E}{\kappa}\bigg)\bigg)=\frac{1}{p}E^{1/2}\,U_{p-1}\bigg(\bigg(1+\frac{E}{\kappa}\bigg)^{1/2}\bigg)\,,
\end{equation}
where the even Chebychev polynomials give an expansion in positive integer powers of $1+E/\kappa$. With $p=3$ and $t_2=2/\kappa$ this becomes our example. These models have known dilaton gravity interpretations \cite{Mertens:2020hbs,douglasunpublished,Goel:2020yxl}. Using the first expression one finds (the same follows from Lagrange inversion)
\begin{equation}
    E(f)=\frac{4}{t_2}\sinh^2\bigg(\frac{1}{3}\text{arcsinh}\bigg(\frac{3t_2^{1/2}}{2}f\bigg)\bigg)\,.\label{241}
\end{equation}
The exact connected spectral form factor in the $\tau$-scaling limit becomes
\begin{equation}
     Z(\beta+\i T,\beta-\i T)_\text{conn}=\frac{e^{\S}}{4\pi\beta}\int_0^{E(\tau)^{1/2}}\d w\, \partial_w f(w)\,e^{-2\beta w^2}=\frac{e^{\S}}{8\pi\beta}\int_{-E(\tau)^{1/2}}^{+E(\tau)^{1/2}}\d w\,(1+t_2w^2)\,e^{-2\beta w^2}\,.
\end{equation}
A nice thing about the $(2,p)$  minimal models is that we know $E(f)$ as meromorphic functions, not just as Taylor series so we can plot the exact answer of this integral, See Fig. \ref{fig:minimalplateau}.

The resulting Taylor expansion
\begin{figure}[t]
    \centering
    \begin{tikzpicture}[baseline={([yshift=-.5ex]current bounding box.center)}, scale=0.7]
 \pgftext{\includegraphics[scale=1]{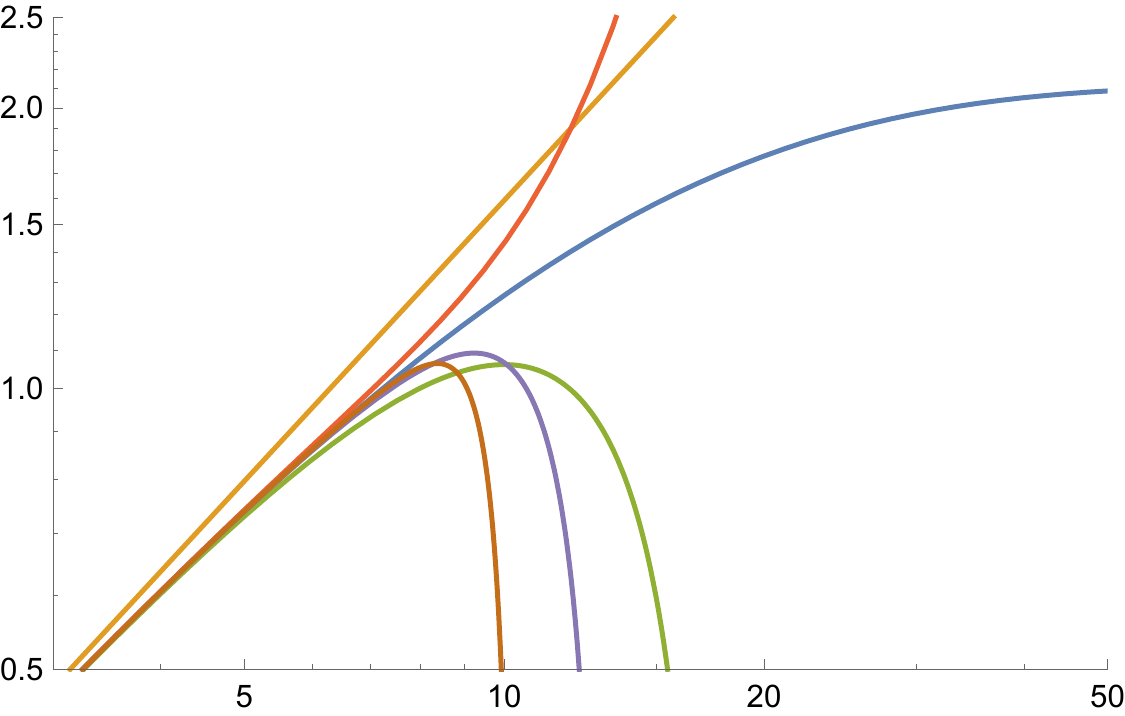}} at (0,0);
     \draw (1.4, 3) node {$0$};
     \draw (-0.3, 3) node {$2$};
     \draw (4.5, 1.9) node {all genus};
     \draw (5, -2.7) node {$T$};
     \draw (-1.2, -2.7) node {$7$};
     \draw (0.5, -2.7) node {$3$};
     \draw (1.4, -2.7) node {$1$};
  \end{tikzpicture}\quad
    \caption{The $\tau$-scaling limit of the sum over wormholes in the $(2,3)$ minimal model \eqref{minmal} with $t_2=1$, $e^{\S}=10$ and $\beta=1/2$. The series expansion converges to the exact answer for $T<2e^{\S}/3t_2^{1/2}=20/3$ and diverges otherwise, but the resummed series has a unique analytic continuation which reproduces the exact answer for all $T>0$.}
    \label{fig:minimalplateau}
\end{figure}
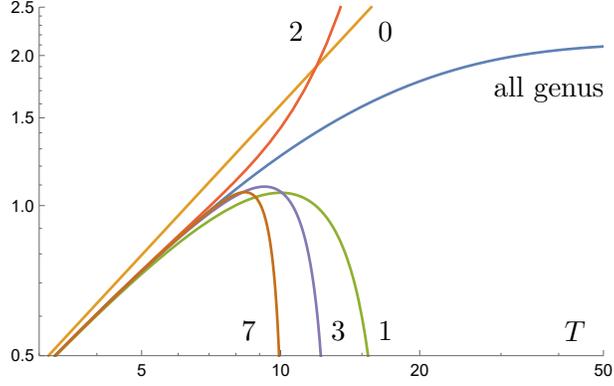
\begin{equation}
    Z(\beta+\i T,\beta-\i T)_\text{conn}=\frac{1}{4\pi\beta}\,T-\frac{1}{6\pi}\,T^3\,e^{-2\S}+\frac{3\beta+2 t_2}{30\pi}\,T^5\, e^{-4\S}-\frac{6\beta^2+12 t_2 \beta+7t_2^3}{126
    \pi}\,T^7\, e^{-6\S}+\dots
\end{equation}
is again reproduced by the relevant volumes, which we emphasize requires $t_2$-dependent cancellations. Much like in the JT example, the series expansion has some finite radius of convergence $Te^{-\S}<2/3t_2^{1/2}$ because $E(f)$ has a branchcut when $\i f<2/3t_2^{1/2}$. For $t_2>0$ though $E(f)$ is smooth so the series will converge to the exact answer. This is the generic case for monotonic spectra, since there will be several point where $\d f/\d E=0$, indicating singular behavior in $E(f)$ and hence a finite radius of convergence. Nevertheless $E(\tau)$ for $\tau>0$ is smooth for monotonic spectra, thus the series uniquely continues to the exact answer. As mentioned below \eqref{expaexpa}, turning on $t_3$ only affects amplitudes for $g\geq 3$, for instance for generic $t_k$ in \eqref{spectrum} the expansion one obtains from either the integral \eqref{expaexpa} or topological recursion is
\begin{equation}
    Z(\beta+\i T,\beta-\i T)_\text{conn}=\dots+\frac{3\beta+2 t_2}{30\pi}\,T^5\, e^{-4\S}-\frac{30\beta^2+60 t_2 \beta+6t_3+35t_2^2}{630
    \pi}\,T^7\, e^{-6\S}+\dots
\end{equation}

\subsection*{Non-monotonic spectra}\label{sect:non-monotonic}
For non-monotonic spectra in general there are several solution $E_1(f)\dots E_n(f)$ for the inverse spectrum and we get other corrections to for instance \eqref{28}. 

\begin{figure}[t]
    \centering
    \begin{tikzpicture}[baseline={([yshift=-.5ex]current bounding box.center)}, scale=0.7]
 \pgftext{\includegraphics[scale=1]{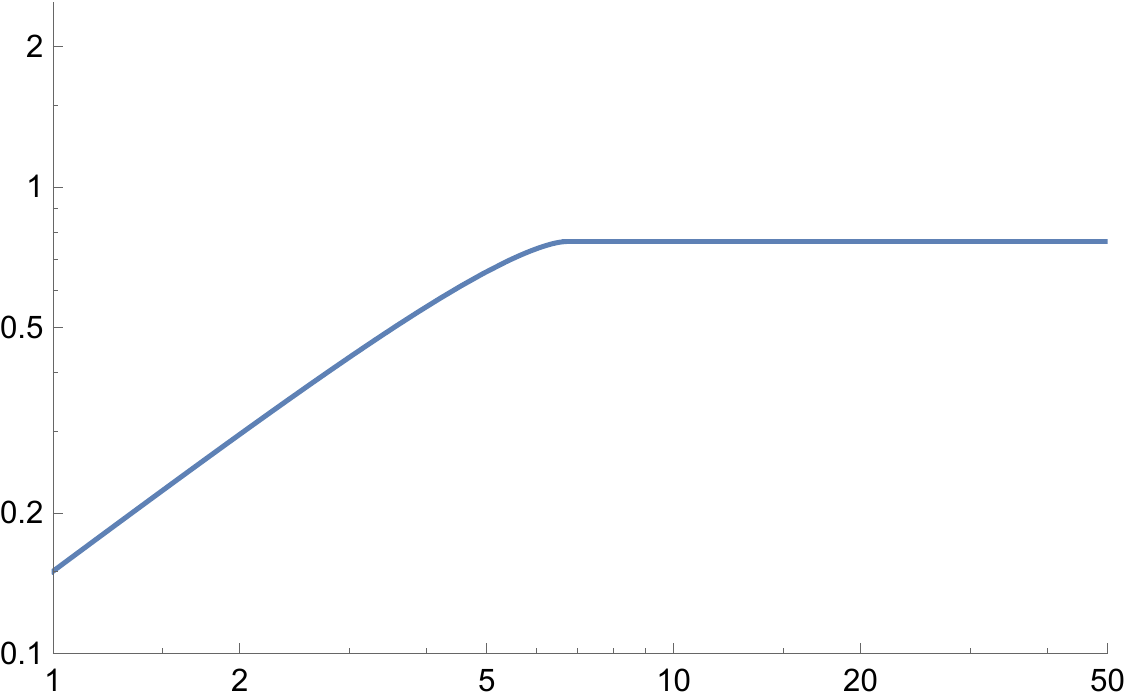}} at (0,0);
     \draw (4.5, 1.7) node {exact};
     \draw (5, -2.6) node {$T$};
  \end{tikzpicture}\quad
    \caption{For non-monotonic spectra the tau-scaled connected spectral form factor has discontinuities at certain points, in the case \eqref{2499} shown here one has a discontinuity in the first derivative. In the case \eqref{252} relevant for black holes the discontinuity is in the second derivative, we used $x_2=1$, $e^{\S}=10$ and $\beta=1/2$.}
    \label{fig:nonmonotonic}
\end{figure}

For concreteness we can first consider the case \eqref{minmal} but with $t_2=-x_2<0$
\begin{equation}
    f(E)=2\pi e^{-\S} \rho_0(E)=E^{1/2}-\frac{x_2}{3}E^{3/2}=\quad \begin{tikzpicture}[baseline={([yshift=-.5ex]current bounding box.center)}, scale=0.7]
 \pgftext{\includegraphics[scale=1]{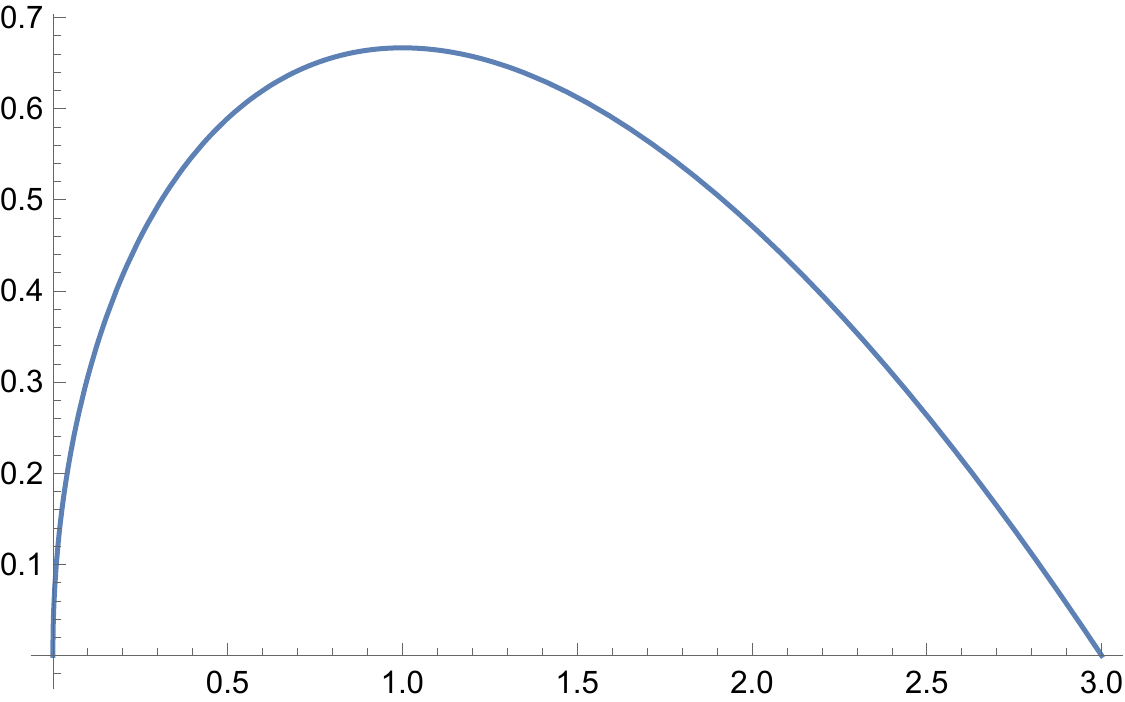}} at (0,0);
 \draw [dashed] (-1.6,3.3) -- (-1.6,0);
     \draw (-1.6, -0.5) node {$E_\text{max}=x_2=1$};
     \draw (4,-2.5) node {$E$};
  \end{tikzpicture}\,.
\end{equation}
This attains a maximum $2/3x_2^{1/2}$ at $E_\text{max}=1/x_2$ and vanishes beyond $E_\text{edge}=3/x_2$. Starting from the first line in \eqref{27} one now obtains
\begin{equation}
    Z(\beta+\i T,\beta-\i T)_\text{conn}=\frac{e^{\S}}{4\pi\beta}\int_0^{\text{min}(\tau.\tau_\text{max})}\d f\,e^{-2\beta E_1(f)}-\frac{e^{\S}}{4\pi\beta}\int_0^{\text{min}(\tau,\tau_\text{max})}\d f\,e^{-2\beta E_2(f)}\,,\label{2499}
\end{equation}
where the two solutions of the inverse spectrum are the two sheets of the analytic continuation of \eqref{241}
\begin{equation}
    E_1(f)=\frac{4}{x_2}\sin^2\bigg(\frac{1}{3}\text{arcsin}\bigg(\frac{3x_2^{1/2}}{2}f\bigg)\bigg)\,,\quad E_2(f)=\frac{4}{x_2}\sin^2\bigg(\frac{\pi}{3}-\frac{1}{3}\text{arcsin}\bigg(\frac{3x_2^{1/2}}{2}f\bigg)\bigg)\,,
\end{equation}
where we restricted the arcsin to the principle branch. The point now is that the gravitational expansion will only reproduce the first integral in \eqref{2499}. This is obvious from the discussion below \eqref{241}, since the genus $g$ Euclidean wormhole amplitudes have trivial analytic continuation to $t_2<0$. The radius of convergence of that expansion in this case is precisely $T=e^{\S}\tau_\text{max}$ for obvious reasons. One noteworthy new feature of \label{249} is a discontinuity in the first derivative at $\tau_\text{max}$, an effect that the Taylor series does not capture, see Fig. \ref{fig:nonmonotonic}

This example was not particularly relevant for black hole physics, where we want a spectrum with Cardy growth at large energies, not one that develops a second edge $E_\text{edge}$, as was the case above. The simplest non-monotonic case for which $\rho_0(E)$ grows for large $E$ has a maximum and a minimum, as in the simple example with $t_2=-5$ and $t_3=-15$
\begin{equation}
     f(E)=2\pi e^{-\S} \rho_0(E)=E^{1/2}-\frac{5}{3}E^{3/2}+E^{5/2}=\quad\begin{tikzpicture}[baseline={([yshift=-.5ex]current bounding box.center)}, scale=0.7]
 \pgftext{\includegraphics[scale=1]{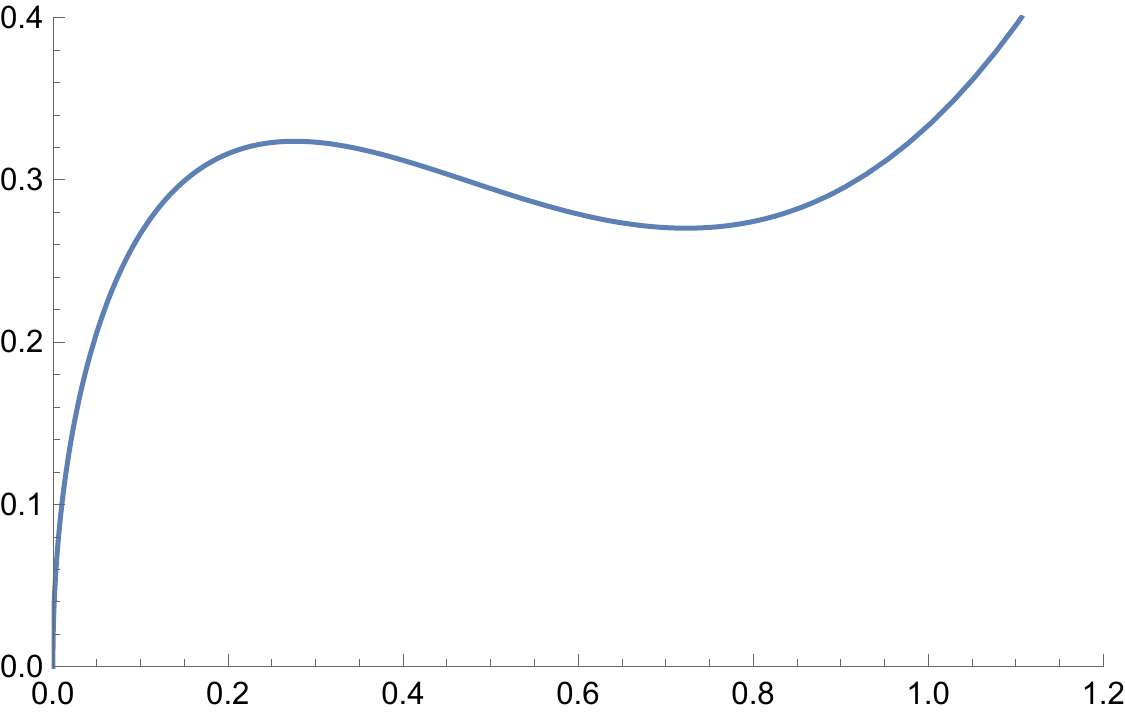}} at (0,0);
 \draw [dashed] (1.2,1.8) -- (1.2,-1.5);
     \draw (1.5, -2) node {$E_\text{min}=0.72\dots$};
     \draw [dashed] (-0.2,1.25) -- (3.2,1.25);
     \draw (5.4, 1.25) node {$\tau_\text{min}=0.27\dots$};
     \draw [dashed] (-3.8,2.2) -- (-0.8,2.20);
     \draw (1.4, 2.2) node {$\tau_\text{max}=0.32\dots$};
     \draw (-4, -1) node {$E_1(f)$};
     \draw (-2, 1.4) node {$E_2(f)$};
     \draw (5, 2.4) node {$E_3(f)$};
     \draw (5,-2.5) node {$E$};
  \end{tikzpicture}\,.
\end{equation}
Let us name the local maximum and minimum of $f$ $\tau_\text{max}$ and $\tau_\text{min}$. In these cases one obviously finds three real solutions $E_1(f)\dots E_3(f)$ for times $\tau_\text{min}<\tau<\tau_\text{max}$, and the exact answer for the tau-scaled connected spectral form factor that one obtains concordantly has three terms
\begin{align}
    Z(\beta+\i T,\beta-\i T)_\text{conn}&=\frac{e^{\S}}{4\pi\beta}\int_0^{\text{min}(\tau,\tau_\text{max})}\d f\,e^{-2\beta E_1(f)}\nonumber\\&\quad -\frac{e^{\S}}{4\pi\beta}\int_{\tau_\text{min}}^{\text{min}(\tau,\tau_\text{max})}\d f\,e^{-2\beta E_2(f)}+\frac{e^{\S}}{4\pi\beta}\int_{\tau_\text{min}}^{\tau}\d f\,e^{-2\beta E_3(f)}\,.\label{252}
\end{align}
Again, one checks case by case via explicit computations that the genus $g$ wormhole amplitude exactly reproduce the first of these integrals. For $T<e^{\S}\tau_\text{min}$ this is the full answer, so the sum over geometries reproduces the ramp-plateau exactly, for those early times. The corrections on the second line are non-perturbative corrections, and the salient feature is again that they come bearing a discontinuity in the second derivative of the tau-scaled spectral form factor at $T=e^{\S}\tau_\text{min}$. One can see this because near the minimum we have an expansion starting with (where $c>0$ an order one number)
\begin{equation}
    E_2(f)=E_\text{min}-c\,(f-\tau_\text{min})^{1/2}+\dots\,,\quad E_3(f)=E_\text{min}+c\,(f-\tau_\text{min})^{1/2}+\dots
\end{equation}
Thus the leading corrections from the second line in \eqref{252} look like (again $C>0$ an order one number)
\begin{align}
Z(\beta+\i T,\beta-\i T)_\text{conn}=\text{perturbative} -C\,(T-T_\text{min})^{3/2}\,\theta(T-T_\text{min})\,e^{-\S/2}+\dots 
\end{align}
This is non-perturbative, since the Heaviside makes these corrections not contribute in a Taylor series around $T=0$. These corrections have no obvious geometric explanation, for instance one would require a fractional Euler character for a dual geometry.


\section{From powers of time to multi-boundary cancellations}\label{app:reverse}
In this appendix we argue that the arrow in \eqref{iff} also works the other way, thus the universal scaling $T^{2g+1}$ for all dilaton gravities implies also \emph{all} of the cancellations of \cite{eynard2021natural}. In a generic background we have
\begin{equation}
    \mathcal{F}(x_1,x_2)_{\g_k}=\sum_{d_1,d_2}x_1^{d_1}x_2^{d_2}\average{\tau_{d_1}\tau_{d_2}}_{\g_k}\,,\label{b1}
\end{equation}
The deformations we are interested in have $\g_0=\g_1=0$. Suppose now we Taylor expand the exponential in \eqref{4.22} in $\g_k$
\begin{align}
    \mathcal{F}_g(x_1,x_2)_{\g_k}&=\sum_{d_1=0}^{3g-1}x_1^{d_1}x_2^{3g-1-d_1}\average{\tau_{d_1}\tau_{3g-1-d_1}}+\g_2\sum_{d_1=0}^{3g-2}x_1^{d_1}x_2^{3g-2-d_1}\average{\tau_{d_1}\tau_{3g-2-d_1}\t_2}+\dots\nn\\&=\sum_{m=0}^{d/2}e_2^m e_1^{d-2m}\sum_{p=0}^m
\average{\tau_p\tau_{d-p}}\frac{d-2p}{d-p-m}\frac{(d-p-m)!}{(m-p)!(d-2m)!}(-1)^{p+m}\\&+\g_2\sum_{m=0}^{d/2-1/2}e_2^m e_1^{d-1-2m}\sum_{p=0}^m
\average{\tau_p\tau_{d-1-p}\tau_2}\frac{d-1-2p}{d-1-p-m}\frac{(d-1-p-m)!}{(m-p)!(d-1-2m)!}(-1)^{p+m}+\dots\nn
\end{align}
In each term in the expansion one can then track the power of $e_2$ and $e_1$ to get more generic cancellation relations, namely the power of $e_2$ may never exceed $g$, because we should have $T^{2g+1}$ for all $\g_k$. Hence
\begin{equation}
    \sum_{p=0}^m
\average{\tau_p\tau_{3g-2-p}\tau_2}\frac{3g-2-2p}{3g-2-p-m}\frac{(d-1-p-m)!}{(m-p)!(d-1-2m)!}(-1)^{p}=0\,,\quad m>g\,.\label{g2cancellations}
\end{equation}
More generally the terms multiplying $\g_k$ for $k\geq 2$ result in the new constraints
\begin{equation}
    \sum_{p=0}^m
\average{\tau_p\tau_{2m_\text{max}-p}\tau_k}\frac{2m_\text{max}-2p}{2m_\text{max}-p-m}\frac{(2m_\text{max}-p-m)!}{(m-p)!(2m_\text{max}-2m)!}(-1)^{p}=0\,,\quad 2m_\text{max}=3g-k\,,\quad m>g\,.\label{glincancellations}
\end{equation}
The term where $m=m_\text{max}$ is most symmetric and becomes
\begin{equation}
    \sum_{p=0}^{3g-k}\average{\tau_p\tau_k\tau_{3g-p-k}}(-1)^p=0\,.
\end{equation}

We want to prove \eqref{iff} by using open-closed duality for generic backgrounds $\g_k$
\begin{equation}
    \mathcal{F}_{g}(x_1,x_2)_{\g_k}=\sum_{m=0}^\infty \frac{1}{m!}\sum_{i_1\dots i_m}\mathcal{F}_g(x_1,x_2,\a_{i_1}\dots \a_{i_m})=\sum_{m_1,m_2=0}^{m_1+2m_2=3g-1}e_1(x)^{m_1}e_2(x)^{m_2}\,C_{g}(m_1,m_2)_{\g_k}\,,\label{duality}
\end{equation}
combined with the fact that the maximal scaling $T^{2g+1}$ implies
\begin{equation}
    C_{g}(m_1,m_2)_{\g_k}=0\,,\quad m_2>g\,.\label{simplerconjecture}
\end{equation}

The idea is to expand each term in the middle expression of \eqref{duality} into powers of $e_1(x)$, $e_2(x)$ and $e_n(\alpha)$, and collecting the terms where all those powers are fixed numbers. Then we demand from \eqref{simplerconjecture} that when the power $q$ of $e_2(x)$ exceeds $g$, that the function multiplying it must vanish. This function will depend on $e_1(x)$ and $e_n(\alpha)$, all of which are independent free parameters, indeed the infinite number of free parameters $\g_k$ maps to $\infty$ free parameters $\a_i$ or equivalently to free parameters $e_n(\alpha)$. Therefore the vanishing of the functions multiplying $e_2(x)^q$ requires the vanishing of the coefficients of each term in the expansion of those functions in powers of $e_1(x)$ and $e_n(\alpha)$, giving a bunch of constraints. Because the terms for each $m$ in \eqref{duality} are homogeneous polynomials of inequivalent degree $3g-1+m$, terms with inequivalent $m$ cannot contribute terms with the same powers of $e_1(x)$, $e_2(x)$ and $e_n(\alpha)$. Therefore the constraint that the coefficients multiplying $e_2(x)^q$ must vanish for $q>2$ in the left of \eqref{duality} implies the same property on each of the terms in the middle of \eqref{duality}, so 
\begin{align}
     C_{g}(m_1,m_2)_{\g_k}=0\,,\quad m_2>g\quad \Rightarrow\quad  \text{maximal power }e_2(x)^g\text{ in }\mathcal{F}_{g}(x_1,x_2,\a_1\dots\a_m)\,.
\end{align}
We also used that each term in the sum over $i_1\dots i_m$ is linearly independent, as there are $\infty$ independent free parameters $\a_i$. We claim that for each fixed number of boundaries $2+n$ we have
\begin{align}
    \text{maximal power }e_2(x)^g\text{ in }\mathcal{F}_{g}(x_1,x_2,\a_1\dots\a_n)\quad\Rightarrow\quad C_g(m_2\dots m_{n+2})=0\,,\quad \sum_{j=2}^{j+n}m_j>g\,.
\end{align}

We now prove this for the three boundary case $n=1$. In general, one can use the following property of elementary symmetric polynomials
\begin{equation}
    e_n(x\oplus \a)=e_n(\a)+e_1(x)e_{n-1}(\a)+e_2(x)e_{n-2}(\a)\,,
\end{equation}
to expand $\mathcal{F}_{g}(x_1,x_2,\a_1\dots\a_n)$ into the building blocks discussed above. For $\mathcal{F}_{g}(x_1,x_2,\a)$ one requires
\begin{equation}
    e_1(x\oplus \a)=e_1(x)+\a\,,\quad e_2(x\oplus \a)=e_1(x)\a+e_2(x)\,,\quad e_3(x\oplus\a)=e_2(x)\a\,.
\end{equation}
Inserting this in the general equation, and absorbing some combinatorial prefactors in $C_g(m_2,m_3)$ for visual purposes, one finds
\begin{align}
    \mathcal{F}_{g}(x_1,x_2,\a)&=\sum_{m_2,m_3=0}^{2m_2+3m_3=3g}\sum_{q=m_3}^{m_3+m_2}\sum_{p=m_2+2m_3-q}^{3g-m_2-m_3-q}\alpha^p e_1(x)^{3g-p-2q}e_2(x)^q\,C_g(m_2,m_3)\,.
\end{align}
The terms with $q>g$ can be rearranged as
\begin{equation}
    \mathcal{F}_{g}(x_1,x_2,\a)\supset \sum_{q=g+1}^{3g/2}\sum_{p=0}^{3g-2q}\alpha^p e_1(x)^{3g-p-2q}e_2(x)^q\sum_{m_3=0}^p \sum_{k=0}^{\text{min}(p-m_3\,,\,3g/2-q-m_3/2)}C_g(q-m_3+k,m_3)\,.
\end{equation}
The constraint that one should impose is that the coefficients of $\a^p e_2(x)^q$ vanish for all $p$ and for $q>g$
\begin{equation}
    \sum_{m_3=0}^p \sum_{k=0}^{\text{min}(p-m_3\,,\,3g/2-q-m_3/2)}C_g(q-m_3+k,m_3)=0\,,\quad q>g\,.
\end{equation}
Clearly having the correct summation ranges is crucial. For $p=0$ there is one term and the constraint becomes
\begin{equation}
    C_g(q,0)=0\,,\quad q>g\,,\label{245}
\end{equation}
which is just the old $n=0$ two boundary constraint. To prove \eqref{iff} it is key to impose the constraints in the correct order, working by induction in $p$. When we now consider $p=1$, the term with $m_3=0$ already vanishes because of \eqref{245}. When $m_3=1$ only $k=0$ contributes and we find the new constraint
\begin{equation}
    C_g(q-1,1)=0\,,\quad q>g\,.
\end{equation}
Since $q>g$ implies $q+k>g$ this means all terms with $m_3=1$ will vanish for arbitrary $p$. This process will continue, at fixed $p$ we can prove that the terms with $m=p$ vanish such that we always only need to deal with one term at $p+1$, this is obvious using induction. Therefore we have obtained the set of constraints
\begin{equation}
    C_g(q-r+k,r)=0\,,\quad g<q\leq 3g/2\,,\quad 0\leq r\leq 3g-2q\,,\quad k\geq 0\,.\label{248}
\end{equation}
Because $2m_2+3m_3\leq 3g$ we have $m_3\leq g$ and therefore one checks that \eqref{248} covers all cases in \eqref{iff} for $n=1$
\begin{equation}
    C_g(m_2,m_3)=0\,,\quad m_2+m_3>g\,,\quad 2m_2+3m_3\leq 3g\,.
\end{equation}
It seems highly plausible that this proof extends to generic $n$, though it requires some combinatorics.

\section{Intersection theory and volumes of moduli space}\label{app:c}

The heart of the paper is concerned with intersection theory and how various intersection numbers are associated with the volumes of moduli space of Riemann surfaces. The intersection theory we will need is a fascinating branch of mathematics that produced many beautiful and deep results. We now dive a bit into these mathematics, we will try to keep it as low brow as possible, combining various insights from \cite{zvonkine2012introduction, hain2008lectures, Dijkgraaf:1991qh, do2008intersection, KontsevichModel}. 

First, let us start with a bit of motivation as to why we would want to study an intersection theory in the context of JT gravity. This has a rather extended history, but we can roughly summarize it as follows. The moduli space of Riemann surfaces of genus $g$ and $n$ punctured is a well motivated and studied object. To understand this space better, one can wonder what happens when one intersects the moduli space with various other subspaces with different co-dimensions. A particularly simple example of this is to consider an intersection such that generically the intersection is just a bunch of points. By counting these points one can learn some more about the geometry of the moduli space. In particular, since subspaces are dual to forms by Poincaré duality, such intersection numbers can even tell us the volume of the moduli space as we will explain below. These volumes are crucial for the understanding of higher-genus corrections to the JT path integral \cite{Saad:2019lba,Eynard:2007fi,Dijkgraaf:2018vnm}. 

Another motivation as to why intersection numbers are interesting is that they have deep relation with a matrix integral, as conjectured by Witten \cite{Witten:1991mn} and proven by Kontsevich in \cite{KontsevichModel}. This matrix integral describes the low energy sector of the matrix model dual to JT gravity and can also be extended to describe the full JT theory. In fact, as explained in section \ref{sect:3.1}, these intersection numbers compute the volumes of moduli space of trivalent ribbon graphs. These ribbons are the propagators of a matrix integral a la t'Hooft \cite{tHooft:1973alw} (the relevant matrix integral is the Kontsevich matrix integral, which is graph dual or color-flavor dual to the standard matrix integral that for instance \cite{Saad:2019lba} use).
\subsection{What is being intersected?}

Let us consider a Riemann surface $\Sigma_{g,n}$ with genus $g$ and $n$ punctures at fixed $x_i$. Such a surface can be obtained by considering a polygon in the hyperbolic disk and identifying the sides in the appropriate way. For instance by identifying the opposite sides of a hyperbolic rectangle one gets a punctured torus. To understand what is being intersected, we will start off with defining various two-forms. These have a concrete definition, resulting in certain limiting cases in simple explicit expressions \eqref{3.6}.

We start with our geometry $\Sigma_{g,n}$ and construct the cotangent space $T_x^*\Sigma_{g,n}$ of one-forms at $x = x_i$. This one-complex-dimensional space depends on the moduli of $\Sigma_{g,n}$ and so we can consider the collection of all cotangent spaces at $x = x_i$ as a function of the moduli of $\Sigma_{g,n}$. This collection is a complex line bundle, which we denote by $\mathcal{L}_i$. We can now choose an element of this bundle and compute its curvature two form, this two-form is usually referred to as the first Chern class of $\mathcal{L}_i$, and denoted $\psi_i = c_1(\mathcal{L}_i)$. Notice that by Poincare duality $\psi_i$ corresponds to codimension two submanifolds of $\mathcal{M}_{g,n}$ and it is these submanifolds whose intersections one can count.

The intersection number of two co-dimension two submanifolds in some manifold can be represented by the wedge product of the two forms Poincare dual to the submanifolds. Likewise, wedge products of $\psi_i$ integrated over $\mathcal{M}_{g,n}$ count the intersection of codimension two submanifolds in $\mathcal{M}_{g,n}$. This means that if we consider integrals of a top-dimensional form
\be \label{psi_intersection}
\int_{\overline{\mathcal{M}}_{g,n}} \psi_1^{d_1} \cdots \psi_n^{d_n}\,,\quad \sum_{i=1}^n d_i = 3g - 3 + n\,,
\ee
the integral describes the intersection of the $n$ submanifolds (or rather cycles) dual to $\psi_i^{d_i}$. These are the intersection numbers discussed in the main text.\footnote{For this to be a bona fide intersection number this should be a topological invariant and for that it is crucial to view the $\psi_i$ as representatives of classes of the second cohomology group, and similarly the codimension two submanifolds should be viewed as codimension two cycles in the relevant homology group. For instance, see Bézout's theorem for the simplest example of intersection numbers of curves in the real projective plane.}

Before proceeding we make two comments. 
\begin{enumerate}
    \item Instead of $\mathcal{M}_{g,n}$ (which is non-compact) one considers the so-called Deligne-Mumford compactification $\overline{\mathcal{M}}_{g,n}$ which includes degenerate Riemann surfaces at the boundary of moduli space. These surfaces have regions where locally the surface looks like two discs with their centers identified. See Fig. \ref{fig:ModuliSpaceStuff} (left) for an example that is counted in the moduli space of punctured tori $\overline{\mathcal{M}}_{1,1}$.
    \begin{figure}
        \centering
        \begin{tikzpicture}[baseline={([yshift=-.5ex]current bounding box.center)}, scale=0.7]
 \pgftext{\includegraphics[scale=1]{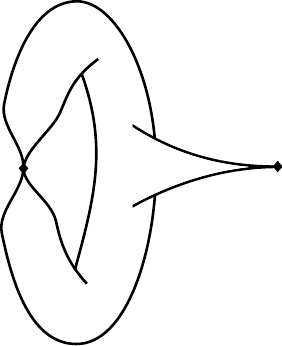}} at (0,0);
  \end{tikzpicture}\qquad\qquad\begin{tikzpicture}[baseline={([yshift=-.5ex]current bounding box.center)}, scale=0.7]
 \pgftext{\includegraphics[scale=1]{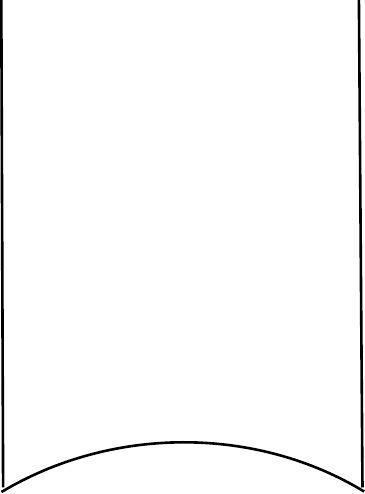}} at (0,0);
     \draw (-2.4, 2) node {$A$};
     \draw (2.4, 2) node {$A'$};
     \draw (-2.4, -2.5) node {$B$};
     \draw (2.4, -2.5) node {$B'$};
     \draw (0, -2.5) node {$C$};
     \draw (0, -1.6) node {$\pi$};
     \draw (-1.4, -1.7) node {$\frac{\pi}{3}$};
     \draw (1.4, -1.7) node {$\frac{\pi}{3}$};
  \end{tikzpicture}
        \caption{Degenerate once punctured torus that is included in the Deligne-Mumford compactification $\overline{\mathcal{M}}_{1,1}$ of $\mathcal{M}_{1,1}$ (left). The orbifold $\mathcal{M}_{1,1}=\mathbb{H}/\text{SL}(2,\mathbb{Z})$ is the interior of the keyhole with extra identification of the half-lines $AB$ and $A'B'$ and of the arcs $CB$ and $CB'$ \cite{zvonkine2012introduction} (right). There is an $\alpha=\pi$ total angle when one circles around $C$ and an $\alpha=2\pi/3$ total angle when one travels around $B=B'$. So, $\mathcal{M}_{1,1}$ is not a smooth manifold, but indeed an orbifold. 
        }
        \label{fig:ModuliSpaceStuff}
    \end{figure}
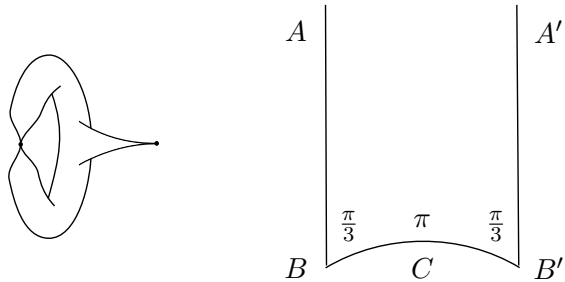
    \item The spaces $\mathcal{M}_{g,n}$ are not manifolds, but orbifolds. We can understand this because we can view $\mathcal{M}_{g,n}$ as Teichmuller space $\mathcal{T}_{g,n}\sim \mathbb{H}^{3g-3+n}$ modulo identifications under the mapping class group, and these identifications can create $\mathbb{Z}_m$ conical singularities in the resulting space $\mathcal{M}_{g,n}$, making them orbifolds. For instance, the keyhole $\mathcal{M}_{1,1}=\mathbb{H}/\text{SL}(2,\mathbb{Z})$ has one $\alpha=\pi$ and $\alpha=2\pi/3$ conical singularity, see Fig. \ref{fig:ModuliSpaceStuff} (right) and \cite{zvonkine2012introduction}.
    
    For orbifolds Chern numbers need not be integers, in general they will be rational numbers. This makes the terminology intersection ``number'', which one thinks of as an integer, odd. In general they are rational numbers for the compact orbifolds $\overline{\mathcal{M}}_{g,n}$, just like their Euler character.
\end{enumerate}

\subsection{Kontsevich and Witten}\label{app:c2}

The objective in intersection theory on the moduli space of Riemann surfaces is to calculate \eqref{psi_intersection}. This is a complicated endeavour, not only because at large genus and $n$ there are many different integrals to consider, but also because the $\psi_i$ itself are hard to find since $\mathcal{M}_{g,n}$ is complicated in general. However in the early nineties the work of Witten and Kontsevich \cite{Witten:1990hr, KontsevichModel} showed that there is a combinatorial solution to the problem involving ribbon graphs. We have already explained how this works in section \ref{sect:intersect}, here we provide an alternative (and historically more chronological) story.

One starts by considering the generating functional of the correlators \eqref{Finter}
\be \label{tau_correlators}
F=\bigg\langle\exp\bigg(\sum_{k=0}^\infty t_k \tau_k\bigg)\bigg\rangle= \sum_{d_i}\prod_{i=0}^\infty\frac{t_i^{d_i}}{d_i!}\bigg\langle \prod_{j=0}^\infty \tau_j^{d_j}\bigg\rangle\,,\quad \braket{\tau_{k_1} \cdots \tau_{k_n}}_g = \int_{\overline{\mathcal{M}}_{g,n}} \psi_1^{d_1} \cdots \psi_n^{d_n}\,.
\ee
Witten conjectured \cite{Witten:1990hr} that this $F$ is the so-called $\tau$-function of the KdV hierarchy, which is a fancy way to say that it satisfies the differential equations \eqref{kdv}. As we explained in the main text, one can use this to get recursion relations \eqref{523} for the intersection numbers, which can be used to compute all intersection numbers given the seeds of the recursion
\be \label{simpletaucorr}
\braket{\tau_0 \tau_0 \tau_0}_{0} = 1,\quad \braket{\tau_1}_1 = \frac{1}{24}.
\ee
The first of these correlators is just a reflection of the fact that the moduli space of the three punctured sphere is a so the correlator can be normalized to one. The second correlator is a statement about the moduli space of the punctured torus and can be understood in various ways, for instance see \cite{zvonkine2012introduction}. 

The proof of this conjecture was given by Kontsevich \cite{KontsevichModel}, who found a way to simply compute the function $F$ and checked that it was indeed a $\tau$-function. His methods start with the story we presented in section \ref{sect:3.1}. He first considered another generating function \eqref{expvolumes}
\begin{equation}
    V_{g,n}(b_1\dots b_n)=\sum_{d_1=0}^\infty\frac{b_1^{2d_1}}{2^{d_1}d_1!}\dots\sum_{d_n=0}^\infty\frac{b_n^{2d_n}}{2^{d_n}d_n!}\average{\tau_{d_1}\dots \tau_{d_n}}_g=\int_{\overline{\mathcal{M}}_{g,n}}\exp\bigg(\frac{1}{2}\sum_{i=1}^n b_i^2\, \psi_i\bigg)\,,\label{c.4}
\end{equation}
and realized that using Penner coordinates $\ell_j$ for $\overline{\mathcal{M}}_{g,n}$, the Chern classes $\psi_i$ can be brought into the simple form \eqref{3.6}, such that (after some combinatorics) this generating functional is seen to compute the volume of moduli space of ribbon graphs with fixed boundary lengths $b_i$ \eqref{3.4}\footnote{The constant $c_\Gamma$ is the flat, but non-unit measure on the moduli space of ribbon graphs. Suppose the lengths constrains read $b_i=A_{i j}\ell_j$ and we form the square matrix $B$ by using the first $n$ columns of $A$, then $c_{\G}= 2^{2g-2+n}/\det(B)$ \cite{do2008intersection}.}
\begin{equation}
     V_{g,n}(b_1\dots b_n)=\sum_{\G_{g,n}}V_{\G_{g,n}} (b_1\dots b_n)\,,\quad V_{\G_{g,n}} (b_1\dots b_n)=\frac{c_\Gamma}{\abs{\text{Aut}(\G)}}\prod_{j=1}^{6g-6+3n}\int_0^\infty \d \ell_j\,\prod_{i=1}^n \delta(b_i-\sum_{j_i}\ell_{j_i})\,.\label{simple?}
\end{equation}
These Penner coordinates $\ell_j$ (which are more open-string like whereas the Fenchel-Nielsen length-twist coordinates are more like a closed string parameterization) originate from constructing the Riemann surface by gluing together flat fixed-width strips of lengths $\ell_i$ at vertices where all the curvature is then localized. These strips naturally form a ribbon graph. The division by the mapping class group, which makes the moduli space so complicated in general, is avoided since there is no constraint on the lengths of the strips \cite{KontsevichModel}, except for the symmetry factor $1/\abs{\text{Aut}(\G)}$ from overcounting cases where say $\ell_1$ and $\ell_2$ label interchangeable edges, so one should only integrate over $\ell_1>\ell_2$.

Equation \eqref{simple?} looks simple. For instance for $V_{1,1}(b)$ there is only one ribbon graph, see Fig. \ref{fig:ribbon}. There are three edges and so three lengths $\ell_i$. The length $b$ of the boundary is given by $2(\ell_1 + \ell_2 + \ell_3) = b$, since one traverses the edges twice. Since $\text{Aut}(\G)=6$ one finds the correct answer
\begin{align}
V_{1,1}(b) &= \frac{1}{\text{Aut}(\G)}\int_0^{\infty} \d\ell_1 \int_0^{\infty} \d \ell_2 \int_0^{\infty} \d \ell_3 \; \delta(\ell_1 + \ell_2 + \ell_3 - b/2)=\frac{1}{6}\int_0^{b/2}\d \ell_1 \int_0^{b/2 - \ell_1} \d \ell_2= \frac{b^2}{48}\,.
\end{align}
Unfortunately life is not so simple. In practice it is hard to find all topologically distinct ribbon graphs for given $g$ and $n$. Moreover, the linear constraints $b_i=A_{i j}\ell_j$ define a polytope, and it is not in general very simple to compute its volume because of these linear constraints. A simple way to get rid of those constraint is integrating over $b_i$. In particular, as Kontsevich noticed, matters simplify drastically when considering a Laplace transform. Doing the $b_i$ integrals first, the volumes simplify tremendously
\begin{equation}
    \prod_{i=1}^n\int_0^\infty \d b_i\,e^{-z_i b_i}\prod_{j=1}^{6g-6+3n}\int_0^\infty \d \ell_j\,\prod_{i=1}^n \delta(b_i-\sum_{j_i}\ell_{j_i})=\prod_{j=1}^{6g-6+3n}\frac{1}{z_{j_1}+z_{j_2}}\,,
\end{equation}
and applying the same Laplace transform to \eqref{c.4} one obtains Kontsevich's famous formula
\be 
\sum_{\G} \frac{2^{2g-2+n}}{\abs{\text{Aut}(\G)}} \prod_{j=1}^{6g-6+3n}\frac{1}{z_{j_1}+z_{j_2}} = \sum_{k_i=0}^\infty\average{\tau_{k_1}\dots \tau_{k_n}}_g\prod_{i=1}^n \frac{(2k_i)!}{2^{k_i}k_i!} \frac{1}{z_i^{2k_i+1}}.
\ee
At this point we have given the $i$'th boundary some flavor $z_i$. Suppose that we have $N$ flavors $z_a$ at our disposal, then it makes sense to consider a sum of the above sum where we sum over all possible flavors for all boundaries, including a symmetry factor $1/(d_j)!$ in this sum when $d_j$ physically indistinguishable boundaries $\tau_j$ appear. If we then also sum over $n$ and $g$, the right hand side becomes precisely $F$, where the KdV times $t_k$ in \eqref{tau_correlators} are related with the flavors $z_a$ as
\begin{equation}
    t_k = \frac{(2k)!}{2^k k!} \sum_{a=0}^N \frac{1}{z_a^{2k+1}}\,.
\end{equation}
The left hand side becomes a sum over all ribbon graphs $\G$ with any number of boundaries, any genus, and any flavor on any of the boundaries, weighed by a symmetry factor $1/\abs{\text{Aut}(\G)}$ that always appears in Feynman diagrams and with each edge contributing a weight $2/(z_{j_1}+z_{j_2})$.

Kontsevich recognized that these are nothing but the Feynman rules of an $N$ by $N$ matrix integral, but counting only fully connected diagrams. Taking into account the fact that fully connected diagrams exponentiate to the full theory, Kontsevich concluded \cite{do2008intersection}
\begin{equation}
    e^F=\frac{\int \d X \exp\bigg(\frac{i}{6} \Tr X^3 - \frac{1}{2} \Tr X^2 Z\bigg)}{\int \d X \exp\bigg(- \frac{1}{2}\Tr X^2 Z\bigg)}\,,\label{c.10}
\end{equation}
where the eigenvalues of $Z$ are $z_a$, and $X$ is integrated over all Hermitian matrices. The propagator is obviously correct, and the $X^3$ interaction generates trivalent ribbon graphs. Not only does this solve intersection theory, Kontsevich also checked explicitly that this expression for $F$ is a $\tau$-function, proving Witten's conjecture \cite{Witten:1990hr}.

\subsection{Relation between the moduli space of ribbon graphs and Riemann surfaces}\label{app:c4}
We gather for clarity some discussion that already appeared in the main text about the relation between the volumes of ribbon graphs and the volumes of Riemann surfaces. Remember that using Gauss-Bonnet with $R=-2$ and $K=0$
\be 
\chi = \frac{1}{4\pi}\int \sqrt{g}R - \frac{1}{2\pi}\int \sqrt{h}K = -\frac{1}{2\pi} \int \sqrt{g} = -\frac{A}{2\pi}\,,
\ee
that we deduced around \eqref{3.3} that Weil-Petersson volumes reduce to volumes of ribbon graphs for large $b$. Weil-Petersson volumes are calculated by invoking the symplectic structure and using the Weil-Petersson two-form $\Omega_{\rm WP}(b_1,\dots,b_n)$ to get a volume form $e^{\Omega_\text{WP}}$
\be 
V_{g,n}(b_1\dotsb_n) = \int_{\overline{\mathcal{M}}_{g,n}} e^{\Omega_{\rm WP}}
\ee
Using \eqref{c.4} this implies
\be 
\Omega_{\rm WP} \to \sum_{i=1}^n \frac{b_i^2}{2} \psi_i
\ee
Terms linear in $b_i$ can be argued from the SL$(2,\mathbb{R})$ BF gauge theory to not exist as hyperbolic holonomies with positive and negative $b_i$ are conjugate, so everything needs to be even in $b_i$. Therefore (see also \cite{Mirzakhani2006WeilPeterssonVA})
\be 
\Omega_{\rm WP} = \w_{\rm WP} + \sum_{i=1}^n \frac{b_i^2}{2} \psi_i\,,
\ee
with $\w_{\rm WP}$ the Weil-Petersson symplectic form on the moduli space of Riemann surfaces with $b_i=0$. 

This relation thus tells us that Weil-Petersson volumes are computed using intersection theory of not only the $\psi$ classes, but together with the Weil-Petersson symplectic form $\omega_{\rm WP}$. It is this extra class that causes the WP volumes to obtain their intricate polynomials structure with coefficients that have various powers of $\pi$. In fact it is customary to write $\w_{\rm WP}$ as 
\be 
\w_{\rm WP} = 2\pi^2 \k,
\ee
with $\k$ a rather standard two-form one an also consider on the moduli space of Riemann surfaces, but whose definition is more complicated and will not be given here, but see \cite{Witten:1990hr}. 

To understand $\w_{\rm WP}$ more physically, note that when it is absent and we just discuss trivalent ribbon graphs, there is a matrix model \eqref{c.10} that also computes these volumes, with 
\be \label{GeneratingFunctionalKontsevich}
F=\bigg\langle\exp\bigg(\sum_{k=0}^\infty t_k \tau_k\bigg)\bigg\rangle= \sum_{d_i}\prod_{i=0}^\infty\frac{t_i^{d_i}}{d_i!}\bigg\langle \prod_{j=0}^\infty \tau_j^{d_j}\bigg\rangle\,,
\ee
and 
\be 
t_k = \frac{(2k)!}{2^k k!} \Tr Z^{-(2k+1)}
\ee
The asymptotic expansion in $Z$ means that when we want to compute the intersection theory relevant for the volumes of ribbon graphs, we need to take derivatives with respect to the $t_k$'s and then set all $t_k$'s to zero. The theory with all the $t_k = 0$ is the unperturbed matrix model and is precisely the double scaled Gaussian matrix model without any determinants inserted. To get more general double scaled models the only thing one then has to do is to simply not set all the $t_k$ to zero afterwards, but set them to specific values $\g_k$. It turns out that one wants to set them to
\be \label{tkJT}
        \g_{n}=-\frac{(-2\pi^2)^{n-1}}{(n-1)!}\,,
\ee
To see how this connects to $\w_{\rm WP}$, notice that the $b$ dependent part will give us the $\tau$ correlators, which are obtained by taking derivatives of $F$ and so the term $e^{\w_{\rm WP}}$ is equivalent to what we set the $t_k$'s to after we are done with taking derivatives. Thus the role of $\w_{\rm WP}$ is to deform the double scaled gaussian matrix model to one with a more general potential, specifically the one relevant for JT gravity. Thus we have the relation \eqref{520}
\be 
\average{(\dots)e^{\w_{\rm WP}}} = \average{(\dots)e^{\sum_k \g_k \tau_k}}\,.
\ee

To summarize what we have learned so far. We have introduced the notion of intersection numbers on the moduli space of Riemann surfaces. By invoking an open string parameterization of the moduli space we showed that the volume of moduli space of ribbon graphs is related to these intersection numbers. Finally by noticing that for large $b_i$ the bordered Riemann surfaces turn into ribbon graphs, we could make a clear relation between Weil-Petersson volumes and volumes of moduli space of ribbon graphs. In particular that deforming to the JT matrix model is equivalent to freezing some of the $t_k$ to some non-trivial value given by \eqref{tkJT}. 
\subsection{Topological recursion for ribbon graphs}\label{app:c5}
Here we discuss a simplification of Mirzakhani's recursion relation for the Weil-Petersson volumes at large $b_i$ and thus becoming a recursion relation for the volumes of moduli space of ribbon graphs. This was also shown in \cite{do2008intersection}. 

The derivation of Mirzakhani's recursion \cite{mirzakhani2007simple} proceeds by considering an experiment where we take an external geodesic with length $b$ and shoot in geodesics orthogonal to it into the surface. We track where those geodesics end up. The crux of the calculation comes down to doing this experiment within a single three holed sphere with boundaries $b$, $b'$ and $b''$ from which one gets the identify
\begin{equation}
    b=T(b,b',b'')+T(b,b'',b')+D(b,b',b'')
\end{equation}
with symmetry
\begin{equation}
    T(b,b',b'')=T(b',b,b'')\,,
\end{equation}
this measures the length along $b$ of geodesics that end up on $b'$ (or the other way around). The function $D(b,b',b'')$ is the length along $b$ of geodesics that self-intersect before reaching a boundary, or that end up back on the $b$ geodesic. The full recursion relation reads \cite{Stanford:2019vob}
\begin{align}
b\, V_{g, n+1}(b, B)=\,&\frac{1}{2} \int_{0}^{\infty} \d b' b,\,\d b'' b''\,D(b,b',b'')\bigg( V_{g-1,n+2}(b',b'',B)+\sum_{\text{stable}} V_{h_1,n_1}(b',B_1) V_{h_2,n_2}(b'',B_2)\bigg) \nonumber\\
&+\sum_{i=1}^{n} \int_{0}^{\infty} \d b' b'\,(b-T(b,b', b_i))\, V_{g,n-1}(b', B/b_i)\,.
\end{align}
One now just has to compute these functions. For ribbon graphs this is borderline trivial. One obtains a piecewise function
\begin{equation}
    T(b_i,b_j,b_k)=\ell_{ij}\,,\quad D(t_i,t_j,t_k)=2\ell_{i i}\,,
\end{equation}
where $\ell_{i j}$ is the length of the ribbon graph shared between boundaries $b_i$ and $b_j$.
There are essentially two ribbon graphs that the three holed sphere can degenerate into, depending on the sizes $b$, $b'$ and $b''$.
\begin{enumerate}
    \item When $b_i>b_j+b_k$ the Riemann surface becomes
    \begin{equation}
        \begin{tikzpicture}[baseline={([yshift=-.5ex]current bounding box.center)}, scale=0.7]
 \pgftext{\includegraphics[scale=1]{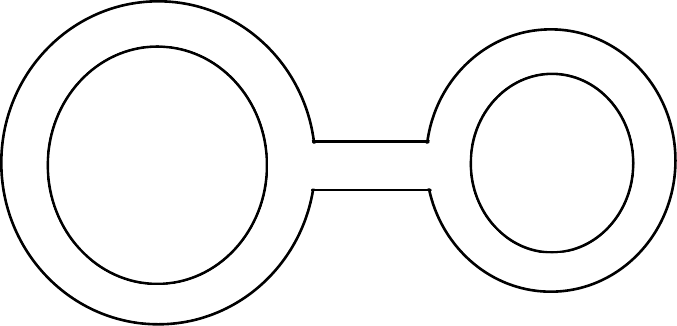}} at (0,0);\draw (0.4, 0.8) node {$b_i$};
 \draw (0.4, -0.8) node {$b_i$};
     \draw (2.2,0) node {$b_j$};
     \draw (-1.8, 0) node {$b_k$};
  \end{tikzpicture}
    \end{equation}
    and one reads off immediately for instance $T(b_i,b_j,b_k)=b_j$ etcetera, because a fraction $b_j$ of the geodesics starting on $b_i$ end up on $b_j$ (or all geodesics starting on $b_j$ end on $b_i$).
    \item When neither of the lengths is bigger than the sum of the two others the Riemann surfaces looks like
    \begin{equation}
        \begin{tikzpicture}[baseline={([yshift=-.5ex]current bounding box.center)}, scale=0.7]
 \pgftext{\includegraphics[scale=1]{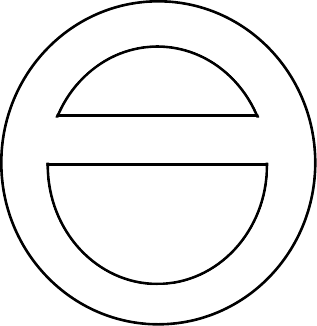}} at (0,0);
     \draw (0, 0.8) node {$b_i$};
     \draw (0, -0.6) node {$b_j$};
     \draw (-2.2, 0) node {$b_k$};
     \draw (2.2, 0) node {$b_k$};
  \end{tikzpicture}
    \end{equation}
    and one can solve for the lengths of the edges, to find for instance $T(b_i,b_j,b_k)=(b_i+b_k-b_k)/2$.
\end{enumerate}
These piecewise linear functions can also be obtained directly as limits of the formulas for finite lengths in \cite{Stanford:2019vob}. One also checks obviously that this reproduces the correct Airy volumes. It is entirely consistent to think of Riemann surfaces with large boundaries as ribbon graphs, for all purposes.

\bibliographystyle{ourbst}
\bibliography{Refs}

\providecommand{\href}[2]{#2}\begingroup\raggedright\begin{thebibliography}{100}

\bibitem{Maldacena:2001kr}
J.~M. Maldacena, ``{Eternal black holes in anti-de Sitter},''
  \href{http://dx.doi.org/10.1088/1126-6708/2003/04/021}{{\em JHEP} {\bfseries
  04} (2003) 021}, \href{http://arxiv.org/abs/hep-th/0106112}{{\ttfamily
  arXiv:hep-th/0106112}}.

\bibitem{Cotler:2016fpe}
J.~S. Cotler, G.~Gur-Ari, M.~Hanada, J.~Polchinski, P.~Saad, S.~H. Shenker,
  D.~Stanford, A.~Streicher, and M.~Tezuka, ``{Black Holes and Random
  Matrices},'' \href{http://dx.doi.org/10.1007/JHEP05(2017)118}{{\em JHEP}
  {\bfseries 05} (2017) 118}, \href{http://arxiv.org/abs/1611.04650}{{\ttfamily
  arXiv:1611.04650 [hep-th]}}.

\bibitem{Saad:2018bqo}
P.~Saad, S.~H. Shenker, and D.~Stanford, ``{A semiclassical ramp in SYK and in
  gravity},'' \href{http://arxiv.org/abs/1806.06840}{{\ttfamily
  arXiv:1806.06840 [hep-th]}}.

\bibitem{Saad:2019pqd}
P.~Saad, ``{Late Time Correlation Functions, Baby Universes, and ETH in JT
  Gravity},'' \href{http://arxiv.org/abs/1910.10311}{{\ttfamily
  arXiv:1910.10311 [hep-th]}}.

\bibitem{Blommaert:2019hjr}
A.~Blommaert, T.~G. Mertens, and H.~Verschelde, ``{Clocks and Rods in
  Jackiw-Teitelboim Quantum Gravity},''
  \href{http://dx.doi.org/10.1007/JHEP09(2019)060}{{\em JHEP} {\bfseries 09}
  (2019) 060}, \href{http://arxiv.org/abs/1902.11194}{{\ttfamily
  arXiv:1902.11194 [hep-th]}}.

\bibitem{Iliesiu:2021ari}
L.~V. Iliesiu, M.~Mezei, and G.~S\'arosi, ``{The volume of the black hole
  interior at late times},'' \href{http://arxiv.org/abs/2107.06286}{{\ttfamily
  arXiv:2107.06286 [hep-th]}}.

\bibitem{Kruthoff:2022voq}
J.~Kruthoff, ``{Higher spin JT gravity and a matrix model dual},''
  \href{http://arxiv.org/abs/2204.09685}{{\ttfamily arXiv:2204.09685
  [hep-th]}}.

\bibitem{Penington:2019kki}
G.~Penington, S.~H. Shenker, D.~Stanford, and Z.~Yang, ``{Replica wormholes and
  the black hole interior},'' \href{http://arxiv.org/abs/1911.11977}{{\ttfamily
  arXiv:1911.11977 [hep-th]}}.

\bibitem{Almheiri:2019qdq}
A.~Almheiri, T.~Hartman, J.~Maldacena, E.~Shaghoulian, and A.~Tajdini,
  ``{Replica Wormholes and the Entropy of Hawking Radiation},''
  \href{http://dx.doi.org/10.1007/JHEP05(2020)013}{{\em JHEP} {\bfseries 05}
  (2020) 013}, \href{http://arxiv.org/abs/1911.12333}{{\ttfamily
  arXiv:1911.12333 [hep-th]}}.

\bibitem{Stanford:2022fdt}
D.~Stanford and Z.~Yang, ``{Firewalls from wormholes},''
  \href{http://arxiv.org/abs/2208.01625}{{\ttfamily arXiv:2208.01625
  [hep-th]}}.

\bibitem{Saad:2019lba}
P.~Saad, S.~H. Shenker, and D.~Stanford, ``{JT gravity as a matrix integral},''
  \href{http://arxiv.org/abs/1903.11115}{{\ttfamily arXiv:1903.11115
  [hep-th]}}.

\bibitem{Blommaert:2019wfy}
A.~Blommaert, T.~G. Mertens, and H.~Verschelde, ``{Eigenbranes in
  Jackiw-Teitelboim gravity},''
  \href{http://arxiv.org/abs/1911.11603}{{\ttfamily arXiv:1911.11603
  [hep-th]}}.

\bibitem{Altland:2020ccq}
A.~Altland and J.~Sonner, ``{Late time physics of holographic quantum chaos},''
  \href{http://arxiv.org/abs/2008.02271}{{\ttfamily arXiv:2008.02271
  [hep-th]}}.

\bibitem{Altland:2022xqx}
A.~Altland, B.~Post, J.~Sonner, J.~van~der Heijden, and E.~Verlinde, ``{Quantum
  chaos in 2D gravity},'' \href{http://arxiv.org/abs/2204.07583}{{\ttfamily
  arXiv:2204.07583 [hep-th]}}.

\bibitem{Haake:1315494}
F.~Haake, S.~Gnutzmann, and M.~Kuś,
  \href{http://dx.doi.org/10.1007/978-3-642-05428-0}{{\em {Quantum Signatures
  of Chaos; 4th ed.}}}
\newblock Springer series in synergetics. Springer, Dordrecht, 2018.

\bibitem{Post:2022dfi}
B.~Post, J.~van~der Heijden, and E.~Verlinde, ``{A universe field theory for JT
  gravity},'' \href{http://dx.doi.org/10.1007/JHEP05(2022)118}{{\em JHEP}
  {\bfseries 05} (2022) 118}, \href{http://arxiv.org/abs/2201.08859}{{\ttfamily
  arXiv:2201.08859 [hep-th]}}.

\bibitem{Anous:2020lka}
T.~Anous, J.~Kruthoff, and R.~Mahajan, ``{Density matrices in quantum
  gravity},'' \href{http://dx.doi.org/10.21468/SciPostPhys.9.4.045}{{\em
  SciPost Phys.} {\bfseries 9} no.~4, (2020) 045},
  \href{http://arxiv.org/abs/2006.17000}{{\ttfamily arXiv:2006.17000
  [hep-th]}}.

\bibitem{Witten:2020wvy}
E.~Witten, ``{Matrix Models and Deformations of JT Gravity},''
  \href{http://dx.doi.org/10.1098/rspa.2020.0582}{{\em Proc. Roy. Soc. Lond. A}
  {\bfseries 476} no.~2244, (2020) 20200582},
  \href{http://arxiv.org/abs/2006.13414}{{\ttfamily arXiv:2006.13414
  [hep-th]}}.

\bibitem{Maxfield:2020ale}
H.~Maxfield and G.~J. Turiaci, ``{The path integral of 3D gravity near
  extremality; or, JT gravity with defects as a matrix integral},''
  \href{http://dx.doi.org/10.1007/JHEP01(2021)118}{{\em JHEP} {\bfseries 01}
  (2021) 118}, \href{http://arxiv.org/abs/2006.11317}{{\ttfamily
  arXiv:2006.11317 [hep-th]}}.

\bibitem{Almheiri:2014cka}
A.~Almheiri and J.~Polchinski, ``{Models of AdS$_{2}$ backreaction and
  holography},'' \href{http://dx.doi.org/10.1007/JHEP11(2015)014}{{\em JHEP}
  {\bfseries 11} (2015) 014}, \href{http://arxiv.org/abs/1402.6334}{{\ttfamily
  arXiv:1402.6334 [hep-th]}}.

\bibitem{Okuyama:2020ncd}
K.~Okuyama and K.~Sakai, ``{Multi-boundary correlators in JT gravity},''
  \href{http://dx.doi.org/10.1007/JHEP08(2020)126}{{\em JHEP} {\bfseries 08}
  (2020) 126}, \href{http://arxiv.org/abs/2004.07555}{{\ttfamily
  arXiv:2004.07555 [hep-th]}}.

\bibitem{workinprogressothergroup}
P.~Saad, D.~Stanford, Z.~Yang, and S.~Yao, ``To appear,''.

\bibitem{Eynard:2007fi}
B.~Eynard and N.~Orantin, ``{Weil-Petersson volume of moduli spaces,
  Mirzakhani's recursion and matrix models},''
  \href{http://arxiv.org/abs/0705.3600}{{\ttfamily arXiv:0705.3600 [math-ph]}}.

\bibitem{Eynard:2007kz}
B.~Eynard and N.~Orantin, ``{Invariants of algebraic curves and topological
  expansion},'' \href{http://dx.doi.org/10.4310/CNTP.2007.v1.n2.a4}{{\em
  Commun. Num. Theor. Phys.} {\bfseries 1} (2007) 347--452},
  \href{http://arxiv.org/abs/math-ph/0702045}{{\ttfamily
  arXiv:math-ph/0702045}}.

\bibitem{Stanford:2019vob}
D.~Stanford and E.~Witten, ``{JT Gravity and the Ensembles of Random Matrix
  Theory},'' \href{http://arxiv.org/abs/1907.03363}{{\ttfamily arXiv:1907.03363
  [hep-th]}}.

\bibitem{Liu:2007ip}
K.~Liu and H.~Xu, ``{New results of intersection numbers on moduli spaces of
  curves},'' \href{http://dx.doi.org/10.1073/pnas.0705910104}{{\em Proc. Nat.
  Acad. Sci.} {\bfseries 104} (2007) 13896--13900},
  \href{http://arxiv.org/abs/0705.3564}{{\ttfamily arXiv:0705.3564 [math.AG]}}.

\bibitem{eynard2021natural}
B.~Eynard, D.~Lewański, and A.~Ooms, ``A natural basis for intersection
  numbers,'' 2021.

\bibitem{workinprogressregensburg}
T.~Weber, F.~Haneder, K.~Richter, and J.-D. Urbina, ``Constraining
  weil-petersson volumes by universal random matrix theory correlations in low
  dimensional quantum gravity,''.

\bibitem{Okuyama:2018gfr}
K.~Okuyama, ``{Eigenvalue instantons in the spectral form factor of random
  matrix model},'' \href{http://dx.doi.org/10.1007/JHEP03(2019)147}{{\em JHEP}
  {\bfseries 03} (2019) 147}, \href{http://arxiv.org/abs/1812.09469}{{\ttfamily
  arXiv:1812.09469 [hep-th]}}.

\bibitem{Mertens:2020hbs}
T.~G. Mertens and G.~J. Turiaci, ``{Liouville quantum gravity -- holography, JT
  and matrices},'' \href{http://dx.doi.org/10.1007/JHEP01(2021)073}{{\em JHEP}
  {\bfseries 01} (2021) 073}, \href{http://arxiv.org/abs/2006.07072}{{\ttfamily
  arXiv:2006.07072 [hep-th]}}.

\bibitem{mehta2004random}
M.~L. Mehta, {\em Random matrices}.
\newblock Elsevier, 2004.

\bibitem{efetov1983supersymmetry}
K.~Efetov, ``Supersymmetry and theory of disordered metals,'' {\em advances in
  Physics} {\bfseries 32} no.~1, (1983) 53--127.

\bibitem{Belin:2021ibv}
A.~Belin, J.~de~Boer, P.~Nayak, and J.~Sonner, ``{Generalized Spectral Form
  Factors and the Statistics of Heavy Operators},''
  \href{http://arxiv.org/abs/2111.06373}{{\ttfamily arXiv:2111.06373
  [hep-th]}}.

\bibitem{Blommaert:2021fob}
A.~Blommaert, L.~V. Iliesiu, and J.~Kruthoff, ``{Gravity factorized},''
  \href{http://arxiv.org/abs/2111.07863}{{\ttfamily arXiv:2111.07863
  [hep-th]}}.

\bibitem{brezin1993exactly}
E.~Brezin and V.~Kazakov, ``Exactly solvable field theories of closed
  strings,'' in {\em The Large N Expansion In Quantum Field Theory And
  Statistical Physics: From Spin Systems to 2-Dimensional Gravity},
  pp.~711--717.
\newblock World Scientific, 1993.

\bibitem{douglas1990strings}
M.~R. Douglas and S.~H. Shenker, ``Strings in less than one dimension,'' {\em
  Nuclear Physics B} {\bfseries 335} no.~3, (1990) 635--654.

\bibitem{Gross1989nonperturbative}
D.~J. Gross and A.~A. Migdal, ``{Nonperturbative Two-Dimensional Quantum
  Gravity},'' \href{http://dx.doi.org/10.1103/PhysRevLett.64.127}{{\em Phys.
  Rev. Lett.} {\bfseries 64} (1990) 127}.

\bibitem{workinprogress}
A.~Blommaert, J.~Kruthoff, and S.~Yao, ``To appear,''.

\bibitem{Eynard:2004mh}
B.~Eynard, ``{Topological expansion for the 1-Hermitian matrix model
  correlation functions},''
  \href{http://dx.doi.org/10.1088/1126-6708/2004/11/031}{{\em JHEP} {\bfseries
  11} (2004) 031}, \href{http://arxiv.org/abs/hep-th/0407261}{{\ttfamily
  arXiv:hep-th/0407261}}.

\bibitem{van2008intersection}
N.~N. Van~Do, ``Intersection theory on moduli spaces of curves via hyperbolic
  geometry,''.

\bibitem{zograf2008large}
P.~Zograf, ``On the large genus asymptotics of weil-petersson volumes,'' {\em
  arXiv preprint arXiv:0812.0544} (2008) .

\bibitem{mirzakhani2015towards}
M.~Mirzakhani and P.~Zograf, ``Towards large genus asymptotics of intersection
  numbers on moduli spaces of curves,'' {\em Geometric and Functional Analysis}
  {\bfseries 25} no.~4, (2015) 1258--1289.

\bibitem{Liu13896}
K.~Liu and H.~Xu, ``New results of intersection numbers on moduli spaces of
  curves,'' \href{https://www.pnas.org/content/104/35/13896}{{\em Proceedings
  of the National Academy of Sciences} {\bfseries 104} no.~35, (2007)
  13896--13900},
  \href{http://arxiv.org/abs/https://www.pnas.org/content/104/35/13896.full.pdf}{{\ttfamily
  https://www.pnas.org/content/104/35/13896.full.pdf}}.

\bibitem{Blommaert:2021gha}
A.~Blommaert and J.~Kruthoff, ``{Gravity without averaging},''
  \href{http://dx.doi.org/10.21468/SciPostPhys.12.2.073}{{\em SciPost Phys.}
  {\bfseries 12} no.~2, (2022) 073},
  \href{http://arxiv.org/abs/2107.02178}{{\ttfamily arXiv:2107.02178
  [hep-th]}}.

\bibitem{Blommaert:2022ucs}
A.~Blommaert, L.~V. Iliesiu, and J.~Kruthoff, ``{Alpha states demystified
  \textemdash{} towards microscopic models of AdS$_{2}$ holography},''
  \href{http://dx.doi.org/10.1007/JHEP08(2022)071}{{\em JHEP} {\bfseries 08}
  (2022) 071}, \href{http://arxiv.org/abs/2203.07384}{{\ttfamily
  arXiv:2203.07384 [hep-th]}}.

\bibitem{Witten:1990hr}
E.~Witten, ``{Two-dimensional gravity and intersection theory on moduli
  space},'' \href{http://dx.doi.org/10.4310/SDG.1990.v1.n1.a5}{{\em Surveys
  Diff. Geom.} {\bfseries 1} (1991) 243--310}.

\bibitem{KontsevichModel}
M.~Kontsevich, ``{Intersection theory on the moduli space of curves and the
  matrix Airy function},'' \href{https://doi.org/}{{\em Communications in
  Mathematical Physics} {\bfseries 147} no.~1, (1992) 1 -- 23}.

\bibitem{Dijkgraaf:2018vnm}
R.~Dijkgraaf and E.~Witten, ``{Developments in Topological Gravity},''
  \href{http://dx.doi.org/10.1142/S0217751X18300296}{{\em Int. J. Mod. Phys. A}
  {\bfseries 33} no.~30, (2018) 1830029},
  \href{http://arxiv.org/abs/1804.03275}{{\ttfamily arXiv:1804.03275
  [hep-th]}}.

\bibitem{Okuyama:2019xbv}
K.~Okuyama and K.~Sakai, ``{JT gravity, KdV equations and macroscopic loop
  operators},'' \href{http://dx.doi.org/10.1007/JHEP01(2020)156}{{\em JHEP}
  {\bfseries 01} (2020) 156}, \href{http://arxiv.org/abs/1911.01659}{{\ttfamily
  arXiv:1911.01659 [hep-th]}}.

\bibitem{penner1987decorated}
R.~C. Penner, ``The decorated teichm{\"u}ller space of punctured surfaces,''
  {\em Communications in Mathematical Physics} {\bfseries 113} no.~2, (1987)
  299--339.

\bibitem{harer1988cohomology}
J.~L. Harer, ``The cohomology of the moduli space of curves,'' in {\em Theory
  of moduli}, pp.~138--221.
\newblock Springer, 1988.

\bibitem{strebel1984quadratic}
K.~Strebel, ``Quadratic differentials,'' in {\em Quadratic Differentials},
  pp.~16--26.
\newblock Springer, 1984.

\bibitem{do2008intersection}
N.~Do, {\em Intersection theory on moduli spaces of curves via hyperbolic
  geometry}.
\newblock PhD thesis, 2008.

\bibitem{mirzakhani2007simple}
M.~Mirzakhani, ``Simple geodesics and weil-petersson volumes of moduli spaces
  of bordered riemann surfaces,'' {\em Inventiones mathematicae} {\bfseries
  167} no.~1, (2007) 179--222.

\bibitem{faber2000logarithmic}
C.~Faber and R.~Pandharipande, ``Logarithmic series and hodge integrals in the
  tautological ring. with an appendix by don zagier.,'' {\em Michigan
  Mathematical Journal} {\bfseries 48} no.~1, (2000) 215--252.

\bibitem{mulase2006mirzakhani}
M.~Mulase and B.~Safnuk, ``Mirzakhani's recursion relations, virasoro
  constraints and the kdv hierarchy,'' {\em arXiv preprint math/0601194} (2006)
  .

\bibitem{dijkgraaf1991topological}
R.~Dijkgraaf, H.~Verlinde, and E.~Verlinde, ``Topological strings in $d < 1$,''
  {\em Nuclear Physics B} {\bfseries 352} no.~1, (1991) 59--86.

\bibitem{Dijkgraaf:1991qh}
R.~Dijkgraaf, ``{Intersection theory, integrable hierarchies and topological
  field theory},'' {\em NATO Sci. Ser. B} {\bfseries 295} (1992) 95--158,
  \href{http://arxiv.org/abs/hep-th/9201003}{{\ttfamily arXiv:hep-th/9201003}}.

\bibitem{fukuma1993continuum}
M.~Fukuma, H.~Kawai, and R.~Nakayama, ``Continuum schwinger-dyson equations and
  universal structures in two-dimensional quantum gravity,'' in {\em The Large
  N Expansion In Quantum Field Theory And Statistical Physics: From Spin
  Systems to 2-Dimensional Gravity}, pp.~820--841.
\newblock World Scientific, 1993.

\bibitem{Witten:1991mn}
E.~Witten, ``{On the Kontsevich model and other models of two-dimensional
  gravity},'' in {\em {International Conference on Differential Geometric
  Methods in Theoretical Physics}}, pp.~176--216.
\newblock 6, 1991.

\bibitem{Dijkgraaf1991loop}
R.~Dijkgraaf, H.~Verlinde, and E.~Verlinde, ``Loop equations and virasoro
  constraints in non-perturbative two-dimensional quantum gravity,'' {\em
  Nuclear Physics B} {\bfseries 348} no.~3, (1991) 435--456.

\bibitem{okounkov2002generating}
A.~Okounkov, ``Generating functions for intersection numbers on moduli spaces
  of curves,'' {\em International Mathematics Research Notices} {\bfseries
  2002} no.~18, (2002) 933--957.

\bibitem{liu2011n}
K.~Liu and H.~Xu, ``The n-point functions for intersection numbers on moduli
  spaces of curves,'' {\em Advances in Theoretical and Mathematical Physics}
  {\bfseries 15} no.~5, (2011) 1201--1236.

\bibitem{Eynard_2009}
B.~Eynard and N.~Orantin, ``Topological recursion in enumerative geometry and
  random matrices,'' \href{https://doi.org/10.1088/1751-8113/42/29/293001}{{\em
  Journal of Physics A: Mathematical and Theoretical} {\bfseries 42} no.~29,
  (Jul, 2009) 293001}.

\bibitem{Johnson:2019eik}
C.~V. Johnson, ``{Nonperturbative Jackiw-Teitelboim gravity},''
  \href{http://dx.doi.org/10.1103/PhysRevD.101.106023}{{\em Phys. Rev. D}
  {\bfseries 101} no.~10, (2020) 106023},
  \href{http://arxiv.org/abs/1912.03637}{{\ttfamily arXiv:1912.03637
  [hep-th]}}.

\bibitem{Johnson:2020exp}
C.~V. Johnson, ``{Explorations of nonperturbative Jackiw-Teitelboim gravity and
  supergravity},'' \href{http://dx.doi.org/10.1103/PhysRevD.103.046013}{{\em
  Phys. Rev. D} {\bfseries 103} no.~4, (2021) 046013},
  \href{http://arxiv.org/abs/2006.10959}{{\ttfamily arXiv:2006.10959
  [hep-th]}}.

\bibitem{Johnson:2020heh}
C.~V. Johnson, ``{Jackiw-Teitelboim supergravity, minimal strings, and matrix
  models},'' \href{http://dx.doi.org/10.1103/PhysRevD.103.046012}{{\em Phys.
  Rev. D} {\bfseries 103} no.~4, (2021) 046012},
  \href{http://arxiv.org/abs/2005.01893}{{\ttfamily arXiv:2005.01893
  [hep-th]}}.

\bibitem{eynard2011recursion}
B.~Eynard, ``Recursion between mumford volumes of moduli spaces,'' in {\em
  Annales Henri Poincar{\'e}}, vol.~12, pp.~1431--1447, Springer.
\newblock 2011.

\bibitem{Gaiotto:2003yb}
D.~Gaiotto and L.~Rastelli, ``{A Paradigm of open / closed duality: Liouville
  D-branes and the Kontsevich model},''
  \href{http://dx.doi.org/10.1088/1126-6708/2005/07/053}{{\em JHEP} {\bfseries
  07} (2005) 053}, \href{http://arxiv.org/abs/hep-th/0312196}{{\ttfamily
  arXiv:hep-th/0312196}}.

\bibitem{Mertens:2019tcm}
T.~G. Mertens and G.~J. Turiaci, ``{Defects in Jackiw-Teitelboim Quantum
  Gravity},'' \href{http://dx.doi.org/10.1007/JHEP08(2019)127}{{\em JHEP}
  {\bfseries 08} (2019) 127}, \href{http://arxiv.org/abs/1904.05228}{{\ttfamily
  arXiv:1904.05228 [hep-th]}}.

\bibitem{Maldacena:2004sn}
J.~M. Maldacena, G.~W. Moore, N.~Seiberg, and D.~Shih, ``{Exact vs.
  semiclassical target space of the minimal string},''
  \href{http://dx.doi.org/10.1088/1126-6708/2004/10/020}{{\em JHEP} {\bfseries
  10} (2004) 020}, \href{http://arxiv.org/abs/hep-th/0408039}{{\ttfamily
  arXiv:hep-th/0408039}}.

\bibitem{DOUGLAS1990176}
M.~R. Douglas, ``Strings in less than one dimension and the generalized kdv
  hierarchies,''
  \href{https://www.sciencedirect.com/science/article/pii/037026939091716O}{{\em
  Physics Letters B} {\bfseries 238} no.~2, (1990) 176--180}.

\bibitem{tan2006generalizations}
S.~P. Tan, Y.~L. Wong, and Y.~Zhang, ``Generalizations of mcshane's identity to
  hyperbolic cone-surfaces,'' {\em Journal of Differential Geometry} {\bfseries
  72} no.~1, (2006) 73--112.

\bibitem{do2009weil}
N.~Do and P.~Norbury, ``Weil--petersson volumes and cone surfaces,'' {\em
  Geometriae Dedicata} {\bfseries 141} no.~1, (2009) 93--107.

\bibitem{do2011moduli}
N.~Do, ``Moduli spaces of hyperbolic surfaces and their weil-petersson
  volumes,'' {\em arXiv preprint arXiv:1103.4674} (2011) .

\bibitem{Cotler:2019nbi}
J.~Cotler, K.~Jensen, and A.~Maloney, ``{Low-dimensional de Sitter quantum
  gravity},'' \href{http://dx.doi.org/10.1007/JHEP06(2020)048}{{\em JHEP}
  {\bfseries 06} (2020) 048}, \href{http://arxiv.org/abs/1905.03780}{{\ttfamily
  arXiv:1905.03780 [hep-th]}}.

\bibitem{Turiaci:2020fjj}
G.~J. Turiaci, M.~Usatyuk, and W.~W. Weng, ``{Dilaton-gravity, deformations of
  the minimal string, and matrix models},''
  \href{http://arxiv.org/abs/2011.06038}{{\ttfamily arXiv:2011.06038
  [hep-th]}}.

\bibitem{Louko:1995jw}
J.~Louko and R.~D. Sorkin, ``{Complex actions in two-dimensional topology
  change},'' \href{http://dx.doi.org/10.1088/0264-9381/14/1/018}{{\em Class.
  Quant. Grav.} {\bfseries 14} (1997) 179--204},
  \href{http://arxiv.org/abs/gr-qc/9511023}{{\ttfamily arXiv:gr-qc/9511023}}.

\bibitem{Witten:2020ert}
E.~Witten, ``{Deformations of JT Gravity and Phase Transitions},''
  \href{http://arxiv.org/abs/2006.03494}{{\ttfamily arXiv:2006.03494
  [hep-th]}}.

\bibitem{joaquinlorenztoap}
L.~Eberhardt and G.~J. Turiaci, ``To appear,''.

\bibitem{Goel:2020yxl}
A.~Goel, L.~V. Iliesiu, J.~Kruthoff, and Z.~Yang, ``{Classifying boundary
  conditions in JT gravity: from energy-branes to $\alpha$-branes},''
  \href{http://dx.doi.org/10.1007/JHEP04(2021)069}{{\em JHEP} {\bfseries 04}
  (2021) 069}, \href{http://arxiv.org/abs/2010.12592}{{\ttfamily
  arXiv:2010.12592 [hep-th]}}.

\bibitem{Mertens:2020pfe}
T.~G. Mertens, ``{Degenerate operators in JT and Liouville (super)gravity},''
  \href{http://dx.doi.org/10.1007/JHEP04(2021)245}{{\em JHEP} {\bfseries 04}
  (2021) 245}, \href{http://arxiv.org/abs/2007.00998}{{\ttfamily
  arXiv:2007.00998 [hep-th]}}.

\bibitem{Yang:2018gdb}
Z.~Yang, ``{The Quantum Gravity Dynamics of Near Extremal Black Holes},''
  \href{http://dx.doi.org/10.1007/JHEP05(2019)205}{{\em JHEP} {\bfseries 05}
  (2019) 205}, \href{http://arxiv.org/abs/1809.08647}{{\ttfamily
  arXiv:1809.08647 [hep-th]}}.

\bibitem{Fateev:2000ik}
V.~Fateev, A.~B. Zamolodchikov, and A.~B. Zamolodchikov, ``{Boundary Liouville
  field theory. 1. Boundary state and boundary two point function},''
  \href{http://arxiv.org/abs/hep-th/0001012}{{\ttfamily arXiv:hep-th/0001012}}.

\bibitem{Ponsot:2001ng}
B.~Ponsot and J.~Teschner, ``{Boundary Liouville field theory: Boundary three
  point function},''
  \href{http://dx.doi.org/10.1016/S0550-3213(01)00596-X}{{\em Nucl. Phys. B}
  {\bfseries 622} (2002) 309--327},
  \href{http://arxiv.org/abs/hep-th/0110244}{{\ttfamily arXiv:hep-th/0110244}}.

\bibitem{Hosomichi:2008th}
K.~Hosomichi, ``{Minimal Open Strings},''
  \href{http://dx.doi.org/10.1088/1126-6708/2008/06/029}{{\em JHEP} {\bfseries
  06} (2008) 029}, \href{http://arxiv.org/abs/0804.4721}{{\ttfamily
  arXiv:0804.4721 [hep-th]}}.

\bibitem{Kostov:2002uq}
I.~K. Kostov, ``{Boundary correlators in 2-D quantum gravity: Liouville versus
  discrete approach},''
  \href{http://dx.doi.org/10.1016/S0550-3213(03)00147-0}{{\em Nucl. Phys. B}
  {\bfseries 658} (2003) 397--416},
  \href{http://arxiv.org/abs/hep-th/0212194}{{\ttfamily arXiv:hep-th/0212194}}.

\bibitem{Okuyama:2021eju}
K.~Okuyama and K.~Sakai, ``{FZZT branes in JT gravity and topological
  gravity},'' \href{http://arxiv.org/abs/2108.03876}{{\ttfamily
  arXiv:2108.03876 [hep-th]}}.

\bibitem{Teschner:2000md}
J.~Teschner, ``{Remarks on Liouville theory with boundary},''
  \href{http://dx.doi.org/10.22323/1.006.0041}{{\em PoS} {\bfseries tmr2000}
  (2000) 041}, \href{http://arxiv.org/abs/hep-th/0009138}{{\ttfamily
  arXiv:hep-th/0009138}}.

\bibitem{Blommaert:2021etf}
A.~Blommaert and M.~Usatyuk, ``{Microstructure in matrix elements},''
  \href{http://arxiv.org/abs/2108.02210}{{\ttfamily arXiv:2108.02210
  [hep-th]}}.

\bibitem{workinprogressmisha}
M.~Usatyuk, ``To appear,''.

\bibitem{Marolf:2022ybi}
D.~Marolf, ``{Gravitational thermodynamics without the conformal factor
  problem: Partition functions and Euclidean saddles from Lorentzian Path
  Integrals},'' \href{http://arxiv.org/abs/2203.07421}{{\ttfamily
  arXiv:2203.07421 [hep-th]}}.

\bibitem{Stanford:2017thb}
D.~Stanford and E.~Witten, ``{Fermionic Localization of the Schwarzian
  Theory},'' \href{http://dx.doi.org/10.1007/JHEP10(2017)008}{{\em JHEP}
  {\bfseries 10} (2017) 008}, \href{http://arxiv.org/abs/1703.04612}{{\ttfamily
  arXiv:1703.04612 [hep-th]}}.

\bibitem{Maldacena:2016hyu}
J.~Maldacena and D.~Stanford, ``{Remarks on the Sachdev-Ye-Kitaev model},''
  \href{http://dx.doi.org/10.1103/PhysRevD.94.106002}{{\em Phys. Rev. D}
  {\bfseries 94} no.~10, (2016) 106002},
  \href{http://arxiv.org/abs/1604.07818}{{\ttfamily arXiv:1604.07818
  [hep-th]}}.

\bibitem{Kitaev:2017awl}
A.~Kitaev and S.~J. Suh, ``{The soft mode in the Sachdev-Ye-Kitaev model and
  its gravity dual},'' \href{http://dx.doi.org/10.1007/JHEP05(2018)183}{{\em
  JHEP} {\bfseries 05} (2018) 183},
  \href{http://arxiv.org/abs/1711.08467}{{\ttfamily arXiv:1711.08467
  [hep-th]}}.

\bibitem{Mertens:2017mtv}
T.~G. Mertens, G.~J. Turiaci, and H.~L. Verlinde, ``{Solving the Schwarzian via
  the Conformal Bootstrap},''
  \href{http://dx.doi.org/10.1007/JHEP08(2017)136}{{\em JHEP} {\bfseries 08}
  (2017) 136}, \href{http://arxiv.org/abs/1705.08408}{{\ttfamily
  arXiv:1705.08408 [hep-th]}}.

\bibitem{douglasunpublished}
D.~Stanford and N.~Seiberg, ``Unpublished,''.

\bibitem{zvonkine2012introduction}
D.~Zvonkine, ``An introduction to moduli spaces of curves and their
  intersection theory,''.

\bibitem{hain2008lectures}
R.~Hain, ``Lectures on moduli spaces of elliptic curves,'' {\em arXiv preprint
  arXiv:0812.1803} (2008) .

\bibitem{tHooft:1973alw}
G.~'t~Hooft, ``{A Planar Diagram Theory for Strong Interactions},''
  \href{http://dx.doi.org/10.1016/0550-3213(74)90154-0}{{\em Nucl. Phys. B}
  {\bfseries 72} (1974) 461}.

\bibitem{Mirzakhani2006WeilPeterssonVA}
M.~Mirzakhani, ``Weil-petersson volumes and intersection theory on the moduli
  space of curves,'' {\em Journal of the American Mathematical Society}
  {\bfseries 20} (2006) 1--23.

\end{thebibliography}\endgroup

\end{document}